\documentclass{emulateapj}
\usepackage{graphicx}
\usepackage{multirow}
\usepackage{textcomp}
\usepackage{epsfig}
\usepackage{amsmath}
\usepackage{amssymb}
\usepackage{amsthm}
\usepackage{xy}

\newcommand{\CXO}{{\it Chandra}}
\newcommand{\chandra}{{\it Chandra}}

\newcommand{\XMM}{{\em XMM--Newton}}
\newcommand{\xmm}{{\em XMM--Newton}}

\newcommand{\rxte}{{\em RXTE}}
\newcommand{\swift}{{\em Swift}}
\newcommand{\fermi}{{\em Fermi}}

\newcommand{\bc}{\begin{center}}
\newcommand{\ec}{\end{center}}
\def\ltsima{$\; \buildrel < \over \sim \;$}
\def\lsim{\lower.5ex\hbox{\ltsima}}
\def\loe{\lower.5ex\hbox{\ltsima}}
\def\gtsima{$\; \buildrel > \over \sim \;$}
\def\gsim{\lower.5ex\hbox{\gtsima}}
\def\goe{\lower.5ex\hbox{\gtsima}}

\def\ltsima{$\; \buildrel < \over \sim \;$}
\def\lsim{\lower.5ex\hbox{\ltsima}}
\def\loe{\lower.5ex\hbox{\ltsima}}
\def\gtsima{$\; \buildrel > \over \sim \;$}
\def\gsim{\lower.5ex\hbox{\gtsima}}
\def\goe{\lower.5ex\hbox{\gtsima}}
\def\nh {$N_{H}$}

\def\ergs {erg\,s$^{-1}$}
\def\ergscm2 {erg\,s$^{-1}$cm$^{-2}$}

\def\cm2 {cm$^{-2}$}
\def\arcsec{$^{\prime\prime}$}

\def\ergs {${\rm erg\, s}^{-1}$}

\def\src {SGR\,0418$+$5729}

\def\nh {$N_{\rm H}$}

\shortauthors{N. Rea et al.}
 
\begin{document}

\title{The outburst decay of the low magnetic field magnetar \src}

\author{N. Rea\altaffilmark{1}, G. L. Israel\altaffilmark{2}, J. A. Pons\altaffilmark{3}, R. Turolla\altaffilmark{4,5}, D. Vigan\`o\altaffilmark{3}, S. Zane\altaffilmark{5}, P. Esposito\altaffilmark{6}, R. Perna\altaffilmark{7}, A. Papitto\altaffilmark{1} \\ G. Terreran\altaffilmark{4,1}, A. Tiengo\altaffilmark{8,9,6}, D. Salvetti\altaffilmark{6,9,10}, J. M. Girart\altaffilmark{1}, Aina Palau\altaffilmark{1}, A. Possenti\altaffilmark{11}, M. Burgay\altaffilmark{11}, \\ E. G{\"o}{\u g}{\"u}{\c s}\altaffilmark{12}, A. Caliandro\altaffilmark{1}, C. Kouveliotou\altaffilmark{13}, D.~G\"otz\altaffilmark{14}, R. P. Mignani\altaffilmark{5,6,15}, E. Ratti\altaffilmark{16},  L. Stella\altaffilmark{2}}
\altaffiltext{1}{Institute of Space Sciences (CSIC--IEEC), Campus UAB, Faculty of Science, Torre C5-parell, E-08193 Barcelona, Spain}
\altaffiltext{2}{INAF/Osservatorio Astronomico di Roma, via Frascati 33, I-00040 Monteporzio Catone, Italy}
\altaffiltext{3}{Department de Fisica Aplicada, Universitat d'Alacant, Ap. Correus 99, E-03080 Alacant, Spain}
\altaffiltext{4}{Universit\`a di Padova, Dipartimento di Fisica e Astronomia, via F.~Marzolo 8, I-35131 Padova, Italy}
\altaffiltext{5}{Mullard Space Science Laboratory, University College London, Holmbury St. Mary, Dorking, Surrey RH5 6NT, UK}
\altaffiltext{6}{INAF/Istituto di Astrofisica Spaziale e Fisica Cosmica - Milano, via E.~Bassini 15, I-20133 Milano, Italy}
\altaffiltext{7}{JILA and Department of Astrophysical and Planetary Sciences, University of Colorado, Boulder, CO 80309, USA}
\altaffiltext{8}{IUSS - Istituto Universitario di Studi Superiori, Piazza della Vittoria 15, I-27100 Pavia, Italy}
\altaffiltext{9}{INFN Ð Istituto Nazionale di Fisica Nucleare, Sezione di Pavia, Via Bassi 6, I-27100 Pavia, Italy}
\altaffiltext{10}{Universit\`a degli Studi di Pavia, Dipartimento di Fisica Nucleare e Teorica, via Bassi 6, I-27100 Pavia, Italy}
\altaffiltext{11}{INAF/Osservatorio Astronomico di Cagliari, localit\`a Poggio dei Pini, strada 54, I-09012 Capoterra, Italy}
\altaffiltext{12}{Sabanc\i\ University, Orhanl\i-Tuzla, 34956 \.Istanbul, Turkey}
\altaffiltext{13}{NASA Marshall Space Flight Center, Huntsville, AL 35812, USA}
\altaffiltext{14}{AIM (UMR 7158 CEA/DSM-CNRS-Universit\'e Paris Diderot) Irfu/Service d'Astrophysique, Saclay,  FR-91191 Gif-sur-Yvette Cedex, France}
\altaffiltext{15}{Kepler Institute of Astronomy, University of Zielona G\'ora, Lubuska 2, 65-265, Zielona G\'ora, Poland}
\altaffiltext{16}{SRON-Netherlands Institute for Space Research, Sorbonnelaan 2, 3584 CA Utrecht, The Netherlands}

\begin{abstract}
We report on the long term X-ray monitoring of the outburst decay of
the low magnetic field magnetar \src\,, using all the available X-ray
data obtained with \rxte, \swift, \chandra, and \xmm\, observations,
from the discovery of the source in June 2009, up to August 2012.
The timing analysis allowed us to obtain the first measurement of the
period derivative of \src: $\dot{P}=4(1)\times 10^{-15}$~s\,s$^{-1}$, significant at
 $\sim3.5\sigma$ confidence level. This leads to a surface dipolar
magnetic field of $B_{\rm dip}\simeq6\times 10^{12}$ Gauss. This
measurement confirms \src\, as the lowest magnetic field
magnetar. Following the flux and spectral evolution from the beginning
of the outburst up to $\sim$1200\,days, we observe a gradual cooling
of the tiny hot spot responsible for the X-ray emission, from a
temperature of $\sim$ 0.9 to 0.3\,keV. Simultaneously, the X-ray flux
decreased by about 3 orders of magnitude: from about $1.4\times10^{-11}$ to $1.2\times10^{-14}$~\ergscm2  . Deep radio, millimeter,
optical and gamma-ray observations did not detect the source
counterpart, implying stringent limits on its multi-band emission, as
well as constraints on the presence of a fossil disk. By modeling the
magneto-thermal secular evolution of \src\,, we infer a realistic age of $\sim$550\,kyr, and a dipolar magnetic field at birth of $\sim10^{14}$\,G. The
outburst characteristics suggest the presence of a thin twisted bundle
with a small heated spot at its base. The bundle untwisted in the
first few months following the outburst, while the hot spot decreases
in temperature and size. We estimate the outburst rate of low magnetic
field magnetars to be about one per year per galaxy, and we briefly discuss the consequences of such result 
in several other astrophysical contexts.

\end{abstract}


\section{Introduction}
\label{intro}

\begin{table*}
\centering

\caption{Journal of all the X-ray observations of SGR\,0418+5729.}\smallskip
\label{obslog}
\begin{tabular}{@{}lccccccc}
\hline
Instrument & Obs. ID & Starting date  & Exp. (ks) & Counts~s$^{-1}$ & Flux$^{c}$  & kT$_{BB}$ (keV)$^{d}$  & BB norm.$^{e}$ \\
\hline

RXTE/PCA$^{a}$  & 94048 & 2009 06-11/11-24 & 194.2 & & -- & -- & -- \\
Swift/XRT  & 00031422001 & 2009-07-08 20:48:01 & 2.9 & 0.229$\pm$0.008 & 13.4$\pm$1.0 & 0.88$\pm$0.05 & 2.2$\pm$0.2 \\
Swift/XRT (PC) & 00031422002 & 2009-07-09 00:04:01 &  10.6 & 0.245$\pm$0.003 & 13.8$\pm$0.7 &  0.94$\pm$0.03 & 1.7$\pm$0.1\\
Swift/XRT (PC) & 00031422003 & 2009-07-10 00:15:01 &   5.6 & 0.179$\pm$0.005 & 11.0$\pm$0.7 &  0.95$\pm$0.05 & 1.33$\pm$0.13 \\
Swift/XRT (WT) & 00031422004  & 2009-07-12 00:27:01 &   7.1 & 0.218$\pm$0.006 & 11.5$\pm$0.7 & 0.91$\pm$0.04 & 1.68$\pm$0.14\\
Chandra/HRC-I$^{a}$ & 10168 &  2009-07-12 06:06:43 &  24.1 & 0.317$\pm$0.005 & -- & -- & -- \\
Swift/XRT (WT)  & 00031422006  & 2009-07-15 00:48:39 &  7.7 & 0.252$\pm$0.006 & 12.7$\pm$1.0  & 0.93$\pm$0.03 & 1.69$\pm$0.12 \\\
Swift/XRT (WT) & 00031422007  & 2009-07-16 00:53:01 &   16.4 & 0.217$\pm$0.004 & 11.7$\pm$0.8 & 0.93$\pm$0.02 &  1.57$\pm$0.08 \\
XMM-Newton/EPIC$^{*}$  & 0610000601 & 2009-08-12 21:09:12  & 67.1 &  1.281$\pm$0.005 & 6.75$\pm$0.07 & 0.897$\pm$0.007 & 1.038$\pm$0.018\\
Swift/XRT  (PC)  & 00031422008 & 2009-09-20 21:09:00 &   9.4 &  0.066$\pm$0.002 & 3.66$\pm$0.30 &  0.82$\pm$0.05 &  0.79$\pm$0.09 \\
Swift/XRT  (PC)  & 00031422009  & 2009-09-22 00:43:00 &  7.6 & 0.072$\pm$0.003 & 3.58$\pm$0.40 &   0.82$\pm$0.05 & 0.79$\pm$0.11 \\
Swift/XRT   (PC) & 00031422010 & 2009-11-08 00:36:01 &  15.1& 0.043$\pm$0.002 & 2.14$\pm$ 0.20 &  0.82$\pm$0.05 & 0.47$\pm$0.05  \\
Swift/XRT  (PC) & 00031422011$^{b1}$   & 2010-01-14 08:06:01 &  3.6 & 0.019$\pm$0.001$^b$ & 1.05$\pm$0.10 &  0.75$\pm$0.07 & 0.32$\pm$0.06 \\
Swift/XRT  (PC) & 00031422012$^{b1}$   & 2010-01-15 13:08:01 & 3.7 &  0.019$\pm$0.001$^b$ & " & " & " \\
Swift/XRT  (PC) & 00031422013$^{b1}$   & 2010-01-16 08:14:01 &  4.0 & 0.019$\pm$0.001$^b$ & " & " & "\\
Swift/XRT  (PC) & 00031422014$^{b1}$   & 2010-01-17 06:47:01 &  3.8 & 0.019$\pm$0.001$^b$ & " & " & " \\
Swift/XRT  (PC) & 00031422015$^{b2}$   & 2010-02-14 17:33:01 &  4.5 & 0.0172$\pm$0.0008$^c$ & 0.76$\pm$0.11 & 0.74$\pm$0.04 & 0.25$\pm$0.03 \\
Swift/XRT  (PC) & 00031422016$^{b2}$  & 2010-02-15 17:37:01 &  4.5 & 0.0172$\pm$0.0008$^c$ & " & " & "\\
Swift/XRT  (PC) & 00031422017$^{b2}$   & 2010-02-16 01:38:01 &  4.6 & 0.0172$\pm$0.0008$^c$ & "& "& " \\
Swift/XRT  (PC) & 00031422018$^{b2}$   & 2010-02-17 09:49:01 &  4.6 & 0.0172$\pm$0.0008$^c$ & " & " & "\\
Swift/XRT  (PC) & 00031422019$^{b2}$   & 2010-02-18 16:14:01 &  3.9 & 0.0172$\pm$0.0008$^c$ & " &  " & " \\
Swift/XRT  (PC) & 00031422020$^{b2}$   & 2010-02-19 00:23:01 & 3.2 & 0.0172$\pm$0.0008$^c$ & " & " &  "\\
Swift/XRT  (PC) & 00031422021$^{b3}$   & 2010-07-09 06:50:01 &  3.6 & 0.0023$\pm$0.0003$^d$ & 0.10$\pm$0.04 &  0.47$\pm$0.13 &  0.23$\pm$0.13\\
Swift/XRT  (PC) & 00031422022$^{b3}$   & 2010-07-10 18:11:00 &  5.2 & 0.0023$\pm$0.0003$^d$ &  "&  "& "\\
Swift/XRT  (PC) & 00031422023$^{b3}$   & 2010-07-11 05:19:01  & 5.0 & 0.0023$\pm$0.0003$^d$ &  "&  "& "\\
Swift/XRT  (PC) & 00031422024$^{b3}$   & 2010-07-11 23:06:01  & 5.4 & 0.0023$\pm$0.0003$^d$ &  "&  "& "\\
Swift/XRT  (PC) & 00031422025$^{b3}$    &  2010-07-13 00:47:01  & 4.9 & 0.0023$\pm$0.0003$^d$ & " &  "& "\\
Chandra/ACIS-S & 12312 &  2010-07-23 15:04:09 &  30.0 &  0.0017$\pm$0.0008 & 0.13$\pm$0.02 & 0.68$\pm$0.04 &  0.061$\pm$0.008\\
XMM-Newton/EPIC & 0605852201 & 2010-09-24 01:54:56  & 34.2 & 0.0370$\pm$0.0020 & 0.16$\pm$0.02 & 0.69$\pm$0.05 & 0.07$\pm$0.01 \\
Chandra/ACIS-S & 13148 &  2010-11-29 05:59:57  &  30.0 & 0.0038$\pm$0.0004 & 0.021$\pm$0.002 &  0.38$\pm$0.11 & 0.12$\pm$0.06 \\
XMM-Newton/EPIC & 0672670201&  2011-03-10 03:15:53 &   35.0 & 0.0071$\pm$0.0007 & 0.015$\pm$0.002 & 0.32$\pm$0.05 &   0.21$\pm$0.08\\
Chandra/ACIS-S & 13235 &  2011-07-20 02:26:12 &  77.0 & 0.0033$\pm$0.0002 & 0.015$\pm$0.003 & 0.37$\pm$0.04 & 0.11$\pm$0.02\\
XMM-Newton/EPIC$^{b4}$ & 0672670401 &   2011-09-09 15:27:23 &  33.0 & 0.0071$\pm$0.0006$^e$ & 0.016$\pm$0.002 &  0.28$\pm$0.05 & 0.34$\pm$0.13\\
XMM-Newton/EPIC$^{b4}$ &  0672670501&   2011-09-11 21:47:41 &  48.5 & 0.0071$\pm$0.0006$^e$ & " & " & " \\
Chandra/ACIS-S & 13236 &  2011-11-26 11:48:02 &  75.0 & 0.0026$\pm$0.0002 & 0.015$\pm$0.002 & 0.35$\pm$0.07 & 0.13$\pm$0.05\\
XMM-Newton/EPIC$^{*}$  &  0693100101 & 2012-08-25 14:18:08 &  78.2 & 0.0058$\pm$0.0004 & 0.012$\pm$0.001 & 0.32$\pm$0.05 & 0.16$\pm$0.05 \\
\hline
\end{tabular}
\begin{list}{}{}
\item[$^{a}$]The \rxte --PCA and the \CXO\, HRC-I were used only for the timing analysis.
\item[$^{b}$]These observations were merged in the timing and spectral analysis to improve statistics. 
\item[$^{c}$]Absorbed flux in the 0.5-10\,keV energy range, and in units of $10^{-12}$\ergscm2 . Errors in the table are at 90\% confidence level.
\item[$^{d}$]Fitted model is: {\tt phabs*bbodyrad}; \nh $=  (1.15\pm0.06)\times10^{21}$\cm2 and $\chi_{\nu}^2 = 1.19$ (for 940
  dof). 
\item[$^{e}$]The BB radius in km is the square root of this BB normalization, times the distance in units of 10\,kpc .
\item[$^{*}$] See \S\ref{sec:spectra} for details on the modeling of these observations.
\end{list}
\end{table*}


Neutron stars showing magnetar-like activity (comprising the anomalous
X-ray pulsars, soft gamma repeaters and a high magnetic field pulsar)
are a small group of X-ray pulsars (about twenty objects) with spin
periods between 0.3--12\,s, whose strong persistent and/or flaring emission are hard
to explain by the common scenarios for rotation powered pulsars or
accreting pulsars. In fact, the very strong X-ray emission of these
objects ($L_{\rm X} \sim10^{35}$\ergs ) is too high and/or variable to be
fed by the rotational energy alone (as in the radio pulsars), and no
evidence for a companion star has been found, hence ruling out accretion in a binary.
Accretion from a fossil disk remnant of the supernova
explosion might be responsible for part of the observational
properties of these objects, but it fails to explain some of their
characteristics, such as the flaring X-ray activity.  Their inferred
magnetic fields, under the typical assumption of magnetic dipolar
losses alone, appear to be as high as $B_{\rm dip} \simeq
3.2\times10^{19} \sqrt{P\dot{P}} \sim 10^{14}- 10^{15}$\,G (see
Mereghetti 2008 for a review). These strong fields are believed to
form either via a dynamo action in a rapidly rotating proto-neutron
star ($<$3\,ms; Thompson \& Duncan 1995), or they are thought to be fossil fields
remnant of a highly magnetic massive star ($\sim$1~kG; Ferrario \&
Wickramasinghe 2006).  Because of these high B fields, the emission of
"magnetars" is thought to be powered by the decay and the instability
of their strong fields (Duncan \& Thompson 1992; Thompson \& Duncan
1993, Thompson, Lyutikov \& Kulkarni 2002). Their powerful X-ray
output is usually well modeled by thermal emission from the neutron
star hot surface, reprocessed in a twisted magnetosphere through
resonant cyclotron scattering (Thompson, Lyutikov \& Kulkarni 2002;
Nobili, Turolla \& Zane 2008; Rea et al. 2008; Zane et al. 2009), a
process favored under these extreme magnetic conditions. On top of
their persistent X-ray emission, magnetars emit very peculiar flares
and outbursts on several timescales, from fractions of a second to
years reaching very high, super-Eddington luminosities ($10^{38}-10^{46}$\ergs ). These flares
are most probably caused by rearrangements of the twisted magnetic
field lines, either accompanied or triggered by fractures of the
neutron-star crust (Thompson \& Duncan 1995; Perna \& Pons 2011).

Transient events are a characteristic signature of magnetar emission,
and one of the main ways to discover new sources of this class and
study their physics. From the discovery of the first transient less
than a decade ago, we now count about a dozen of outbursts, which
increased the number of known magnetars by a third in six years (see
Rea \& Esposito 2011; Rea 2013 for recent reviews). Magnetar outbursts
might involve their multi-band emission resulting in an increased
activity from radio to hard X-ray, usually with a soft X-ray flux
increase of a factor of 10--1000 with respect to the quiescent
level. An associated X-ray spectral evolution is often observed, with
a spectral softening during the outburst decay (Rea et al. 2009). The flux decay
timescale varies substantially from source to source, ranging from a
few weeks to several years (Rea \& Esposito 2011; Pons \& Rea 2012).


\begin{figure*}
\hbox{
\includegraphics[width=12cm,height=10cm]{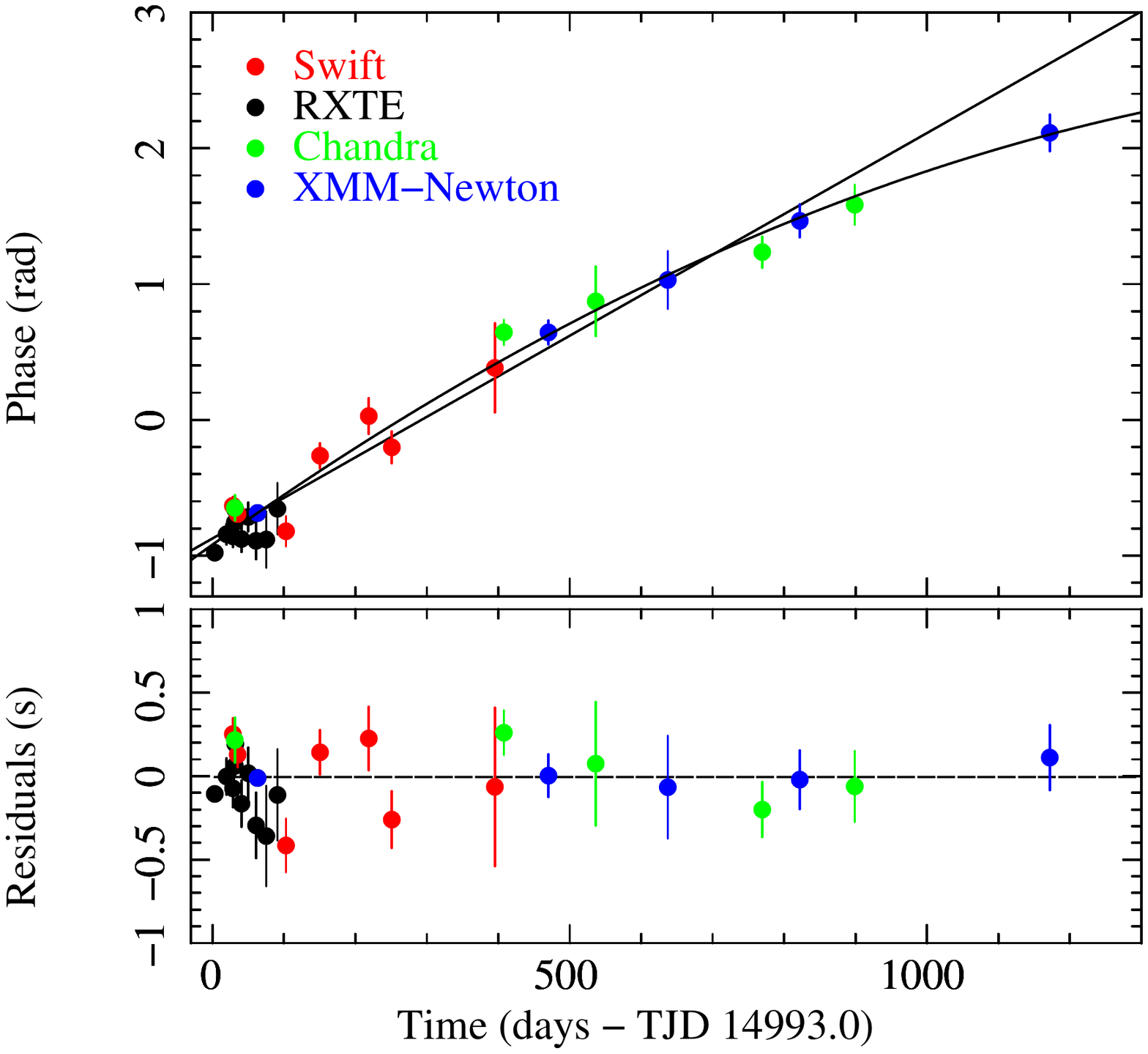}
\hspace{-3.3cm}
\includegraphics[width=10cm,height=8.2cm]{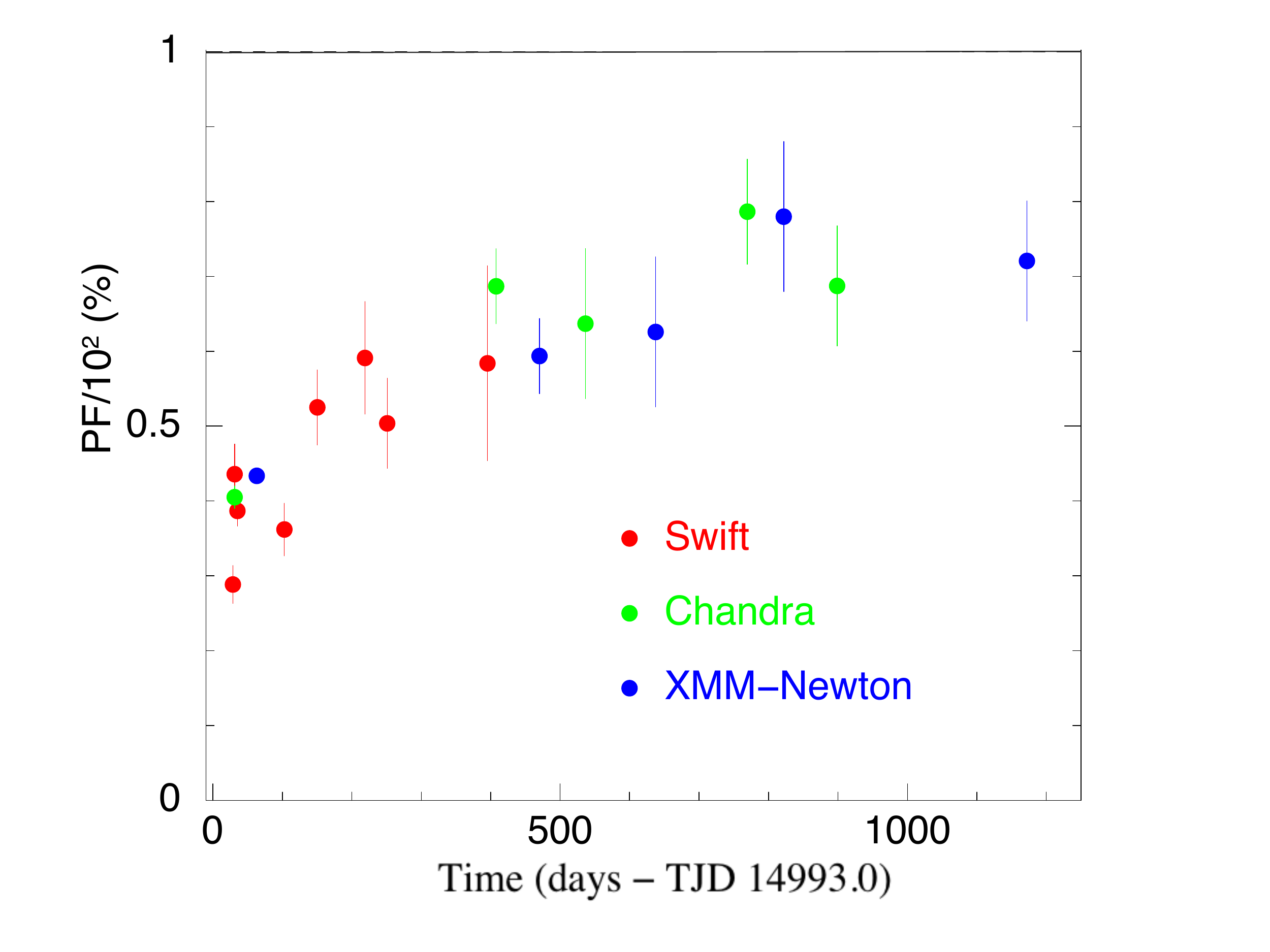}}
\caption{{\em Left panel}: Evolution of the pulse phases with time
  (upper panel). The solid lines represent the timing solution without
  (linear) and with a $\dot{P}$ component (quadratic). The time
  residuals (lower panel) are relative to the quadratic fit. {\em
    Right panel}: Pulsed fraction evolution in the 0.5--10\,keV band
  for the \swift\, (red), \chandra\, (green) and \xmm\,(blue)
  observations.}
\label{timing}
\end{figure*}

The extensive follow-up of magnetars undergoing an outburst yielded
the most unexpected discovery of the past years in the magnetar
field. Prompted by the detection of typical magnetar-like bursts and a
powerful outburst, a new transient magnetar with a spin period of $\sim$9\,s  was discovered in 2009,
namely \src\, (van der Horst et al. 2010, Esposito et
al. 2010). However, after more than 2 years of extensive monitoring,
no period derivative was detected. This led to an upper limit on the
source surface dipolar field of $B_{\rm dip} < 7.5\times10^{12}$\,G
(Rea et al. 2010). For the first time, we detected a magnetar with a
low dipolar magnetic field, showing that a critical magnetic field is
not necessary for a neutron star in order to display magnetar-like
activity. In turn, this means that many seemingly normal pulsars could
turn out as magnetars at anytime (this was supported by the discovery
of a second low-B magnetar followed soon after; Rea et al. 2012;
Scholz et al. 2012). After the discovery of this low dipolar magnetic
field soft gamma repeater, several models were put forward to explain
its puzzling emission. They involve the possible presence of a
fall-back disk slowing down the pulsar up to the current spin period
(Alpar et al. 2011), a tiny inclination angle between the magnetic and
rotational axis resulting in a higher inferred magnetic field (Tong \&
Xu 2012), a pulsar with a strongly magnetized core (Soni 2012), an old
quark nova (Ouyed, Leahy \& Niebergal 2011), or a massive highly
magnetized, rotating white dwarf (Malheiro, Rueda \& Ruffini 2012).  In
Rea et al. (2010) and Turolla et al. (2011), we suggested that a
non-dipolar component of the field, larger than the measured
dipolar one, can be responsible for the behavior of this magnetar, if
it has a relatively old age ($\simeq$1\,Myr).

In this paper we present the complete study of the outburst of the low
dipolar magnetic field magnetar \src\,, from the first outburst phases
until about 3\,years after its onset. This long term monitoring
campaign using several X-ray satellites, allowed us to estimate \src's
period derivative, and follow the cooling of its surface temperature
during the outburst decay up to the (probable) quiescent level. Furthermore, we
inferred limits on its emission in the radio, millimeter, optical and
gamma-ray bands. We discuss our findings in terms of the
magneto-thermal history of this magnetar, discuss the current limits
on the presence of a fossil disk, and present some discussion on the
broader consequences of the discovery of low magnetic field magnetars.


\begin{figure*}
\hbox{
\includegraphics[width=6.7cm,height=6cm]{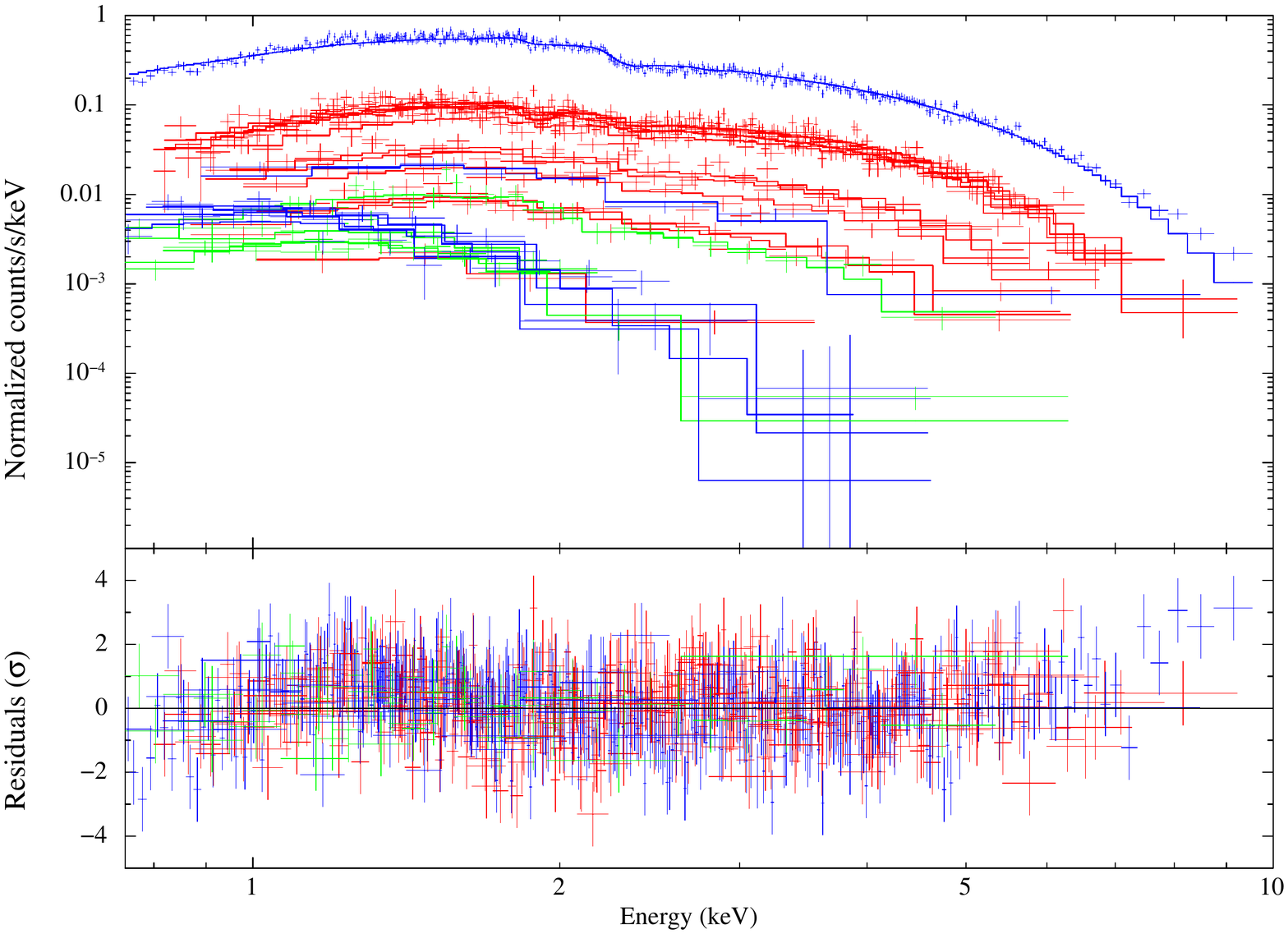}
\hspace{-1cm}
\includegraphics[width=6.7cm,height=6cm]{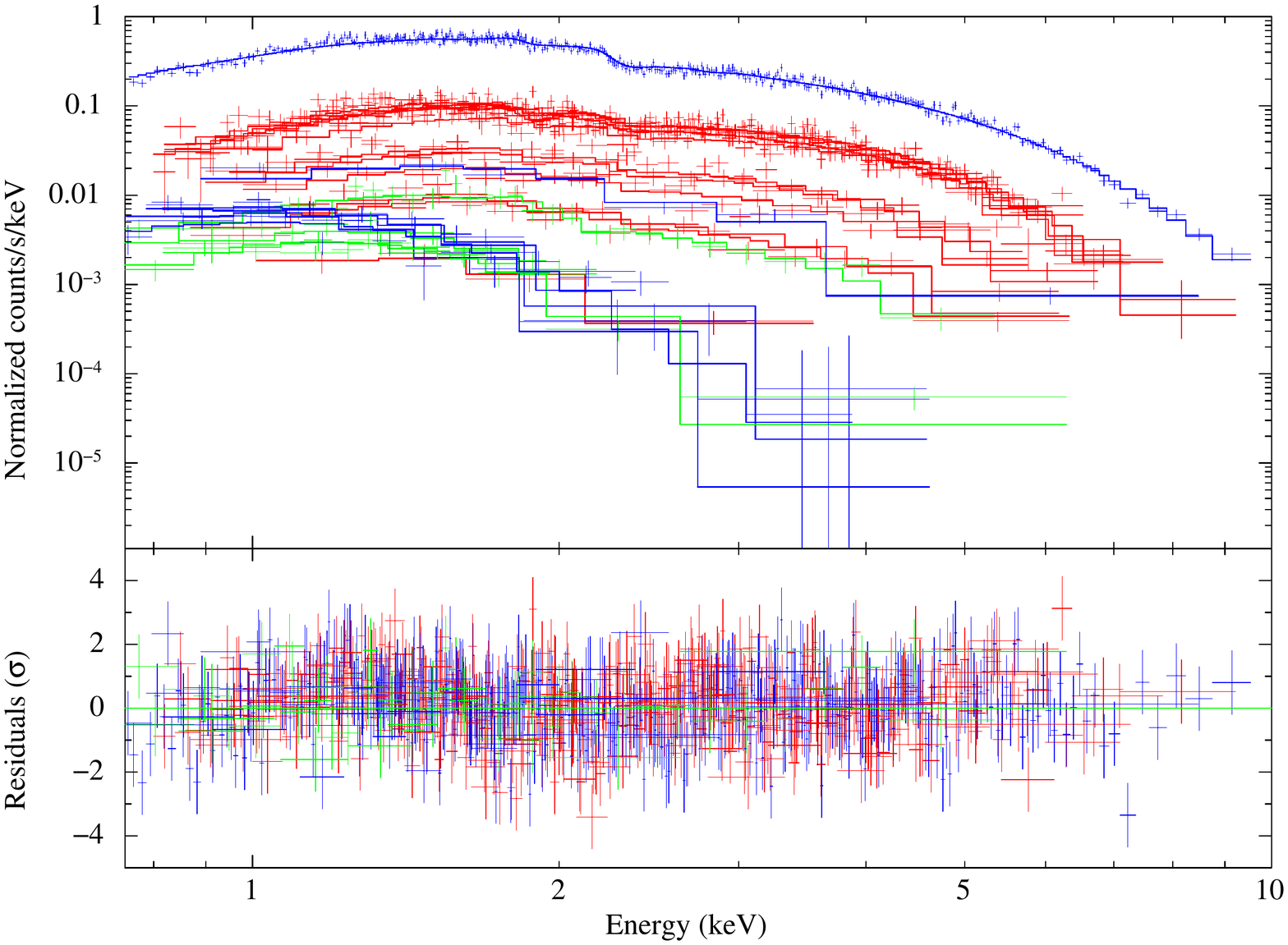}
\hspace{-1cm}
\includegraphics[width=6.7cm,height=6cm]{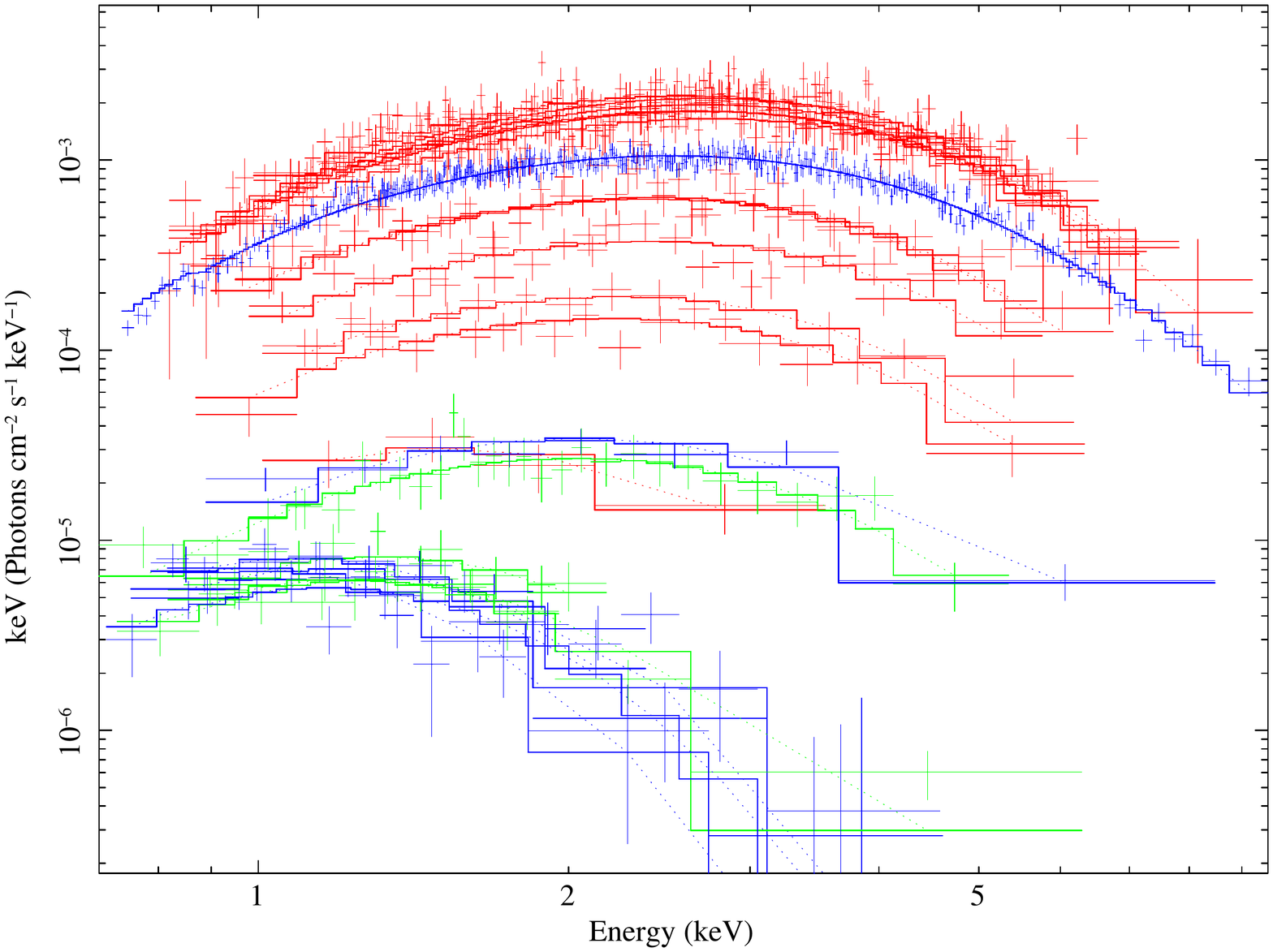}}
\caption{Spectral modeling of all observations listed in Table \,1. Left and middle panels: spectra and residuals for all observations ( \swift\, (red), \chandra\, (green) and \xmm\, (blue))fitted simultaneously with a single blackbody model (left) and using an RCS model only for the first \xmm\, observation (center). Right panel: unfolded spectrum relative to the modeling shown in the central panel.}
\label{spectra}
\end{figure*}

\section{X-ray observations and data reduction}
\label{xrayreduction}

In this study, we used data obtained from several different satellites (see Table~1 for a summary).  We describe below the observations and data analysis. Part of the data we used in this paper were already published by van der Horst et al. (2010), Esposito et al. (2010) and Rea et al. (2010). 

\subsection{Swift data}

The X-Ray Telescope (XRT; \citealt{burrows05}) on-board \swift\ uses a front-illuminated CCD detector sensitive to photons between 0.2 and 10 keV. Two main readout modes are available: photon counting (PC) and windowed timing (WT). PC mode provides two dimensional imaging information and a 2.5073\,s time resolution; in WT mode only one-dimensional imaging is preserved, achieving a time resolution of 1.766 ms.  
The XRT data were uniformly processed with \textsc{xrtpipeline} (version 12, in the \textsc{heasoft} software package version 6.11), filtered and screened with standard criteria, correcting for effective area, dead columns, etc. The source counts were extracted within a 20-pixel radius (one XRT pixel corresponds to about $2\farcs36$). For the spectroscopy, we used the spectral redistribution matrices in \textsc{caldb} (20091130; matrices version v013 and v014 for the PC and WT data, respectively), while the ancillary response files were generated with \textsc{xrtmkarf}, and they account for different extraction regions, vignetting and point-spread function corrections.

\subsection{RXTE data}

The Proportional Counter Array (PCA; \citealt{jahoda96}) on-board \rxte\ consists of five collimated xenon/methane multi-anode Proportional Counter Units (PCUs) operating in the 2--60\,keV energy range.  Raw data were reduced using the \textsc{ftools} package (version 6.11). To study the timing properties of \src, we restricted our analysis to the data in Good Xenon mode, with a time resolution of 1 $\mu$s and 256 energy bins. The event-mode data were extracted in the 2--10\,keV energy range from all active detectors (in a given observation) and all layers, and binned into light curves of 0.1\,s resolution. We use here 46 RXTE/PCA observations of \src , spannig the first 6 months of the outburst, until the source flux decayed below the instrument detection level. The total 194.2\,ks exposure time is divided in observations of 0.6 to 13.6 ks exposure each. See Esposito et al. (2010) for further details on the \swift\, and \rxte\, observations.

\subsection{Chandra data}

The \chandra\, X-ray Observatory monitored \src\, five times during
the past three years. The first one with the High Resolution Imaging
Camera (HRC--I; Zombeck et al. 1995) and the following four
observations with the Advanced CCD Imaging Spectrometer (ACIS-S;
Garmire et al. 2003).  Data were analyzed using standard cleaning
procedures\footnote{http://asc.harvard.edu/ciao/threads/index.html}
and {\tt CIAO} version 4.4. The HRC-I camera does not have a
sufficient spectral resolution, and it was used only for the timing
analysis; it has a timing resolution of $\sim$16\,$\mu$s. All ACIS-S
observations were performed in {\tt VERY FAINT} mode, with only the S7
CCD on, resulting in a timing resolution of 0.44\,s. Photons were
extracted from a circular region with a radius of 3$^{\prime\prime}$
around the source position, including more than 90\% of the source
photons, and background was extracted from a similar region far from
the source position.

\subsection{XMM-Newton data}
\label{xmm}

\src\, was observed six times with \xmm\ \citep{jansen01}. Data have been processed using SAS version 12, and we have employed the most updated calibration files available at the time the reduction was performed (August 2012). Standard data screening criteria are applied in the extraction of scientific products.  For our spectral analysis we used only the EPIC-pn camera (Turner et al. 2001) which provides the spectra with the best statistics, while the MOS cameras (Str\"uder et al. 2001) were added in the timing analysis. The EPIC-pn camera was set in Small Window (timing resolution of 6\,ms) and Full Frame (73\,ms) modes in the first two observations, respectively, and in Large Window mode for all the following ones (48\,ms), with the source at the aim-point of the camera, and the MOS cameras in Small Window mode (0.3\,s). We extracted the source photons from a circular region of 30\arcsec radius, and a similar region was chosen for the background in the same CCD. We restricted our spectral analysis to photons having PATTERN$\leq$4 and FLAG=0 for the EPIC-pn data.


\begin{figure}
\vbox{
\includegraphics[width=8cm,height=4.8cm]{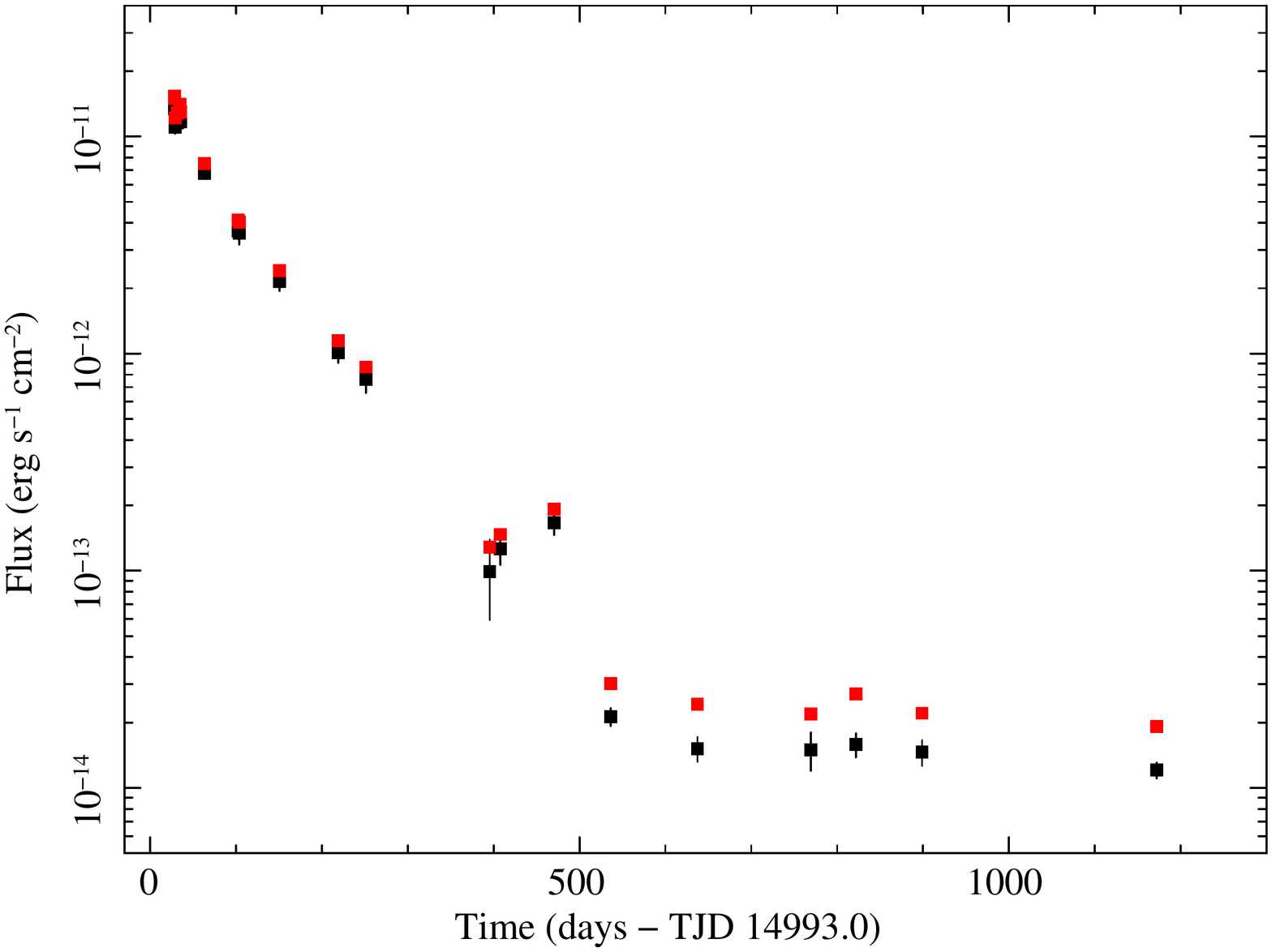}
\includegraphics[width=8cm,height=4.4cm]{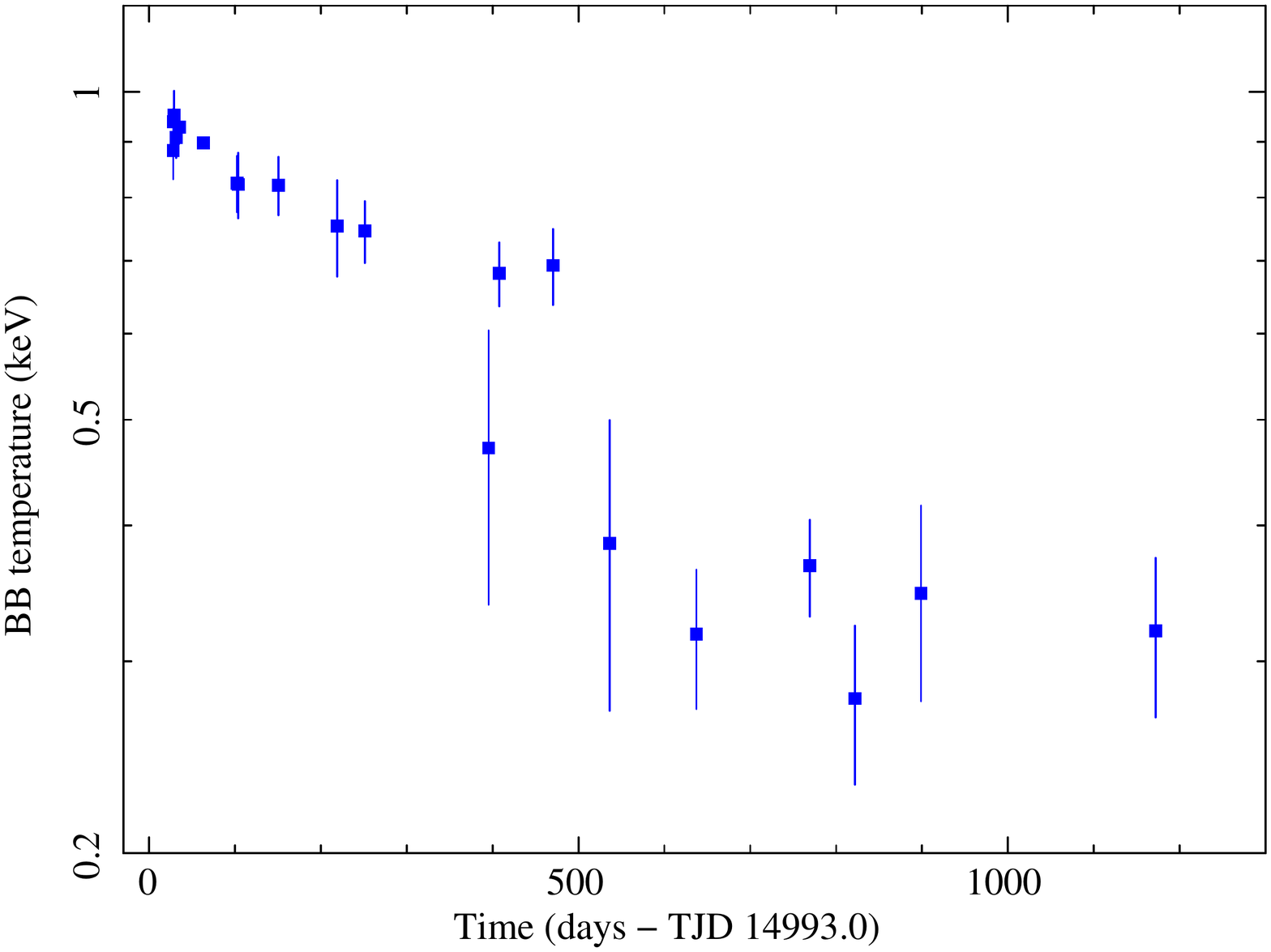}
\includegraphics[width=8cm,height=4.4cm]{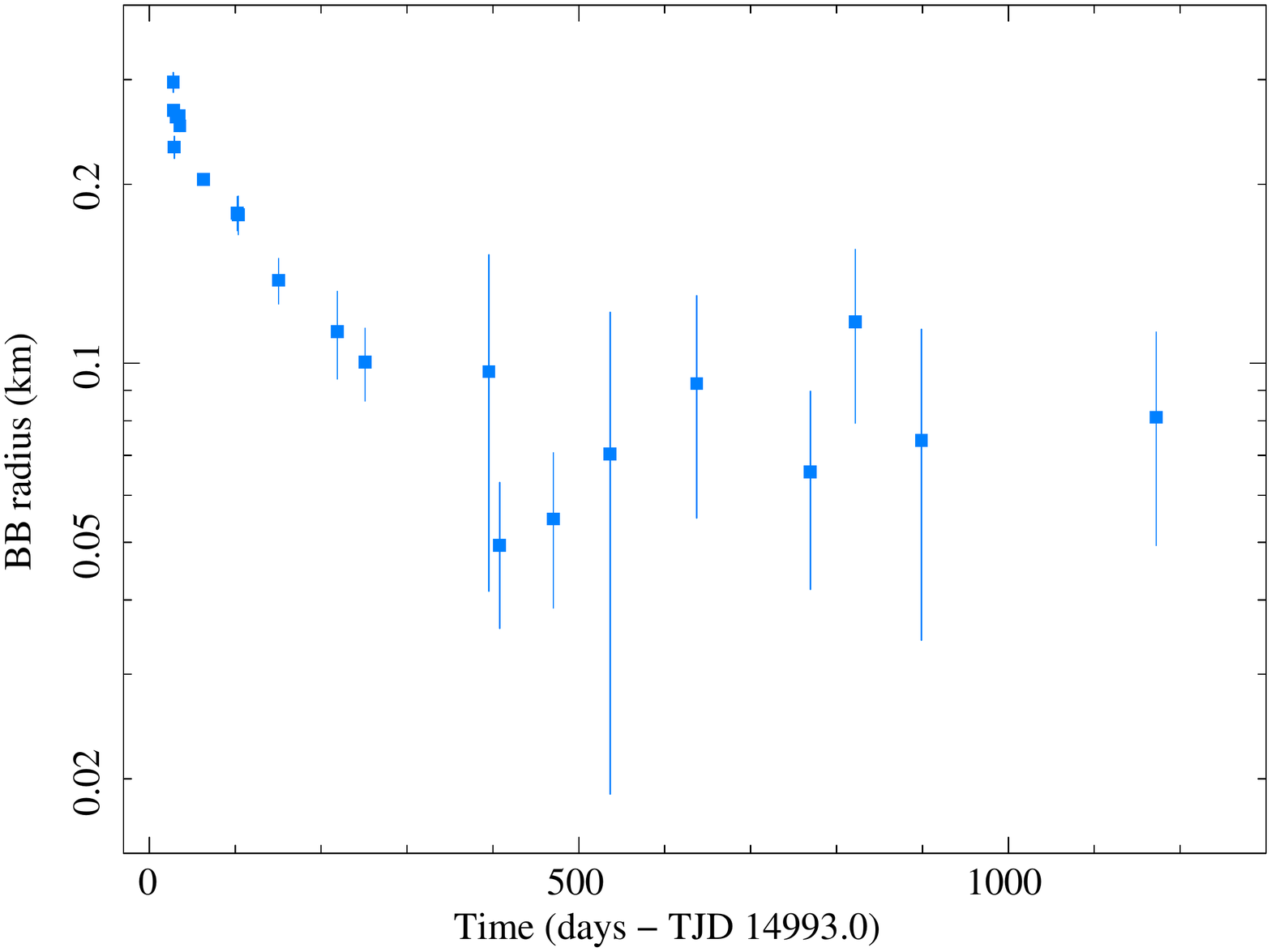}}
\caption{Spectral evolution with time. Top panel: flux evolution for the absorbed 0.5-10\,keV flux (black), and for the bolometric unabsorbed flux (red). Middle and bottom panels: evolution of the blackbody temperature and radius, calculated at infinity (the latter assuming a 2\,kpc distance).}
\label{parameters}
\end{figure}


\begin{figure}
\includegraphics[width=9cm,height=6cm]{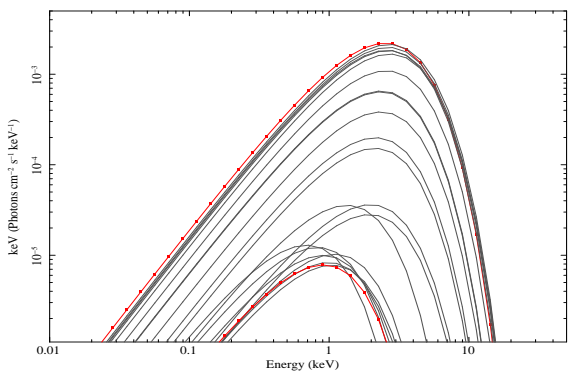}
\caption{Fitted blackbody models (see Table \,1), with the first and last observations labeled as red squares  (see Figure \,3 and text for details).}
\label{bbody}
\end{figure}

\section{Results of the X-ray monitoring}
\label{results}

\subsection{X-ray spectral modeling}
\label{sec:spectra}

For the spectral analysis we used source and background photons from
the \swift, \chandra\, and \xmm\, observations extracted as described
in the previous section (we also checked our results using a larger extraction region for the background spectra). The response matrices were built using ad-hoc
bad-pixel files built for each observation. We used the {\tt XSPEC}
package (version 12.4) for all fittings, and the {\tt phabs}
absorption model with the Anders \& Grevesse (1989) abundances, and
the Balucinska-Church \& McCammon (1992) photoelectric
cross-sections. We restricted our spectral modeling to the
0.7--10\,keV energy band, excluding bad channels when needed. The
\swift\, spectra were binned in order to have at least 30 counts per
spectral bin. \xmm\, spectra were grouped such to have at least 100,
50 and 40 photons per bin in the first three observations,
respectively, and a minimum of 30 counts in the subsequent
observations. On the other hand, all \chandra\, spectra have at least
25 counts per bin. 

\begin{figure*}
\vbox{
\hbox{
\includegraphics[width=4cm,height=4cm]{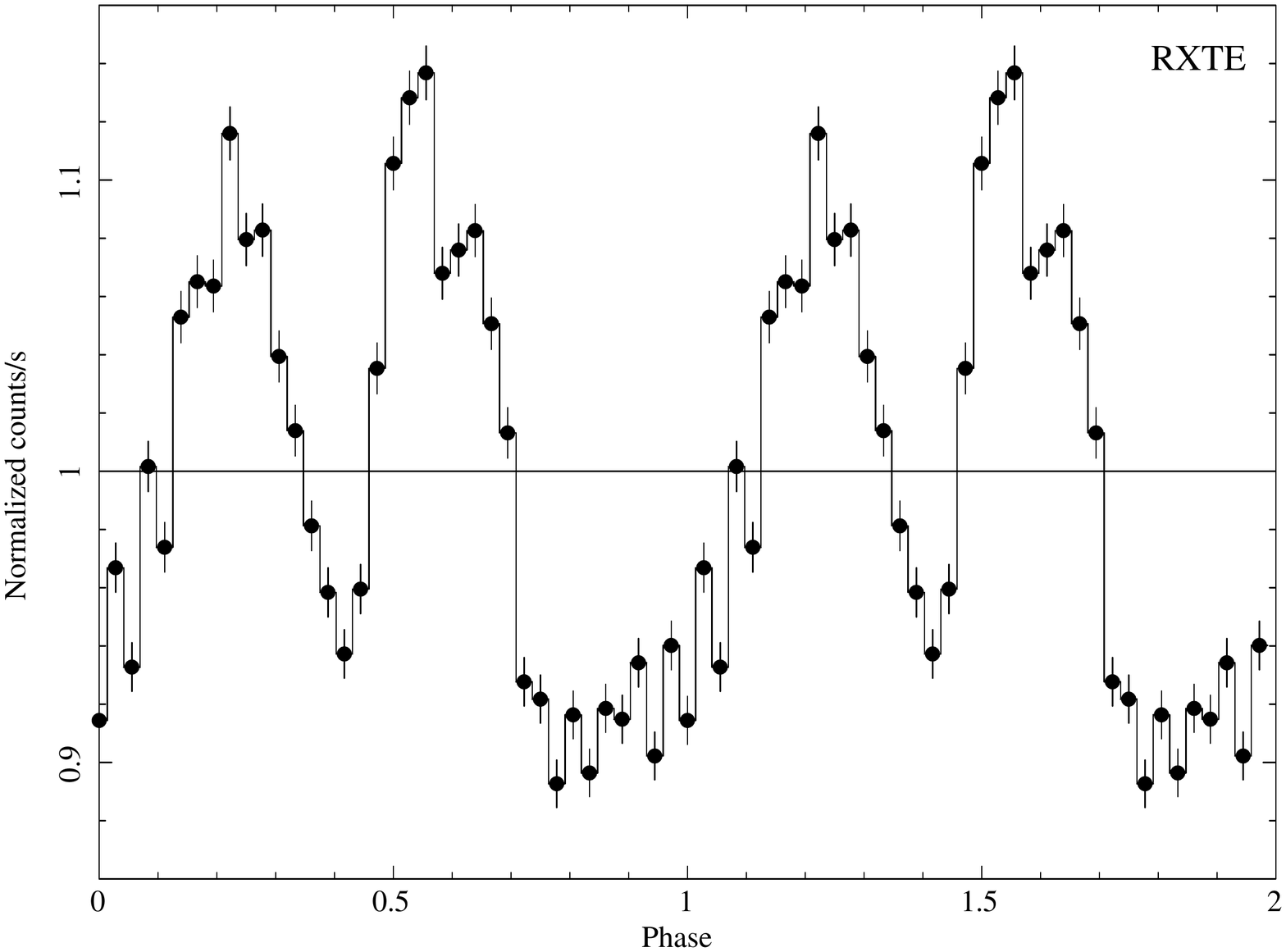}
\hspace{-0.7cm}
\includegraphics[width=4cm,height=4cm]{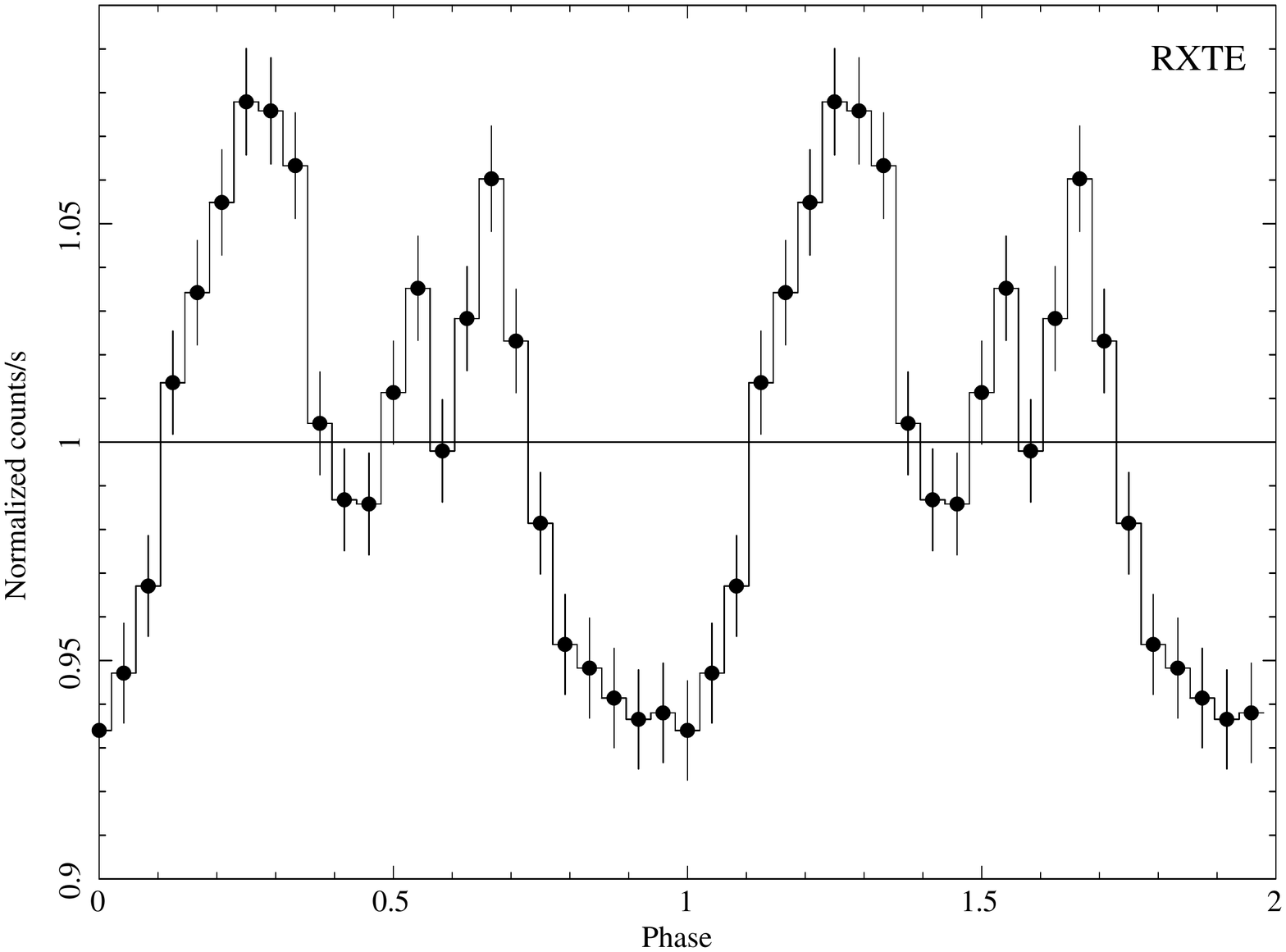}
\hspace{-0.7cm}
\includegraphics[width=4cm,height=4cm]{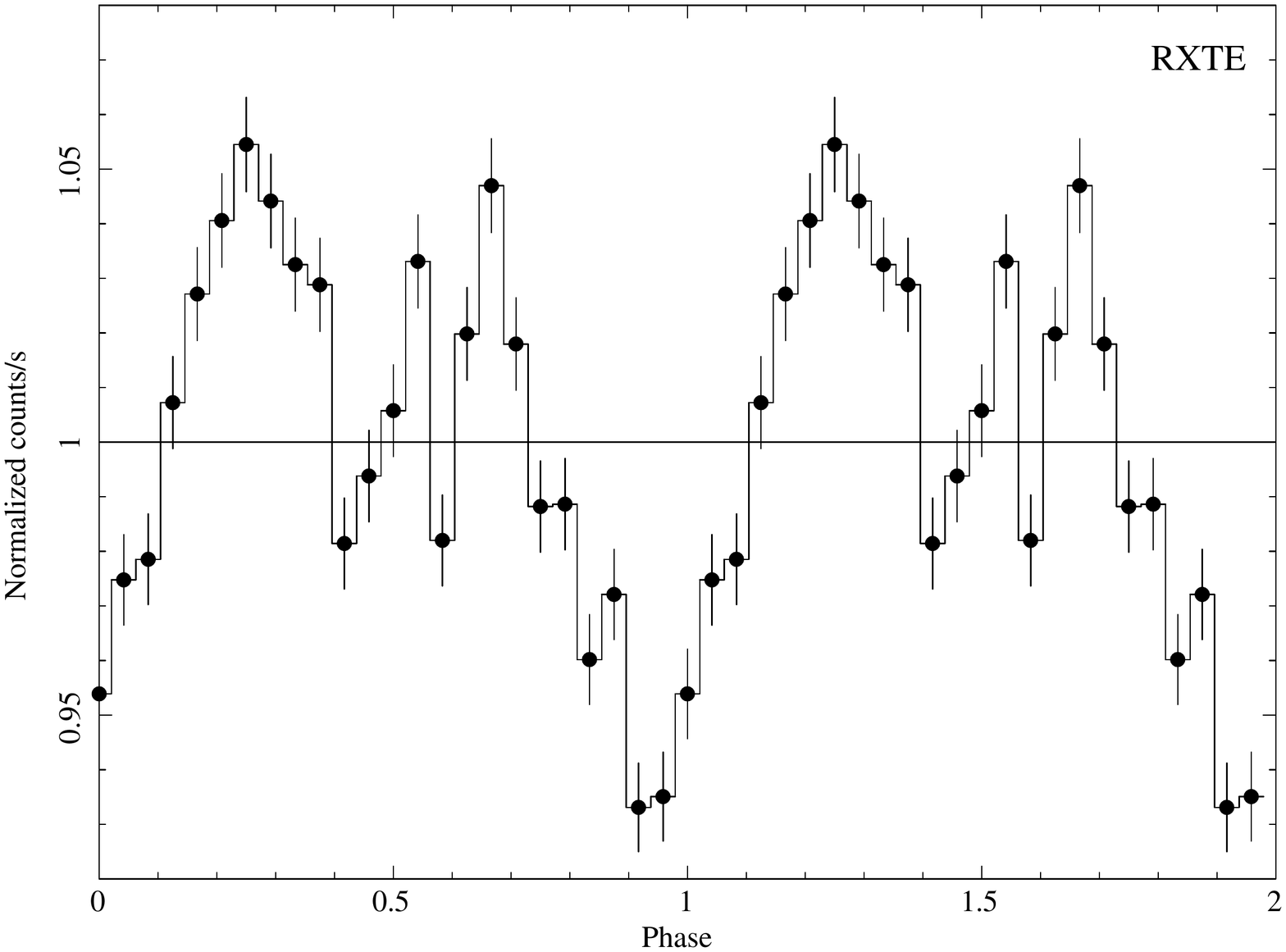}
\hspace{-0.7cm}
\includegraphics[width=4cm,height=4cm]{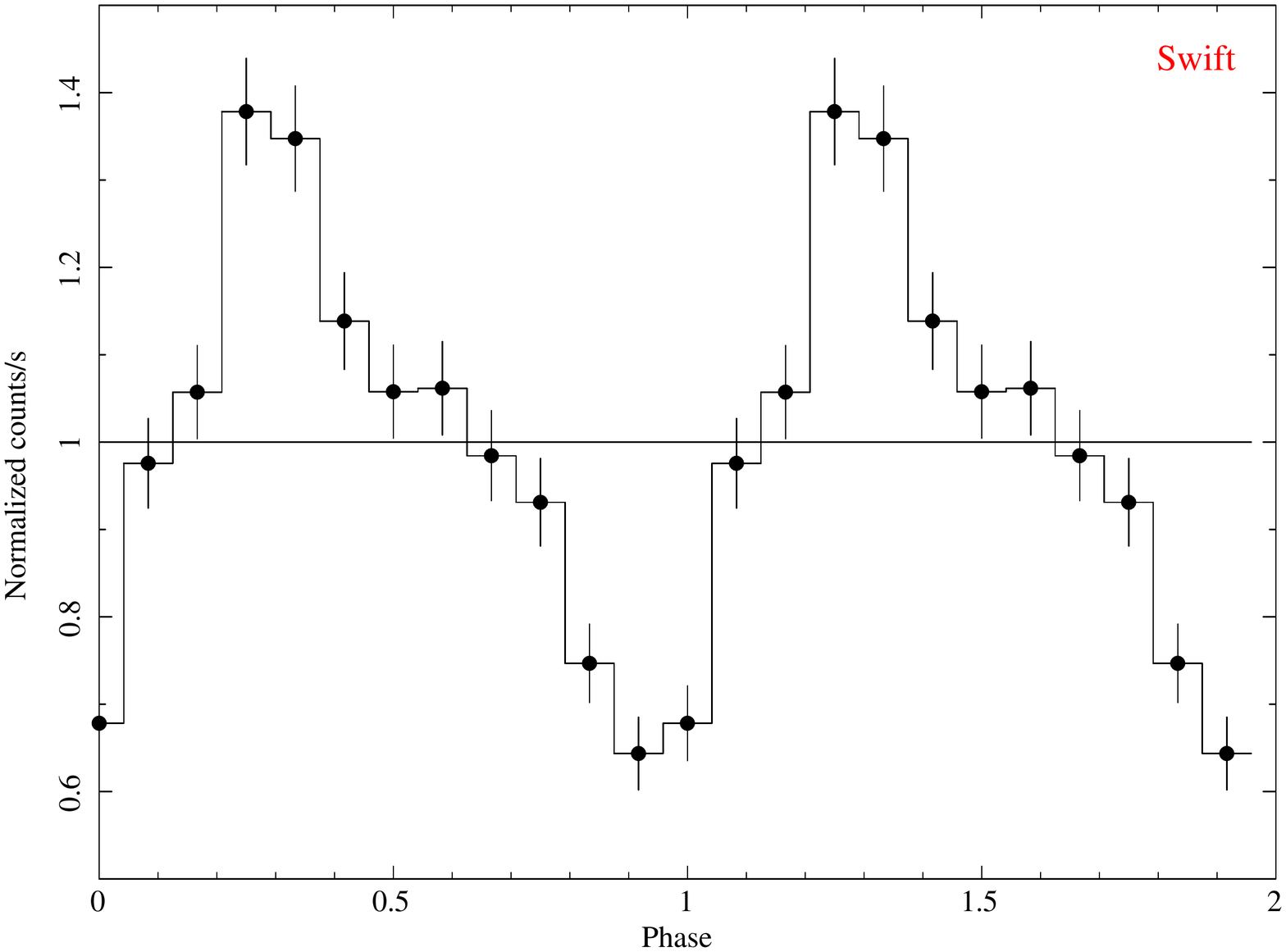}
\hspace{-0.7cm}
\includegraphics[width=4cm,height=4cm]{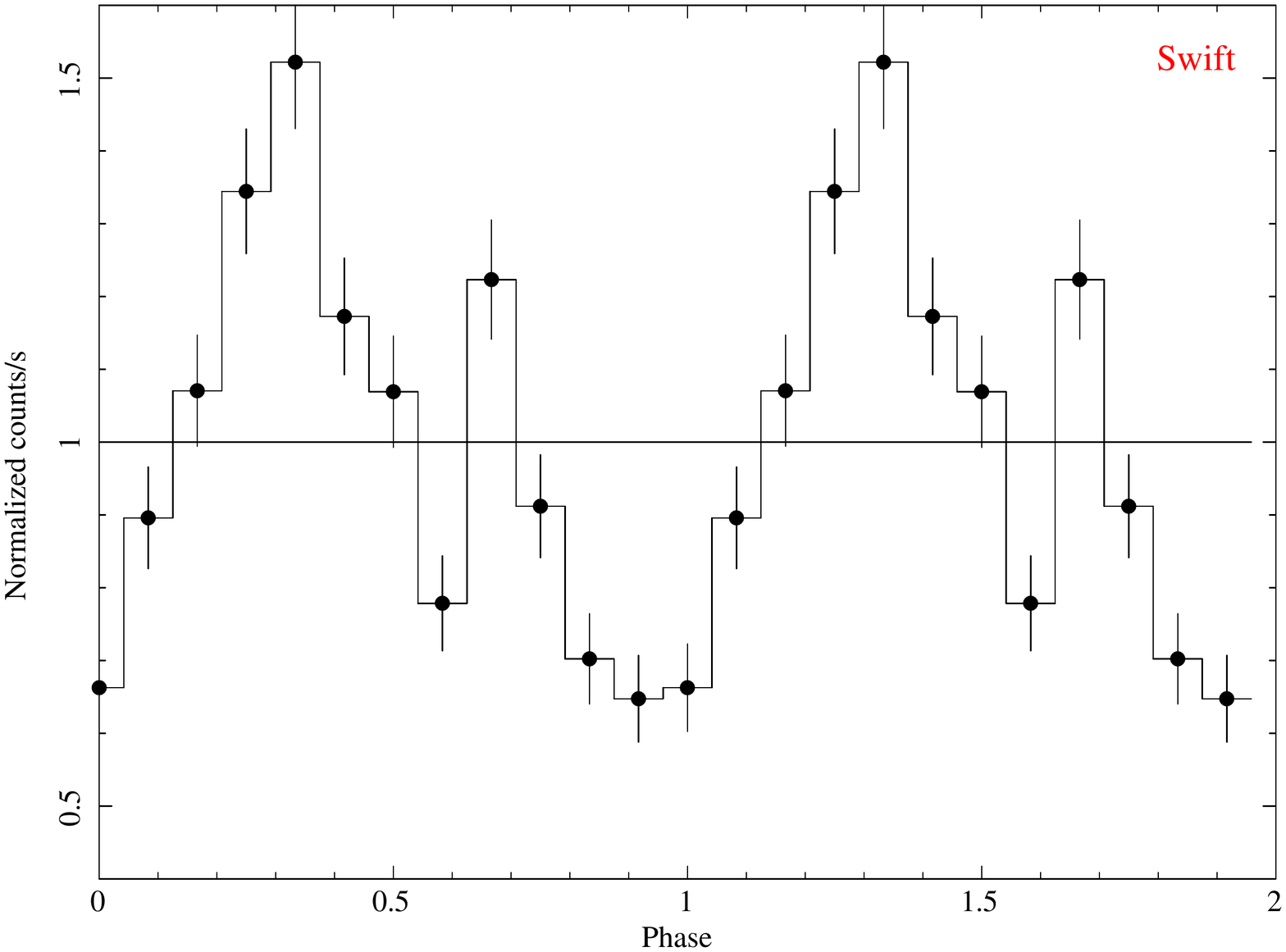}}
\vspace{-0.7cm}
\hbox{
\includegraphics[width=4cm,height=4cm]{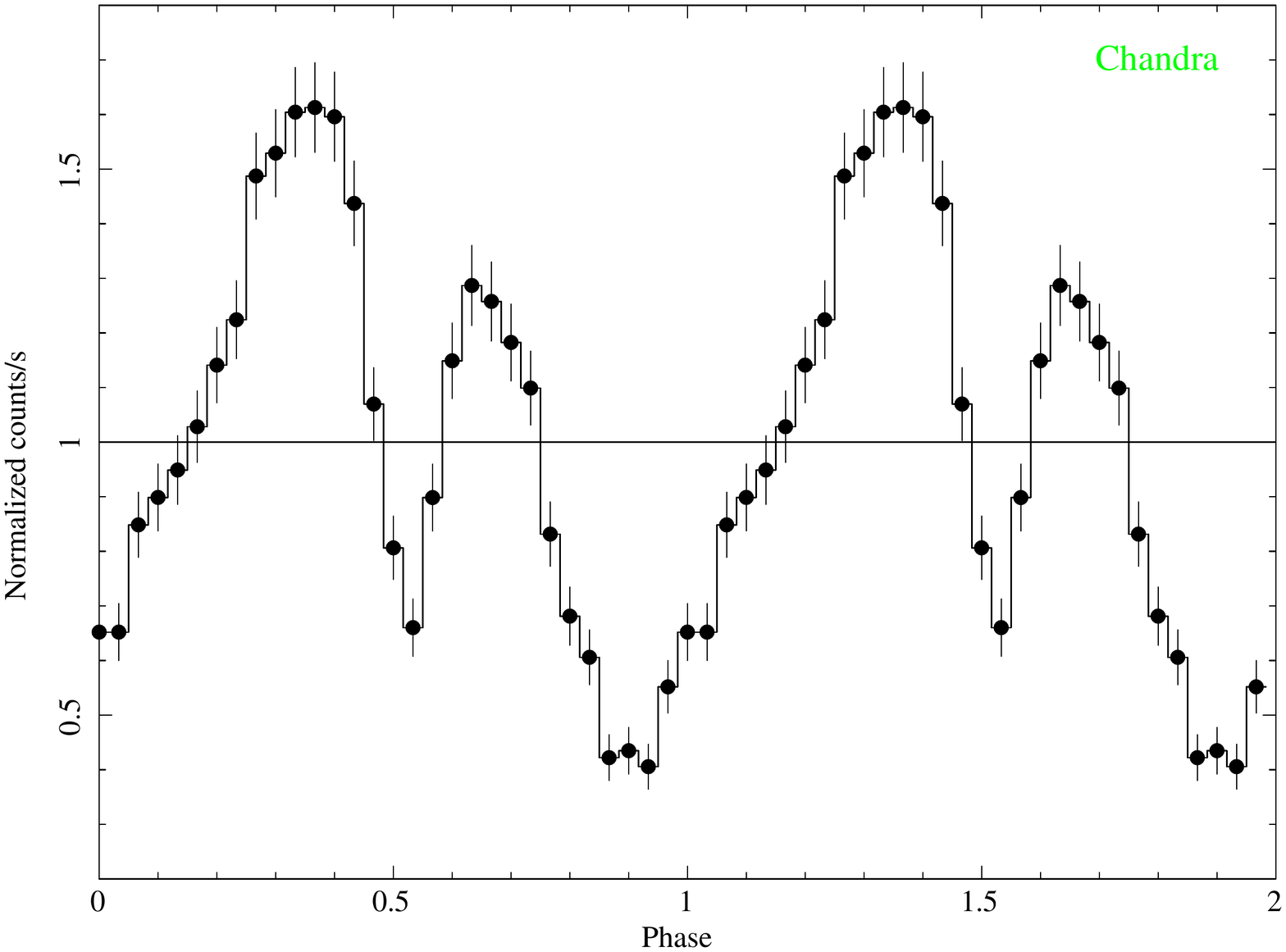}
\hspace{-0.7cm}
\includegraphics[width=4cm,height=4cm]{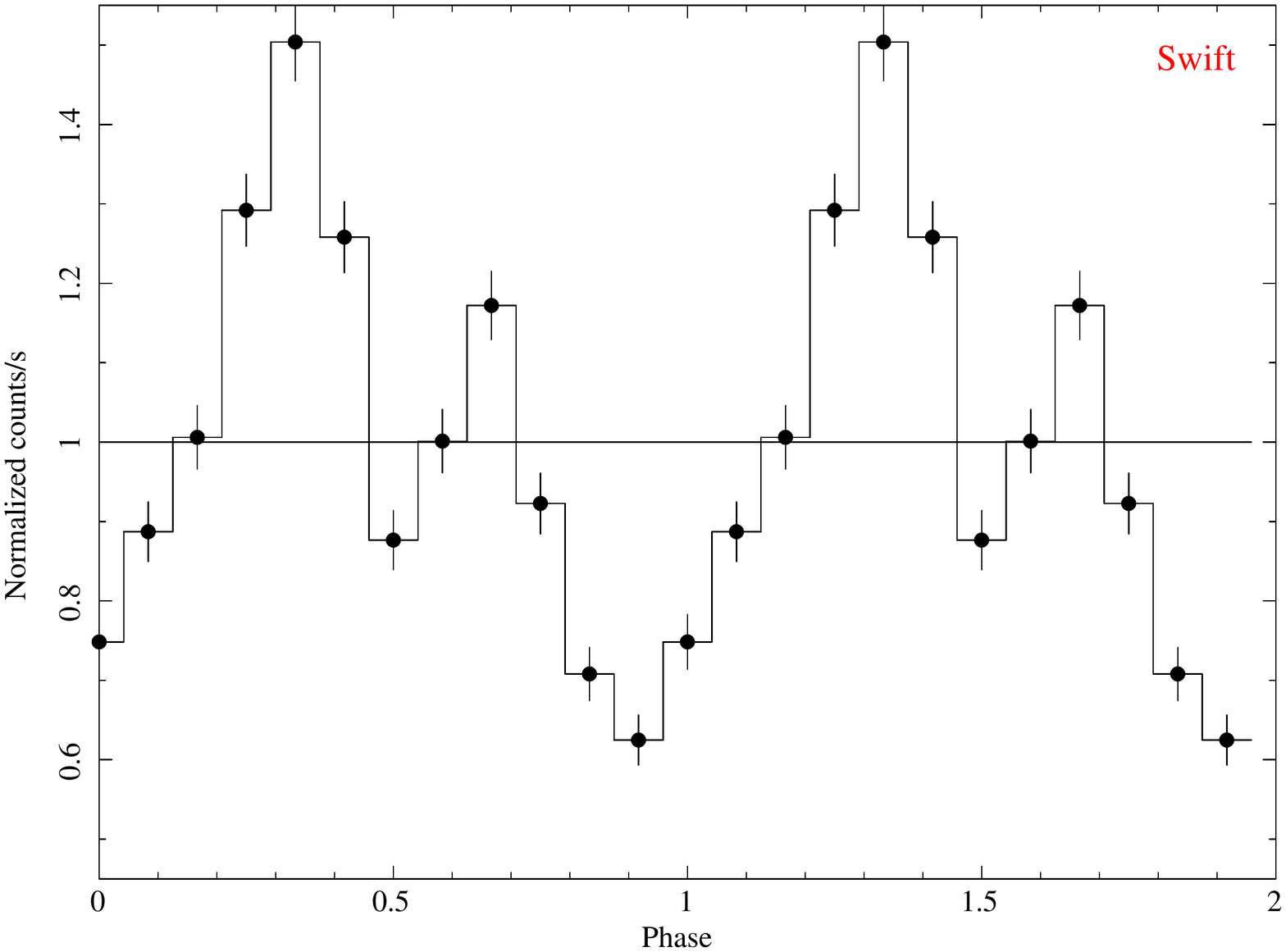}
\hspace{-0.7cm}
\includegraphics[width=4cm,height=4cm]{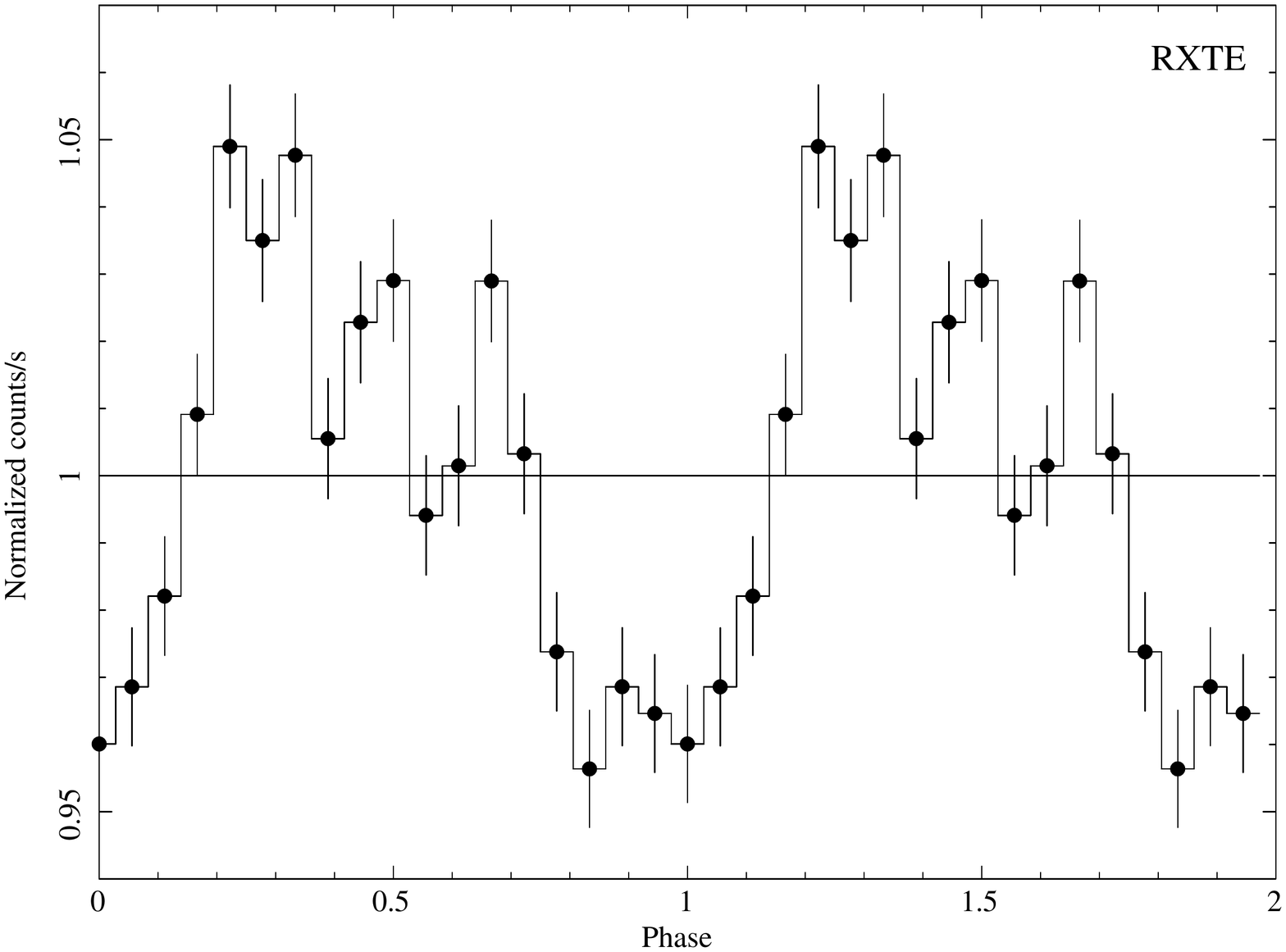}
\hspace{-0.7cm}
\includegraphics[width=4cm,height=4cm]{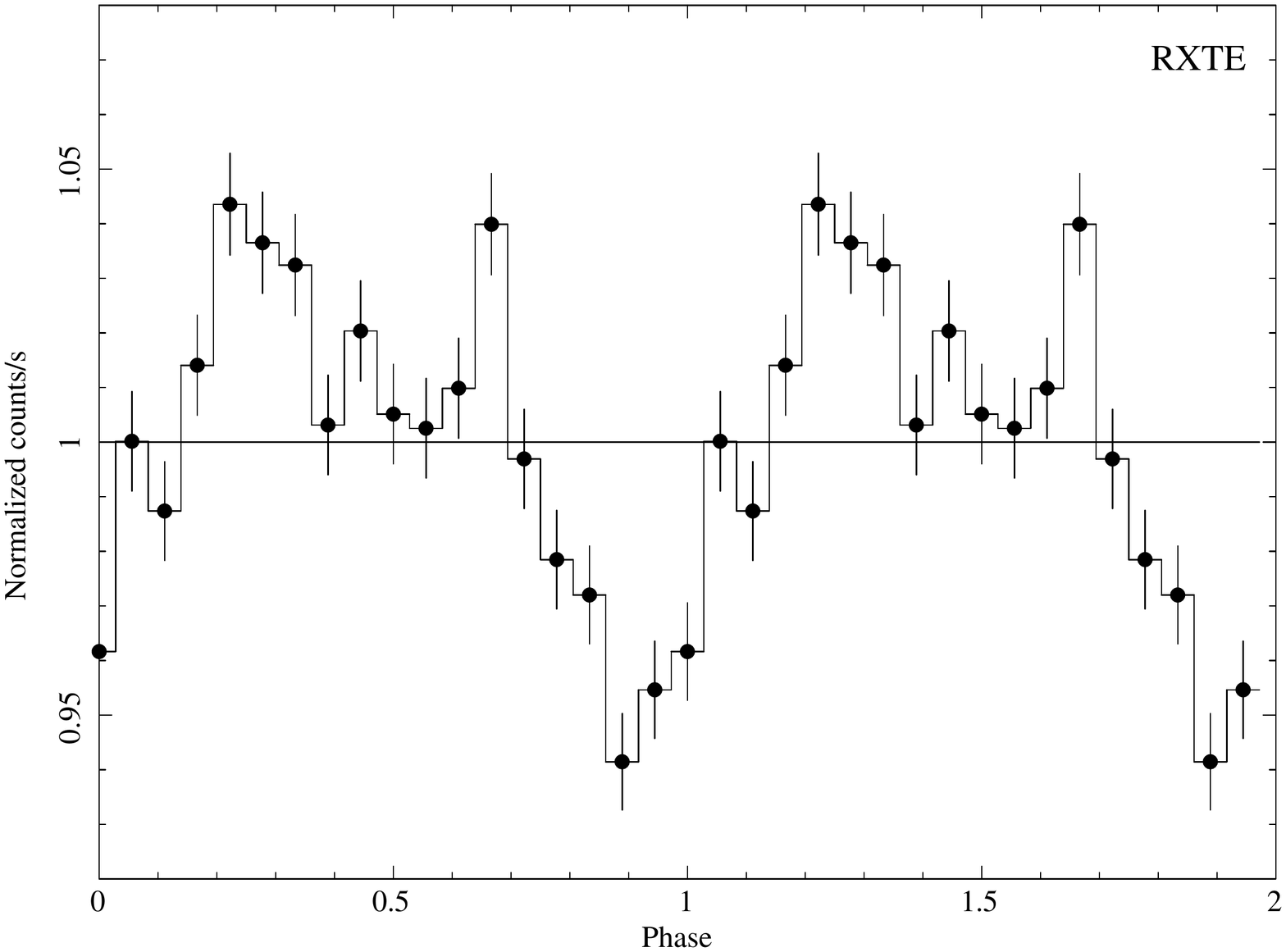}
\hspace{-0.7cm}
\includegraphics[width=4cm,height=4cm]{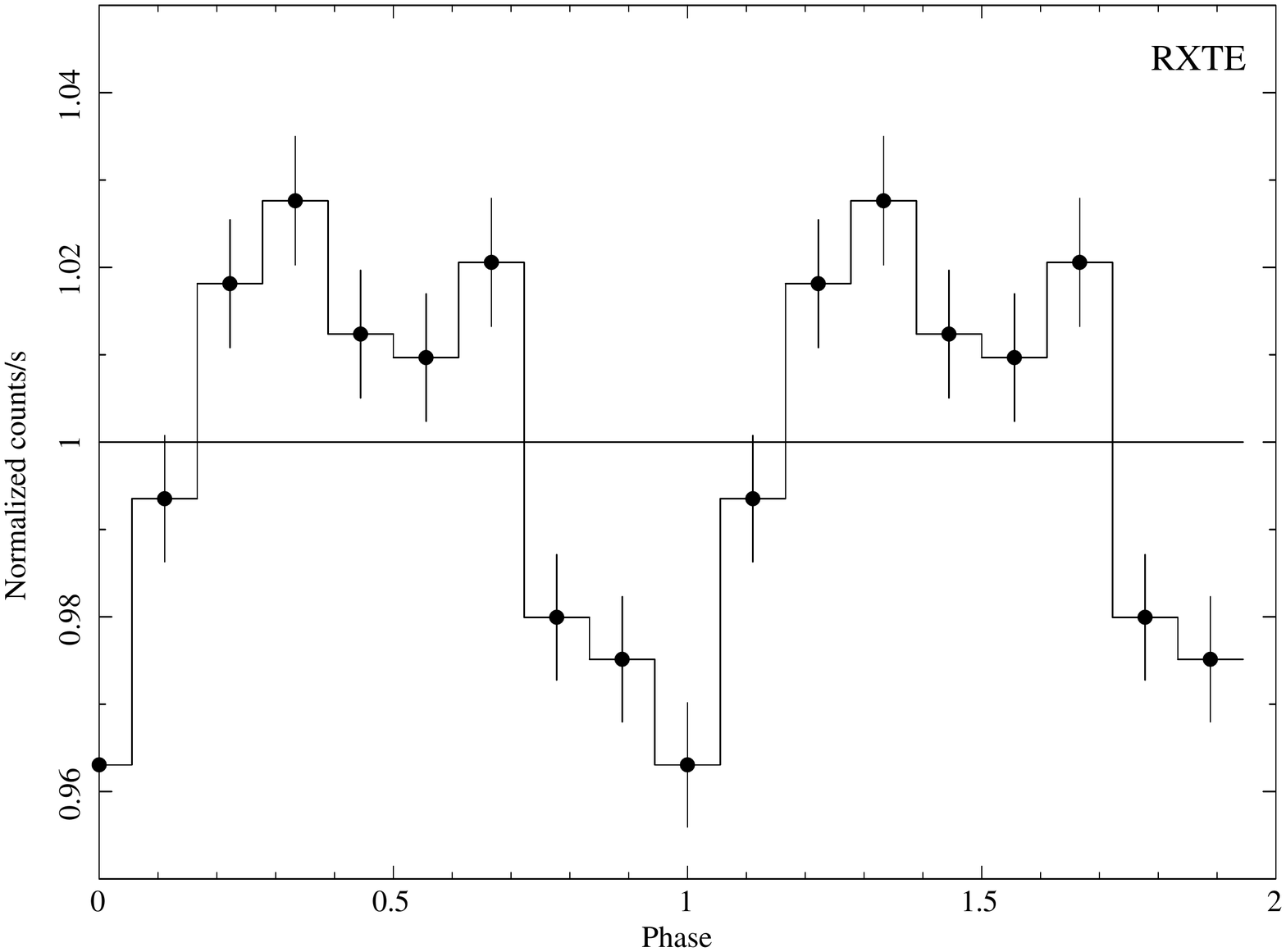}}
\vspace{-0.7cm}
\hbox{
\includegraphics[width=4cm,height=4cm]{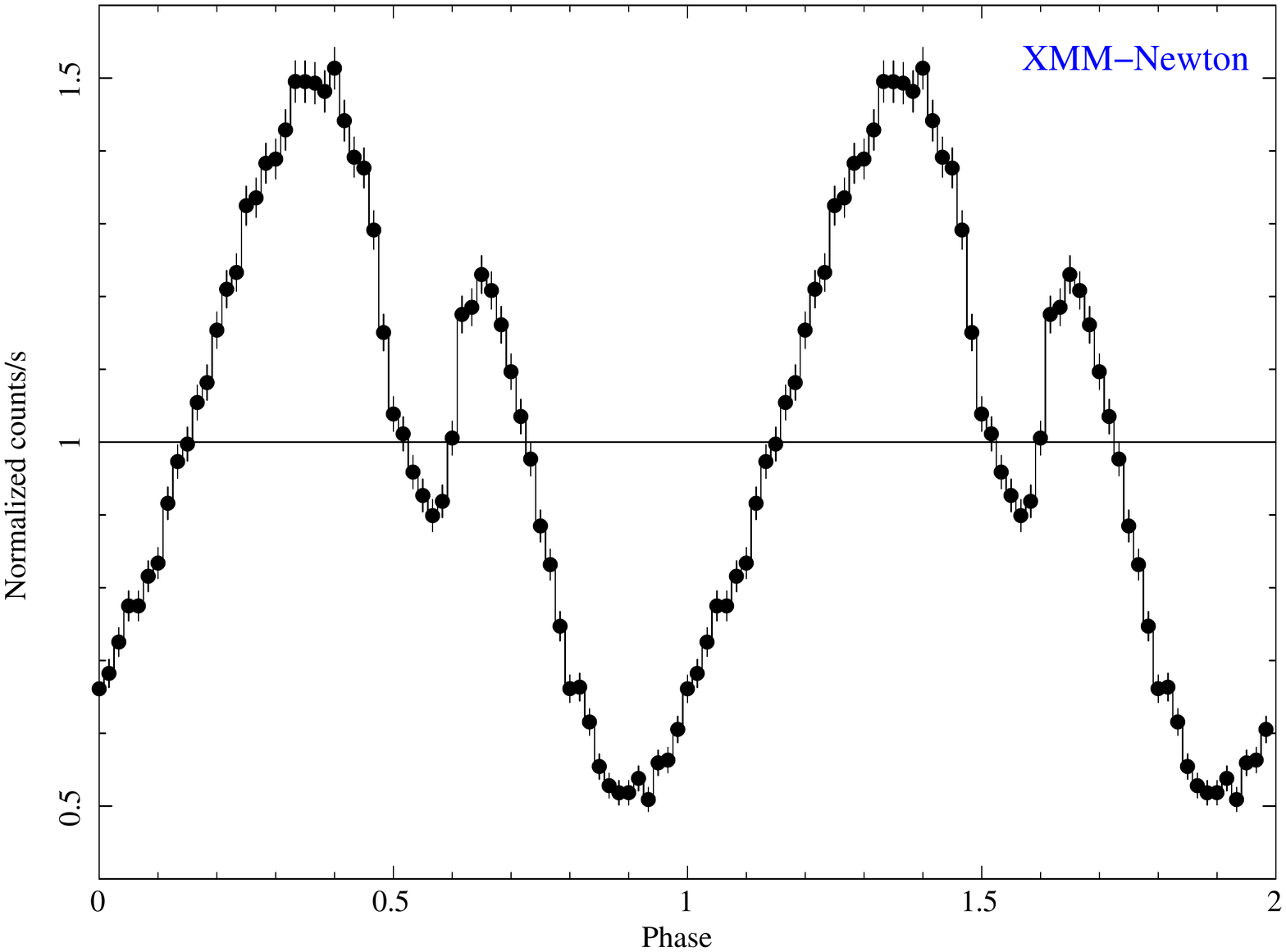}
\hspace{-0.7cm}
\includegraphics[width=4cm,height=4cm]{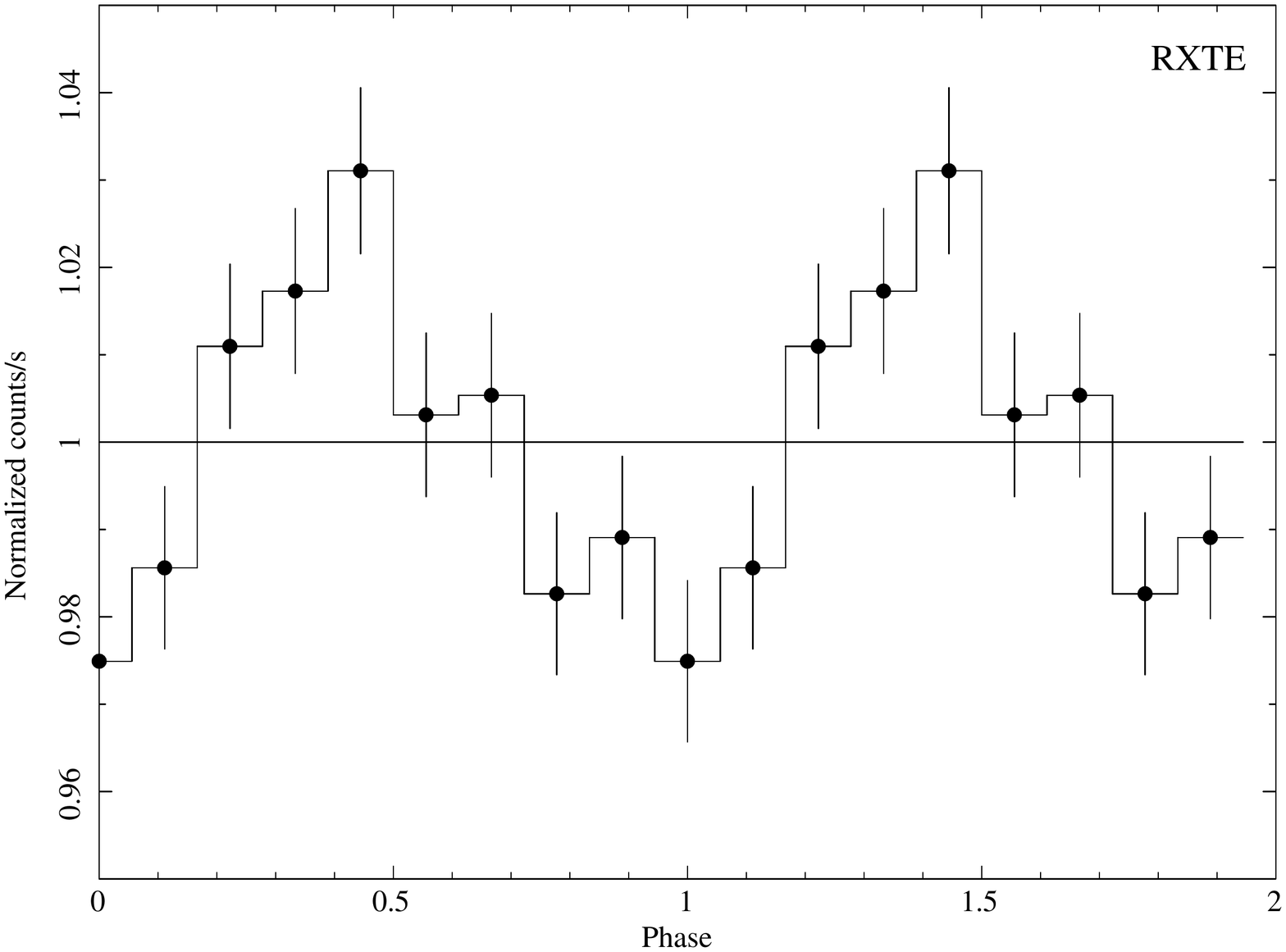}
\hspace{-0.7cm}
\includegraphics[width=4cm,height=4cm]{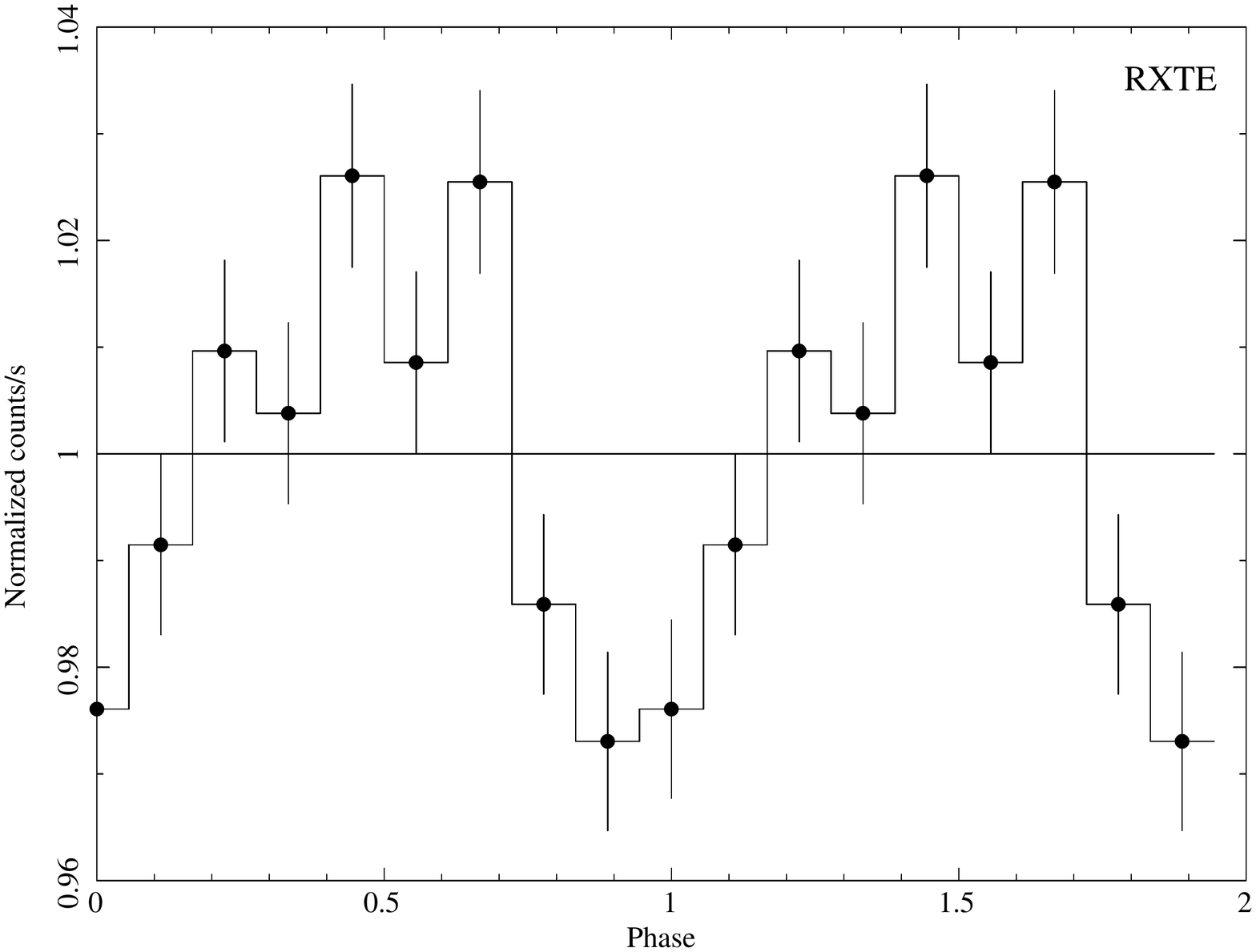}
\hspace{-0.7cm}
\includegraphics[width=4cm,height=4cm]{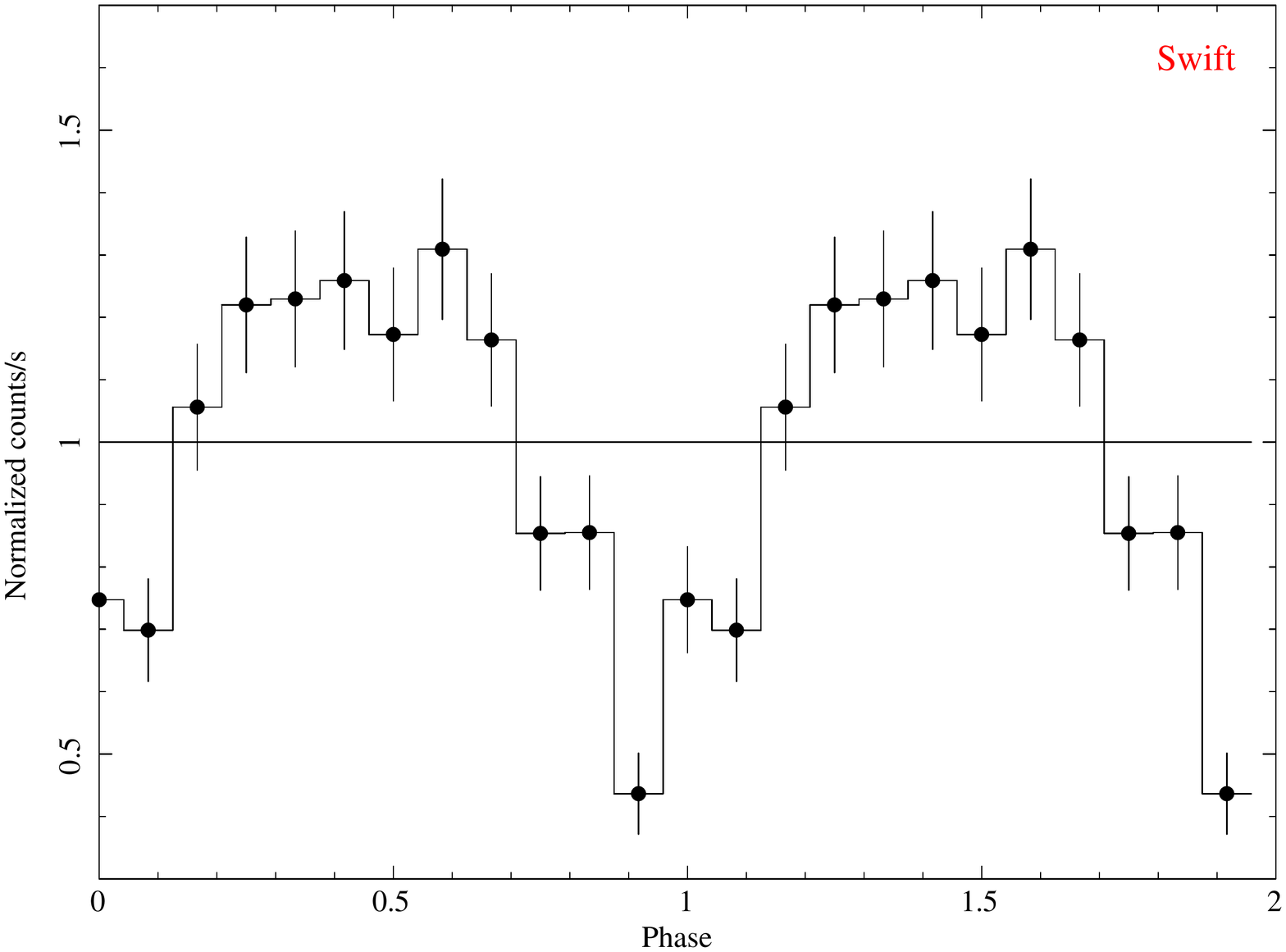}
\hspace{-0.7cm}
\includegraphics[width=4cm,height=4cm]{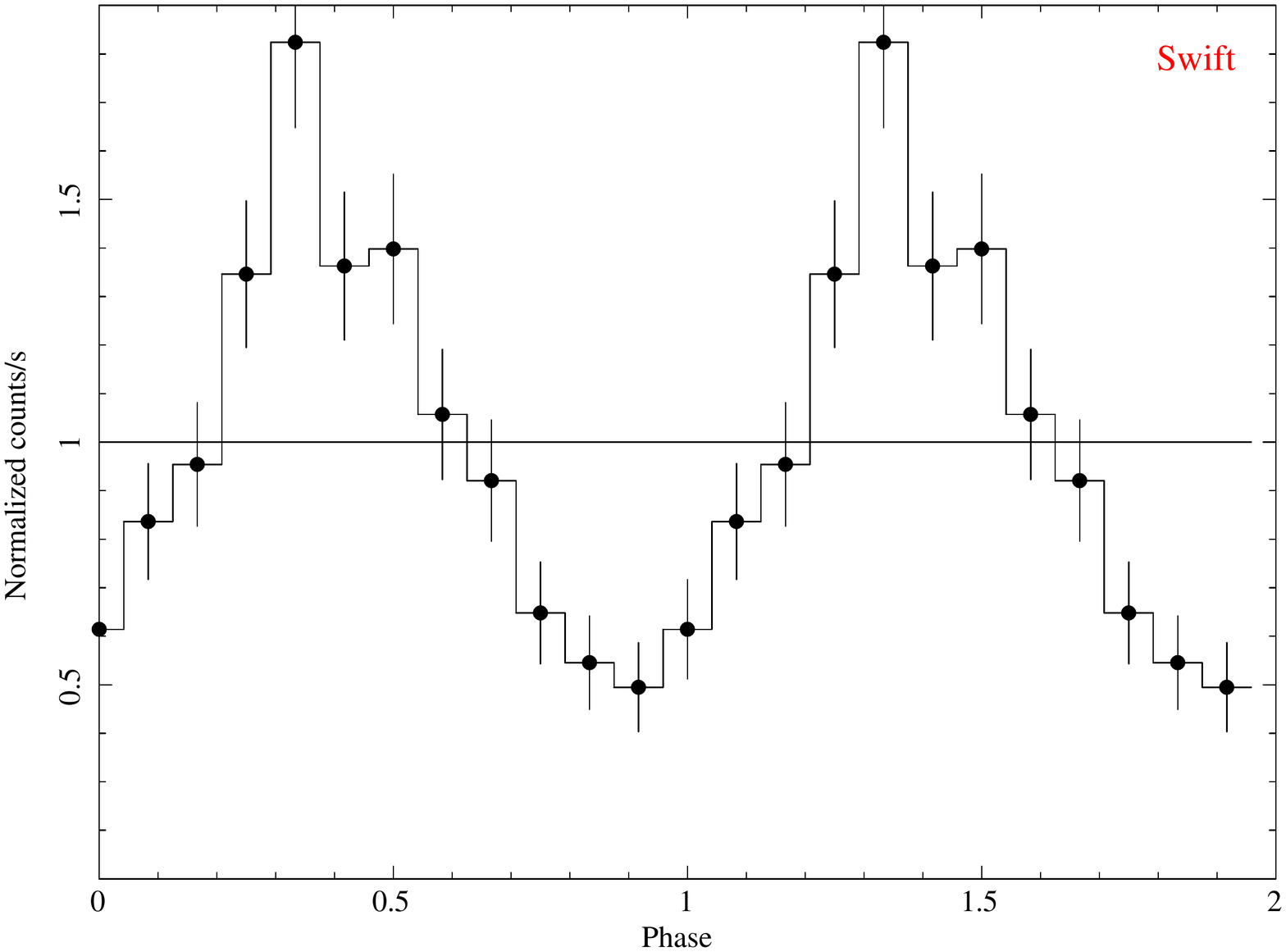}}
\vspace{-0.7cm}
\hbox{
\includegraphics[width=4cm,height=4cm]{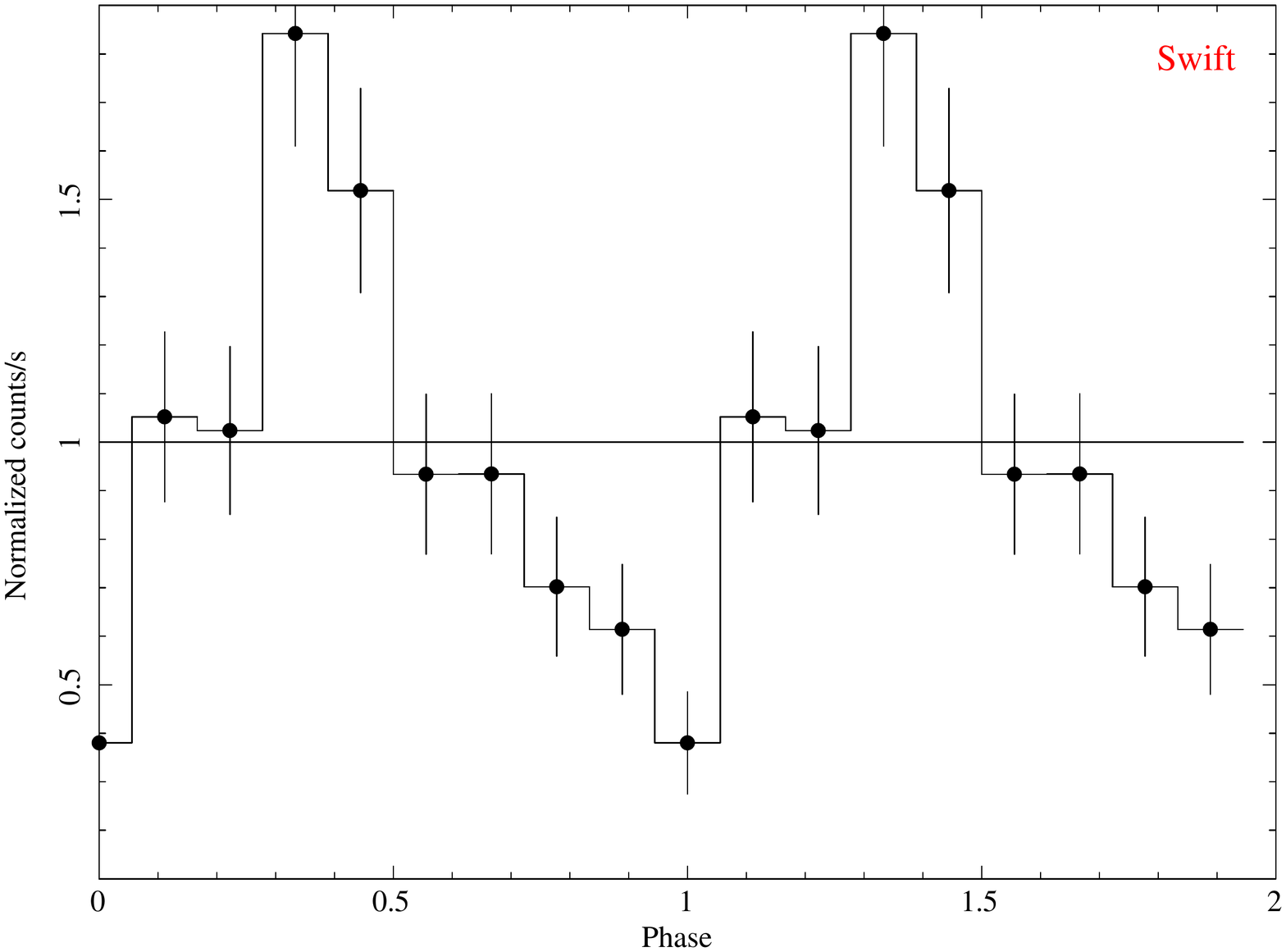}
\hspace{-0.7cm}
\includegraphics[width=4cm,height=4cm]{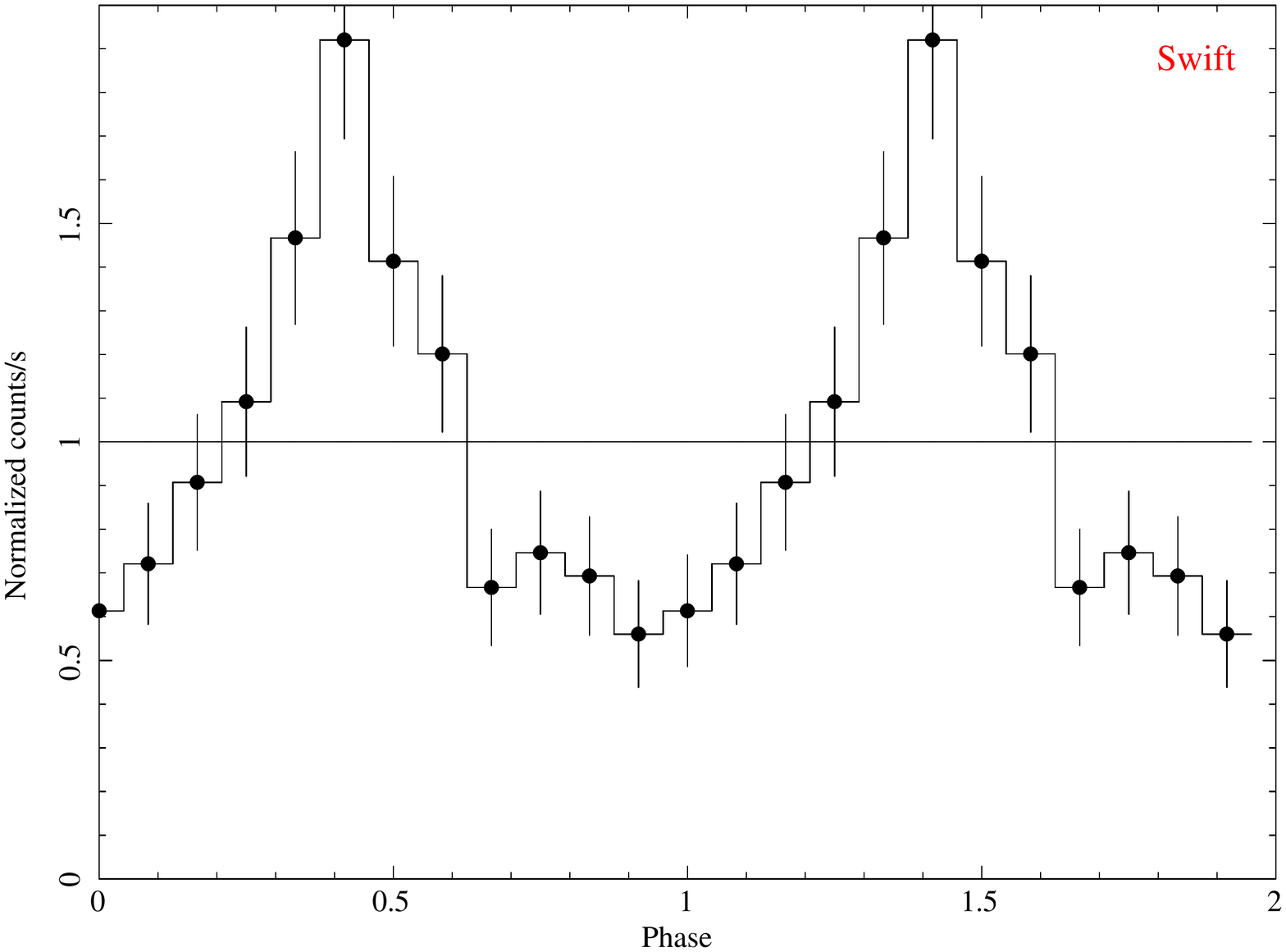}
\hspace{-0.7cm}
\includegraphics[width=4cm,height=4cm]{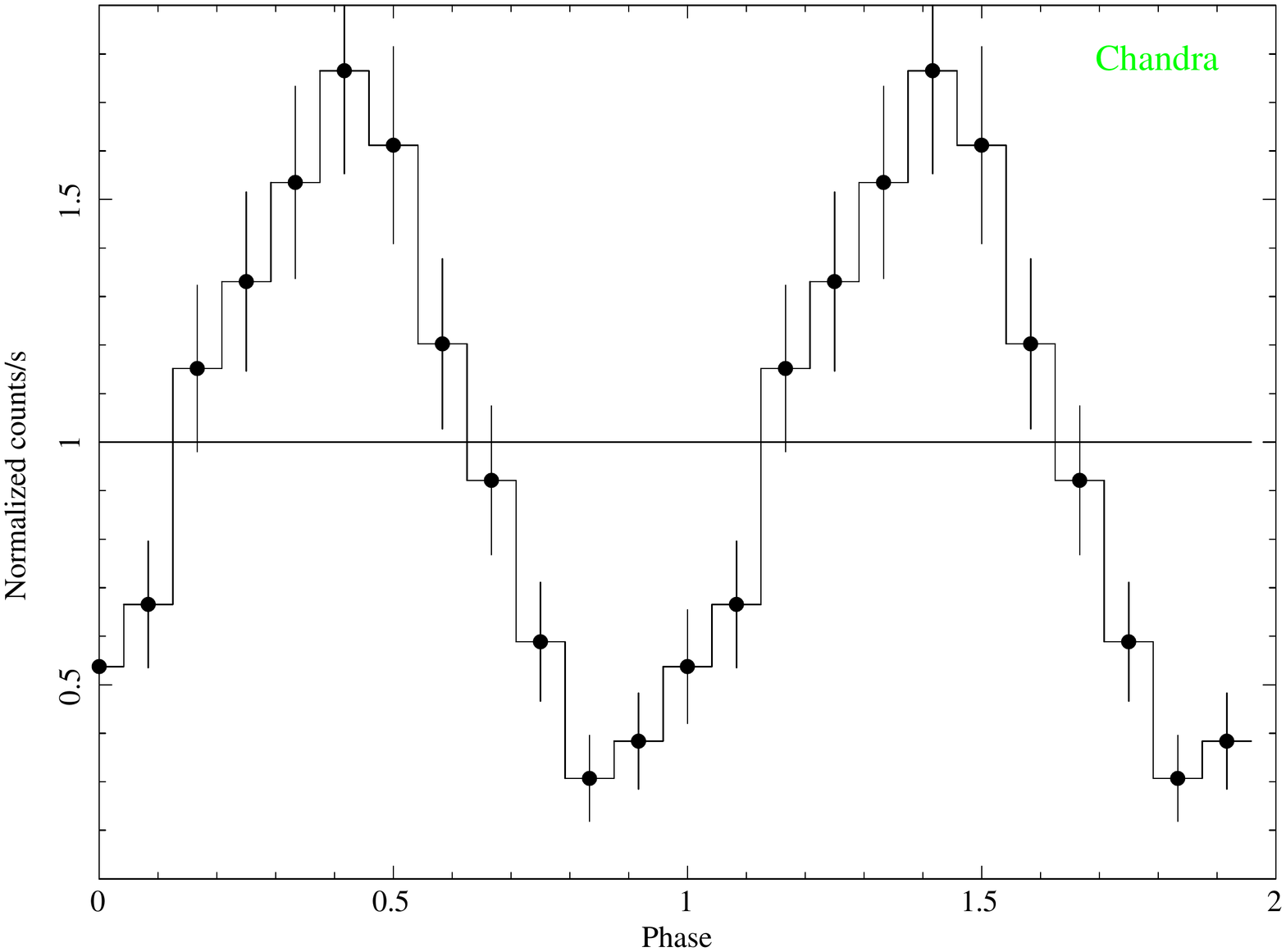}
\hspace{-0.7cm}
\includegraphics[width=4cm,height=4cm]{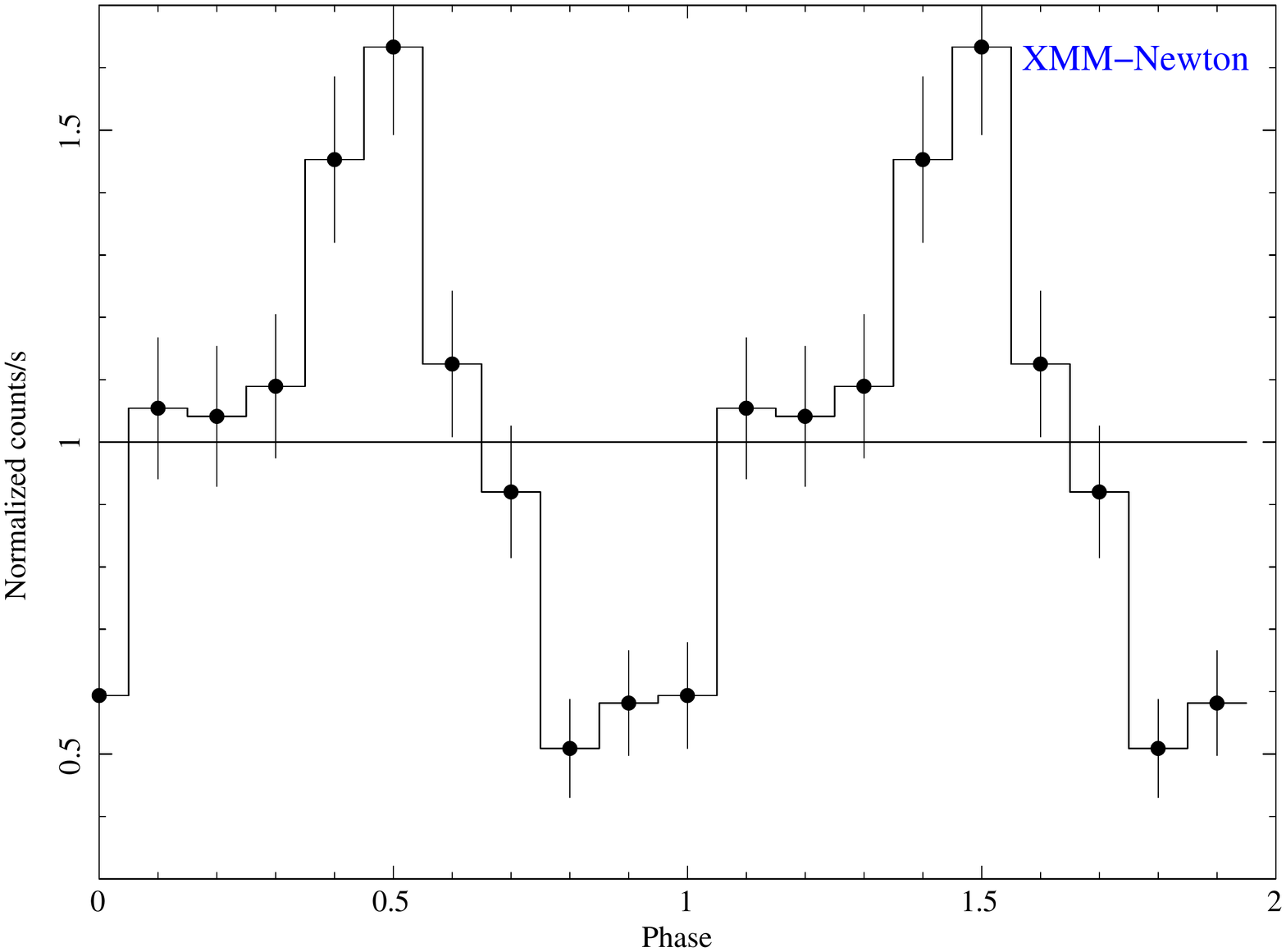}
\hspace{-0.7cm}
\includegraphics[width=4cm,height=4cm]{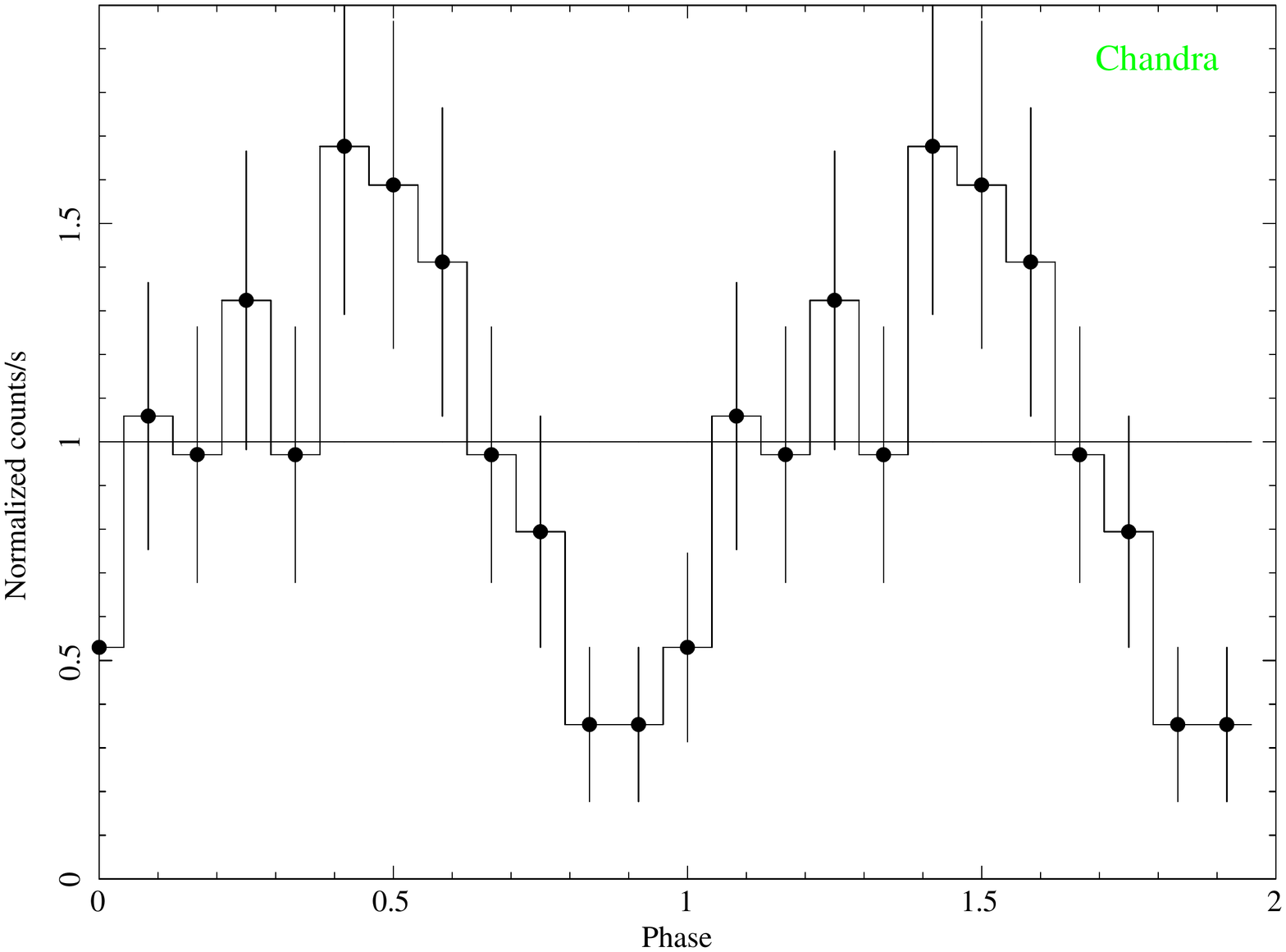}}
\vspace{-0.7cm}
\hbox{
\includegraphics[width=4cm,height=4cm]{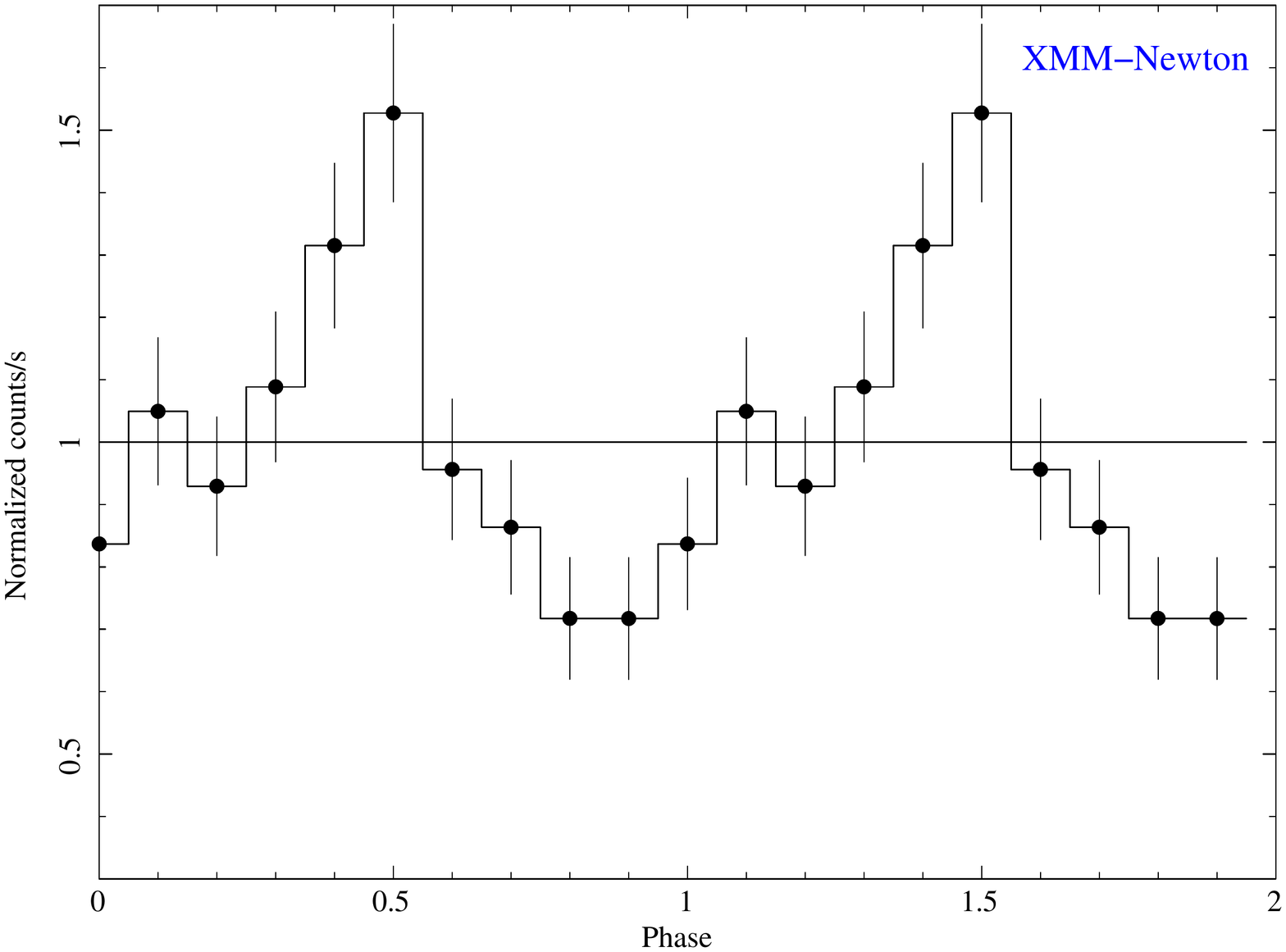}
\hspace{-0.7cm}
\includegraphics[width=4cm,height=4cm]{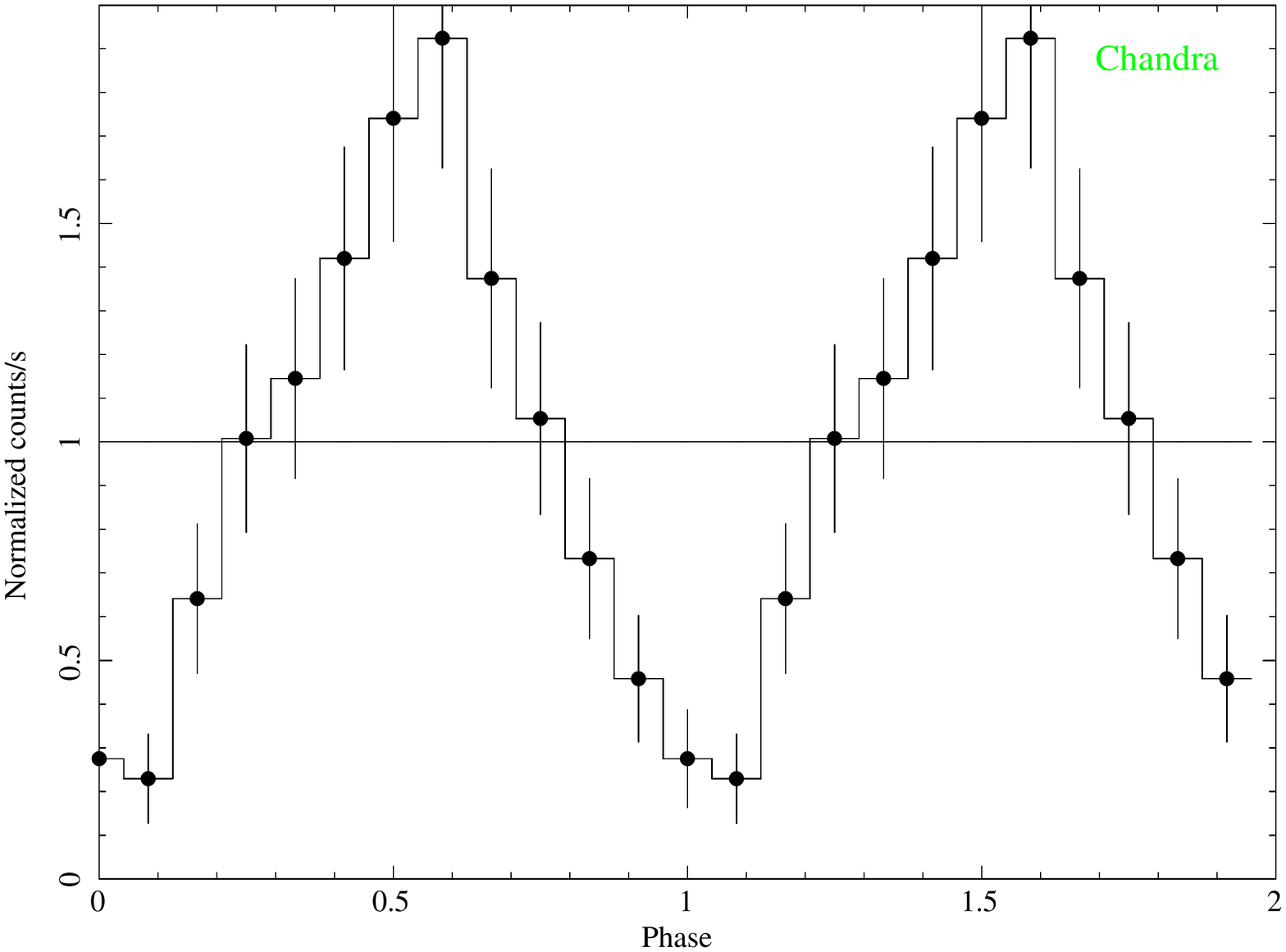}
\hspace{-0.7cm}
\includegraphics[width=4cm,height=4cm]{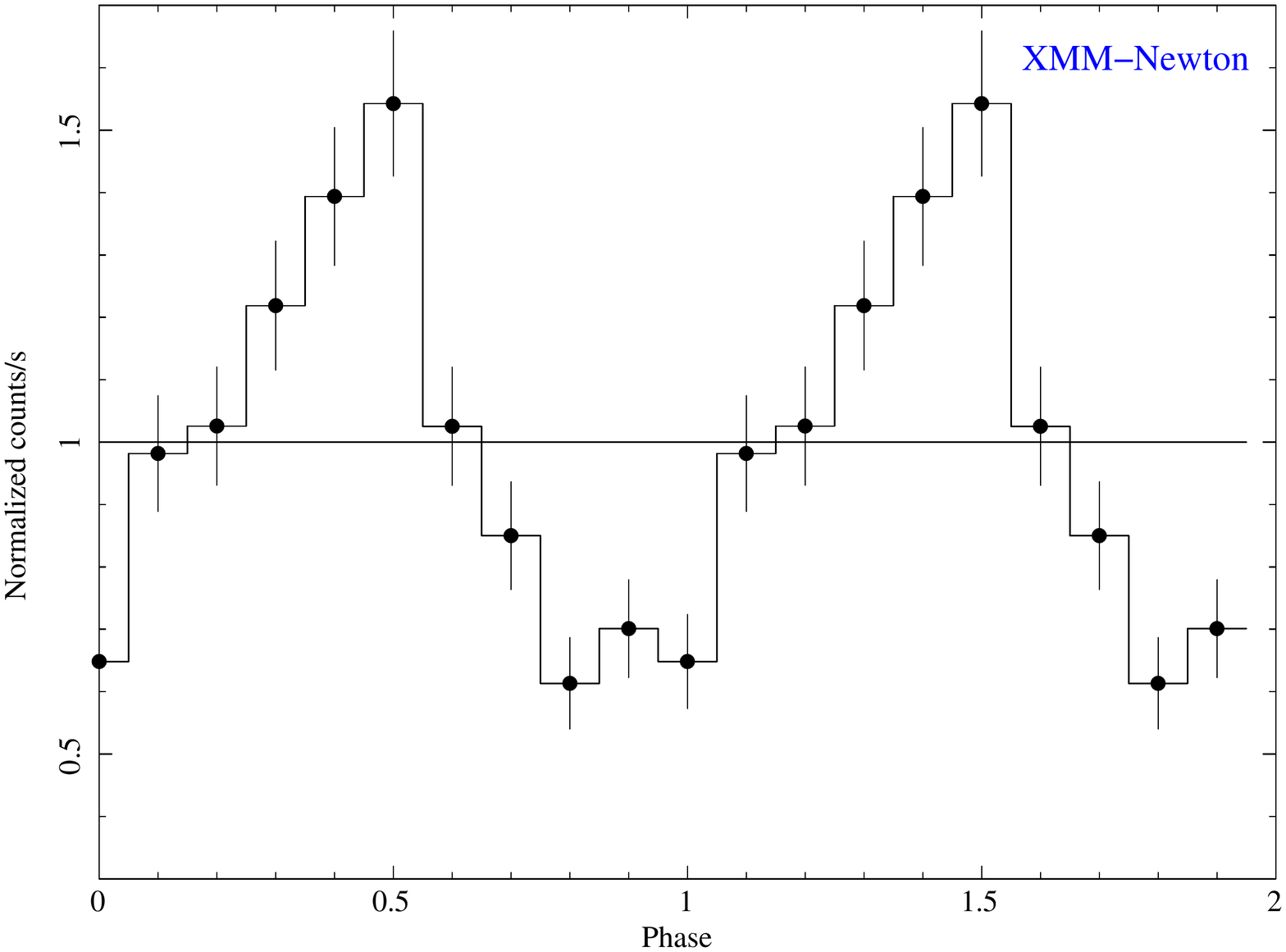}
\hspace{-0.7cm}
\includegraphics[width=4cm,height=4cm]{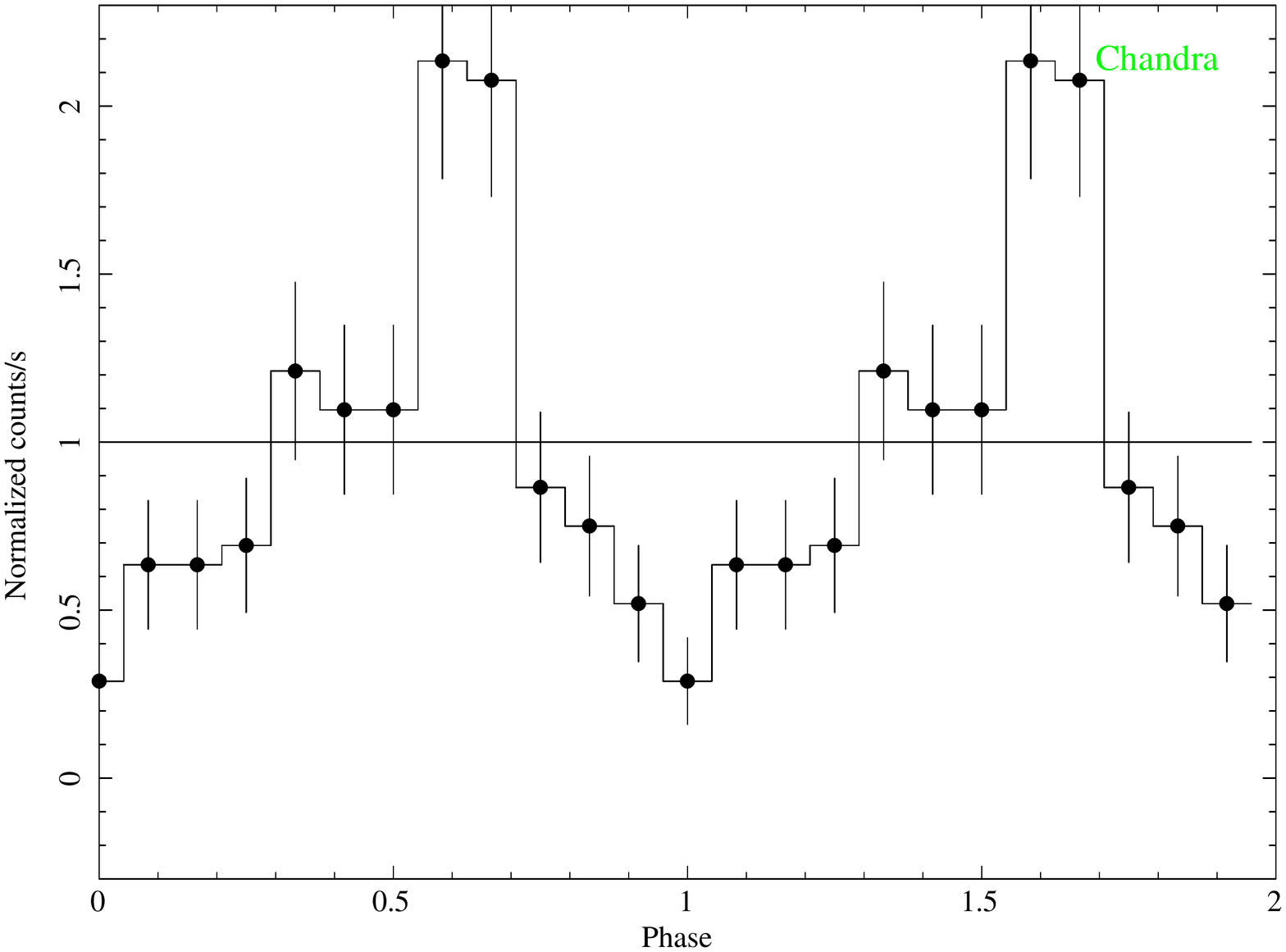}
\hspace{-0.7cm}
\includegraphics[width=4cm,height=4cm]{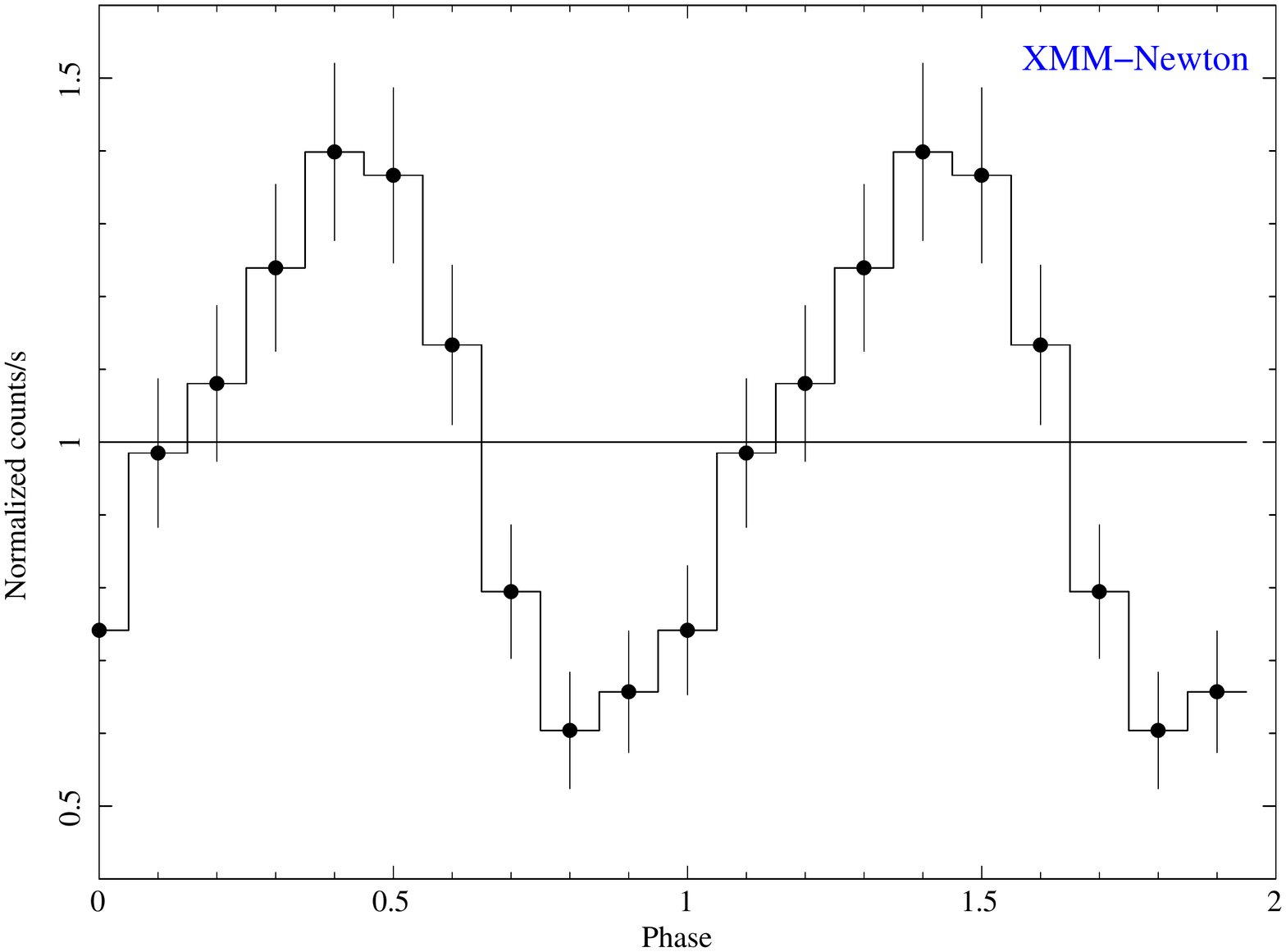}}
}

\caption{Pulse profiles evolution in the 2--10\,keV (\rxte) and
  0.5--10\,keV (\chandra\, and \xmm ) energy ranges, for most of the
  observations reported in Table \,1. Epoch increases from left to
  right, and top to bottom.}
\label{profiles}
\end{figure*}


We started the spectral analysis by fitting all the spectra together (see
Table \,\ref{obslog}) with a single component model: an absorbed blackbody or a powerlaw
model. While the former gave a good fit, a single powerlaw could not
reproduce all the spectra at the same time. Fixing the absorption
value to be the same for all spectra, for a single blackbody model
({\tt phabs*bbodyrad}) we find an acceptable fit with \nh $=
(1.15\pm0.06)\times10^{21}$\cm2 and $\chi_{\nu}^2 = 1.19$ (940 dof;
errors on the spectral parameters are all reported at 90\% confidence
level). However, not unexpectedly, the best collected spectrum (the
first \xmm\, observation on 2009--08--12; see Table\,1) gave bad
residuals at lower and higher energies (see Figure \,\ref{spectra} left panel). We
tried to model this observation alone, and indeed a single absorbed
blackbody or power-law components were not reproducing this spectrum
properly ($\chi_{\nu}^2 > 2$). We then used a composite model. Good
fits were found both using an absorbed blackbody plus a powerlaw ({\tt
  phabs*(bbodyrad + power)}; \nh$=(6.32\pm0.04)\times10^{21}$\cm2 ,
kT$=0.91\pm0.07$\,keV, $\Gamma=2.82\pm0.16$, and $\chi_{\nu}^2 = 0.97$
for 392 dof), and an absorbed resonant cyclotron scattering model
(RCS: Rea et al. (2007, 2008), or NTZ: Zane et al. (2009)). Two blackbodies
were also producing acceptable reduced chi-square values
($\chi_{\nu}^2 = 1.01$ for 392 dof) but with worse residuals at higher
energies (this is compatible with what found in Esposito et al. (2011)
and Turolla et al. (2011)). The parameters we found for the resonant
cyclotron scattering models are: \nh$=(1.9\pm0.3)\times10^{21}$\cm2 ,
$\tau=8.8\pm1.2$, $\beta=0.21\pm0.08$ and kT$=0.63\pm0.11$\,keV
($\chi_{\nu}^2 = 1.08$ for 392 dof) for the RCS model; and
\nh$=(1.8\pm0.2)\times10^{21}$\cm2 , $\Delta\phi=1.9\pm1.0$,
$\beta_{\rm bulk}=0.13\pm0.05$ and kT$=0.88\pm0.07$\,keV
($\chi_{\nu}^2 = 1.11$ for 392 dof) for the NTZ model. 

We then continued our spectral modeling of all spectra together by
adding a further component only for this observation (adding a further
component for all spectra was not significantly changing the goodness
of the fit; $\chi_{\nu}^2 = 1.13$ (938 dof); see Figure\,\ref{spectra}
middle and right panels).  In Table \,\ref{obslog} we report the
values of the single absorbed blackbody model (see also
Figure\,\ref{spectra} left panel, and Figure\,\ref{bbody}), since when
fitting all the spectra by using a composite model only for the first
\xmm\, observation, we find no change in the parameters of the other
spectra with respect to the single blackbody fit. However, although a
blackbody plus powerlaw model gives a good fit when fitting the first
\xmm\, observation alone, it is not so when fitting all data together.
This is because the powerlaw component produces an unrealistic \nh\,
increase, which does not match the value required by all the other
observations modeled by a single blackbody. We then use one of the
resonant cyclotron scattering models, the RCS model, for the joint-fit
as an empiric model for the first \xmm\, observation \footnote{Note
  that both the RCS and NTZ models are built for higher surface
  dipolar fields, hence the fact that they provide a very good fit to the 
  data is probably just an indication of the presence of some magnetospheric
  distorsion. However, no real physical information can be derived
  from the resulting magnetospheric parameters. For the purpose of this
  work, we are mainly interested in the surface thermal cooling of the
  source.}.

In addition to the joint-spectral modeling, we also fitted all the
spectra individually. Beside the first \xmm\, observation discussed
above, the last \xmm\, observation, when fitted alone with a single
blackbody model, did not give a good chi-square ($\chi_{\nu}^2 = 2.2$
for 16 dof). Given the low number of counts in the spectrum of this
observation ($\sim$400 background-subtracted counts), this deviation
from the blackbody model had only a marginal effect on the joint fit.
A better fit was found adding a second blackbody ($\chi_{\nu}^2 = 1.2$
for 14 dof) or (with a slightly worse chi-square) a power-law
component ($\chi_{\nu}^2 = 1.5$ for 14 dof). However, given the
reduced number of counts collected in this observation, a detailed
modeling of the quiescent spectrum of \src\ will be possible only when
more data will be accumulated.

We also tried to: 1) model all the spectra with two blackbodies
leaving one of the blackbodies with a fixed area mimicking the whole
surface emission, and 2) fix one blackbody to the value observed
in the last observation (see Table\,1) and leave the second blackbody
free to vary. In neither of those two cases we could find any
improvement in the modeling of the data. Note that although a joint
two blackbody model can fit the first few observations (Turolla et
al. 2011), this is no longer the case when modeling together all the
data collected in the whole 1200\,days long timespan.


\begin{figure}
\includegraphics[width=8cm]{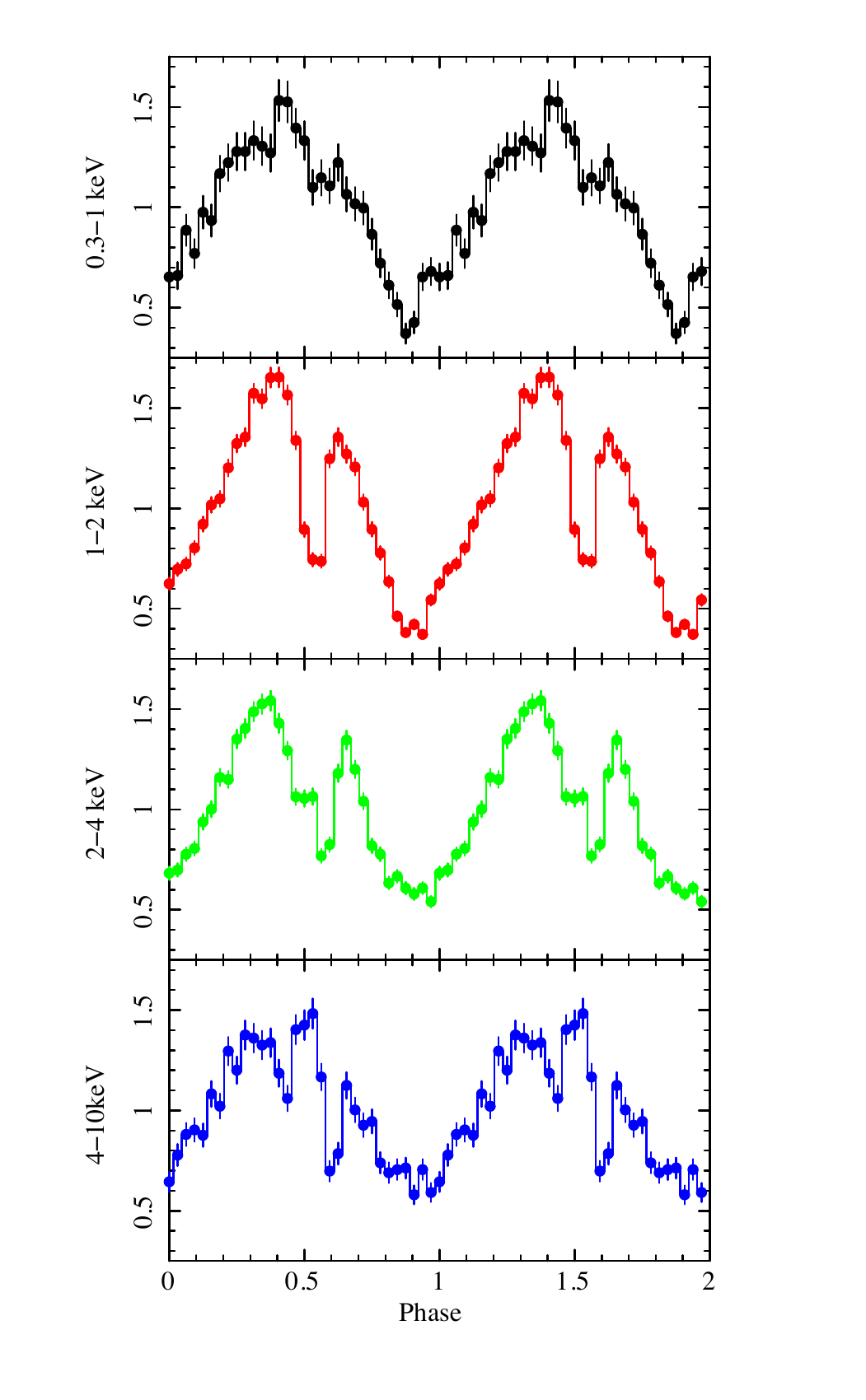}
\caption{Pulse profile (normalized counts/s versus phase) as a function of energy, relative to the first \XMM\, observation (see Table \,1). }
\label{profilesenergy}
\end{figure}

\subsection{X-ray timing analysis}
\label{sec:timing}

All the \chandra\, and \xmm\, event files collected between November
2010 and August 2012 were used in order to extend the coherent timing
solution we derived in Rea et al. (2010): $P= 9.07838827(4)$\,s 90\%
c.l., and 3$\sigma$ first period derivative upper limit of $|\dot{P}|<
6.0\times 10^{-15}$\,s\,s$^{-1}$ at epoch 54993.0 MJD. Photon arrival
times were corrected to the barycentre of the solar system\footnote{We
  have corrected the arrival times of the last \xmm\, observation for
  the 2012 June 30 leap second (see
  http://xmm.esa.int/sas/current/watchout/12.0.0/leapsec\_2012.shtml
  for further details).}. Timing analysis was carried out by means of
a phase-fitting technique (details on this technique are given in
\citealt{dallosso03}; see also Esposito et al. 2010 for further
details on this source).  Given the intrinsic variability of the pulse
shape as a function of time (see Figure\,\ref{profiles}), we inferred
the phase of the modulation by fitting the average pulse shape of each
observation with a number of harmonics, the exact number of which is
variable and determined by requesting that the inclusion of any higher
harmonic is statistically significant (by means of an F-test). All
data reported in Table\,1 were folded using a reference period
9.07838880562798\,s at epoch 54993\,MJD, and fitted with one or more
harmonics. In Figure\,\ref{timing} we plot the phases at which the
fundamental sine function is equal to zero in its ascending part
(positive derivative).

The fit of the resulting pulse phases with a linear component gives a reduced $\chi^2_r \sim 3.2$ for 26 degree of freedom (d.o.f. hereafter). The inclusion of a quadratic term,  corresponding to a first period derivative component, was found to be significant at a confidence level of 3.5$\sigma$ (by means of a F-test).  
The resulting best-fit solution corresponds to $P=9.07838822(5)$\,s (1$\sigma$ c.l., 2 parameters of interest; epoch 54993.0 MJD) and  $\dot{P}=4(1)\times 10^{-15}$\,s\,s$^{-1}$ with a reduced $\chi^2_r \sim 2.1$ (for 25 d.o.f.; see also Figure\,\ref{timing}).  
The new timing solution implies a r.m.s. variability of only 0.2\,s . As depicted above, the time evolution of the phase can be described by a relation of the
form $\phi=\phi_0+2\pi(t-t_0)/P-\pi(t-t_0)2\dot{P}/P^2$.

\begin{table}
\centering
\caption{Pulse phase spectroscopy of the first \XMM\, observation of \src .}
\smallskip
\begin{tabular}{@{}lccc}
\hline
Phase & Counts~s$^{-1}$ & Flux$^{a}$  &  Photon Index$^{b}$  \\
\hline
0.0--0.4 & $1.46\pm0.01$ & $7.5\pm0.1$ ($1.71\pm0.02$)& $2.86\pm0.15$\\
0.4--0.6 & $1.00\pm0.01$ & $5.5\pm0.1$ ($1.20\pm0.02$) & $4.08\pm0.44$\\
0.6--1.0 & $1.26\pm0.01$ & $6.6\pm0.1$ ($1.47\pm0.02$) & $3.11\pm0.21$\\
\hline
\end{tabular}
\begin{list}{}{}
\item[$^{a}$] Absorbed flux in the 0.5--10\,keV energy range, and in units of $10^{-12}$\ergs ($10^{-3}$~photons~cm$^{-2}$s$^{-1}$). See also \S\ref{sec:pps}.
\item[$^{b}$] Fitted model is: {\tt phabs*(bbodyrad+power)}; \nh $= (6.9\pm0.5)\times10^{21}$\cm2 ,  kT$_{BB} = 0.91\pm0.01$~keV, BB norm$=0.82\pm0.05$, and $\chi_{\nu}^2 = 0.96$ (for 680 dof). 
\end{list}
\end{table}


To further assess the significance of the quadratic component,
reflecting the period derivative, we performed detailed Monte Carlo
(MC) simulations assuming as the null model a simple linear relation
(see Protassov et al. 2002 for further details).  By running $10^{5}$
MC simulations we verified that the quadratic component is significant
at $>99.96\%$ confidence level, which is in very good agreement with
what estimated by means of the F-test. We notice that examining the
simulated data with different sampling distributions does not change
significantly the results. Furthermore, we performed the same MC
simulations using the whole set of observations but without
considering the last \XMM\, observation. We find a chance probability
for the addition of a quadratic component of 0.65\% ($<
3\sigma$). Given the strong influence of the last of our observations
in the determination of the period derivative, we will perform further
X-ray observations in the next few years in order to increase the
significance of the current $\dot{P}$ measurement.

Using this $\dot{P}$ measurement, we infer a surface dipolar
magnetic field strength of $B_{\rm dip} = (6\pm2) \times 10^{12}$~G,
calculated at the neutron star equator. This value is fully consistent
with the 3$\sigma$ upper limit reported in \cite{rea10}. We also
estimate a characteristic age of $\tau_{\rm c} \simeq P/2\dot{P} \sim
35$\,Myr, and a rotational power of $\dot{E} \simeq 3.9\times10^{46}
\dot{P}/P^3 \sim 2\times10^{29}$\ergs .

Based on the above phase coherent timing solution, we also studied the
pulse shape and pulsed fraction evolution. Figure\,\ref{timing} shows
the pulsed fraction evolution as a function of time. There is an
evident increase starting soon after the burst detection with a
recovery towards an asympthotic quiescent value which appears to be at
about the 70-80\% level.

\begin{figure*}
\hbox{
\includegraphics[width=8cm,height=7cm]{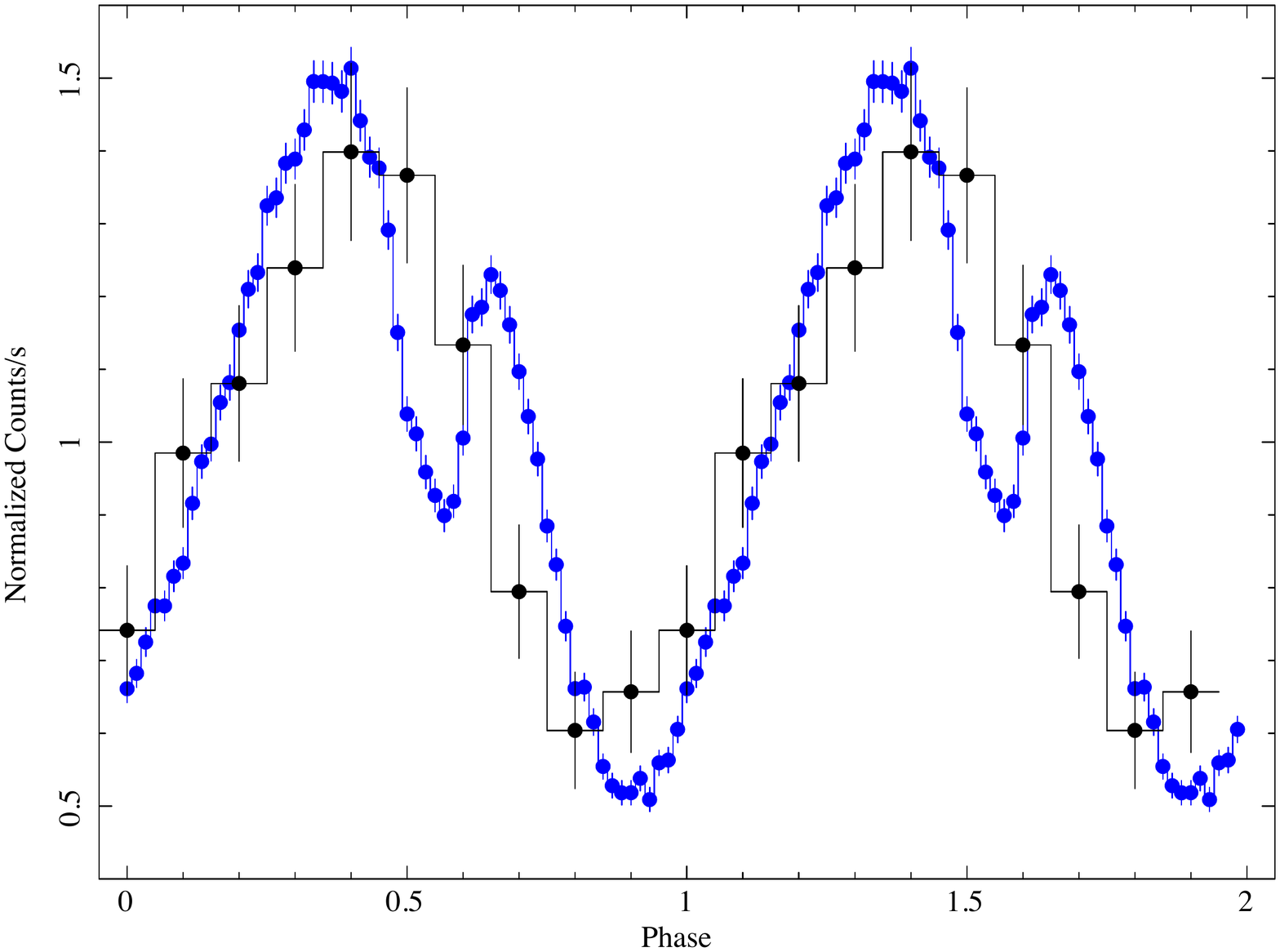}
\includegraphics[width=10cm,height=7cm]{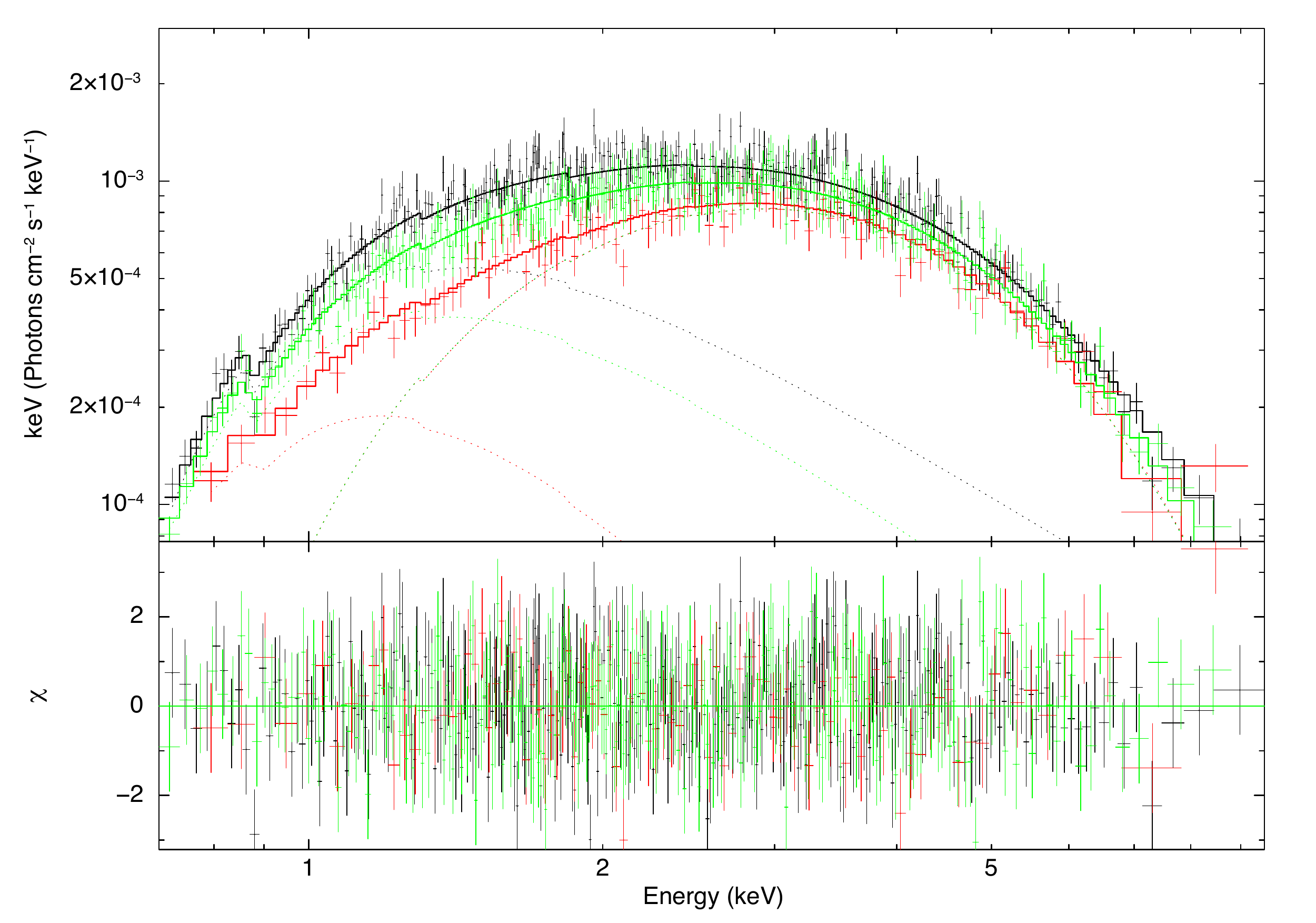}}
\caption{{\em Left panel}: Pulse profile of the first (blue) and last
  (black) \XMM\, observations in the 0.5--10\,keV energy band. {\em
    Right panel}: phase-resolved unfolded spectra for the first \xmm\,
  observation. The spectra are relative to phases: 0.0--0.4 (black),
  0.4--0.6 (red), and 0.6--1 (green). The phase ranges are relative to
  the blue pulse profile in the left panel.}
\label{profilexmm}
\label{pps}
\end{figure*}

In Figures\,\ref{timing}, \ref{profilexmm} and \ref{profilesenergy} we
study in the detail the shape of the pulse profile as it evolves in
time and in energy. By looking at the profile shapes of all X-ray
observations performed so far, the source appears to be switching
among a three/two/single peak shape during the early outburst phases,
with no clear trend in time (Figures\,\ref{timing}). The pulse profile
stabilizes to a single peak about three months after the outburst
onset. However, studying in detail the first and the last \xmm\,
observations, a few key pieces of information can be extracted: a) at lower
energies ($<$1\,keV), the pulse profile is mainly single peaked, while
the second, and possibly also the third peak appears at higher
energies (see Figure\,\ref{profilesenergy}); b) the main component of
the pulse profile continues to be at the same phase over the whole
outburst decay (see Figure\,\ref{profilexmm}).


\subsection{Pulse Phase Spectroscopy}
\label{sec:pps}

We performed a pulse phase spectroscopy of the first \xmm\, observation. A clear pulse phase dependence of the spectrum is already observed by
simply looking at the pulse profile changes as a function of the
energy (see Figure\,\ref{profilesenergy}). In order to quantify the
spectral variability as a function of the rotational phase, we
performed a pulse phase spectroscopy extracting the spectra from
phases 0--0.4, 0.4--0.6, and 0.6--1. These phase intervals were chosen
by looking at Figure\,\ref{profilesenergy} in order to isolate the dip
in the 1--4\,keV pulse profiles at phase $\sim$0.55. In Table\,2 we
report the results of our modeling. The phase-averaged spectrum is not
well fit by either a single blackbody nor a powerlaw.  We then
used an absorbed blackbody plus powerlaw modeling (note that the RCS
and NTZ models are not suited for phase resolved analysis since they
are intrinsically phase-average), using for the photoelectric
absorption model the same cross-section and abundances as for the
phase-average spectrum (see \S\ref{sec:spectra}).  The blackbody
temperature and radius were consistent in all three spectra, hence we
fixed them to be the same for all spectra (kT$_{BB} =
0.91\pm0.01$~keV, BB norm$=0.82\pm0.05$), while a variability
$>3\sigma$ has been observed in the photon index (it changed from
about 2.9 to 4.1 between the spectra of the first peak and the dip).

However, from Figure\,\ref{pps} it is clear that the main difference
in the spectra is at lower energies. In particular, above 5\,keV the
three spectra are very similar, while the 0.4--0.6 phase-resolved
spectrum seems to have less counts than the other below such energy.

\section{Green Bank Telescope radio observations}
\label{gbt}

We observed \src\, using the 101\,m Green Bank Telescope (GBT) on 2012
October 4th, during the return to quiescence. Data were acquired with
the Green Bank Ultimate Pulsar Processing Instrument (GUPPI; DuPlain
et al.  2008) at a central frequency of 2.0\, GHz (with a bandwidth of
800 MHz, integration time of $\sim 5400$ s and sampling time of 655
$\mu$s) and 820\, MHz (with a bandwidth of 200 MHz, integration time
of $\sim 5600$ s, and sampling time of 655 $\mu$s).  To minimize the
dispersive effects of the interstellar medium, the bandwidths were
split into 2048 and 512 channels, respectively.  The working of the
system was checked looking at the pulsar PSR B0450+55. 
A mask was first applied to the full resolution data for reducing
the effects of impulsive RFI and of bad channels. Then the cleaned data
were dowsampled a factor 2 in sampling time, matching
the frequency resolution in order to have a maximum dispersion smearing 
of order 1.3 ms in each channel for a pulsar with dispersion measure (DM)
$\sim 100$ pc cm$^{-3}$ 

The ephemerides acquired from the X-ray observations (see \S3.2), were
used to fold the resulting data in $\sim 3$-min long subintegrations at the
known magnetar period. We also folded the data at half, one third and
a quarter of the nominal period in order to detect putative higher
harmonics components of the intrinsic signal, in case the latter were
deeply contaminated by interference (RFI). Folding was done using
dspsr (van Straten \& Bailes 2011). The sub-integrations and the
frequency channels, cleaned from RFI, were then searched around the
pulsar period P and over a wide range of DM
values (from 0 to 1000 pc cm$^{-3}$) to find the P--DM combination
maximizing the signal-to-noise ratio. No dispersed signal was found in
the data down to a signal-to-noise limit of 10 in both datasets. Given the
parameters of the antenna and of the
receivers\footnote{http://www.gb.nrao.edu/gbtprops/man/GBTpg.pdf}, and
assuming a pulsar with a duty cycle of 10\%, that translates in flux
densities of $\sim 0.02$\,mJy and $\sim 0.05$\,mJy, for the 2\,GHz and
820\,MHz observations, respectively. Data were also blindly searched
for a periodic signal in the Fourier domain, and for single
de-dispersed pulses (within a DM range from 0 to 200 pc cm$^{-3}$). No
signal was found in either the Fourier domain (down to a spectral
signal-to-noise ratio 4) or in the single pulse searches (down to a
signal-to-noise ratio 5 for the individual pulses).

No previous search for pulsed radio emission had been
performed at 2\,GHz, whereas the flux density value at 820\,MHz
improves by $\sim 15\%$ the limit of the observation at
820\,MHz performed on 2009 July 19th (Lorimer et al. 2009, Atel 2096), 
when the source was in the phase of outburst. Assuming a typical pulsar 
spectral index of 1.7, a typical duty cycle $\sim 10\%$ and a distance
of 2 kpc (van der Horst et al. 2009), the observations at 820\, MHz sampled 
more than $97\%$ of the luminosity distribution of the population 
of known ordinary pulsars with rotational period longer than 100 ms, 
as derived from the ATNF pulsar 
catalogue\footnote{http://www.atnf.csiro.au/research/pulsar/psrcat/}.

\section{Plateau de Bure mm observations}
\label{pdb}

\src\, was observed with the Plateau de Bure Interferometer (PdBI) at 1.8~mm 
(166.50 GHz) in the D configuration between June and July  2011 (June 27, and 
July 09, 10, 15, and 16).  This configuration provides baselines between 22.1 and 
95.6\,m. The phase center of the observations was 04:18:33.867, $+$57:32:22.910. 
The dominant track was July 15 (8h-track and excellent weather conditions). The system 
temperatures were typically in the 150 to 200~K range, and the averaged atmospheric 
precipitable water vapor was 2\,mm.  The gain calibration was performed observing
the quasars B0552$+$398 and J0512$+$294. After calibration, the phase $rms$ was 
20--60$^{\circ}$. The bandpass calibrator used was B0851$+$202. The adopted flux 
density for the flux calibrator 3C273 was 16.57 Jy.  Calibration and imaging were 
performed using the standard procedures in the CLIC and MAPPING packages 
of the GILDAS\footnote{GILDAS data reduction package is available at  
http://www.iram.fr/IRAMFR/GILDAS}  software.  The resulting final map, obtained
combining all the data, yields a synthesized beam size of $3\farcs95\times3\farcs16$ 
with a position angle of $PA = 0.0^{\circ}$.  The $rms$ noise achieved using
the full 3.6~GHz provided by the WideX correlator is 60~$\mu$Jy beam$^{-1}$.
The primary beam of the PdBI at 166.50 GHz is $30\farcs3$. 

\begin{figure*}
\hbox{
\includegraphics[width=6.2cm]{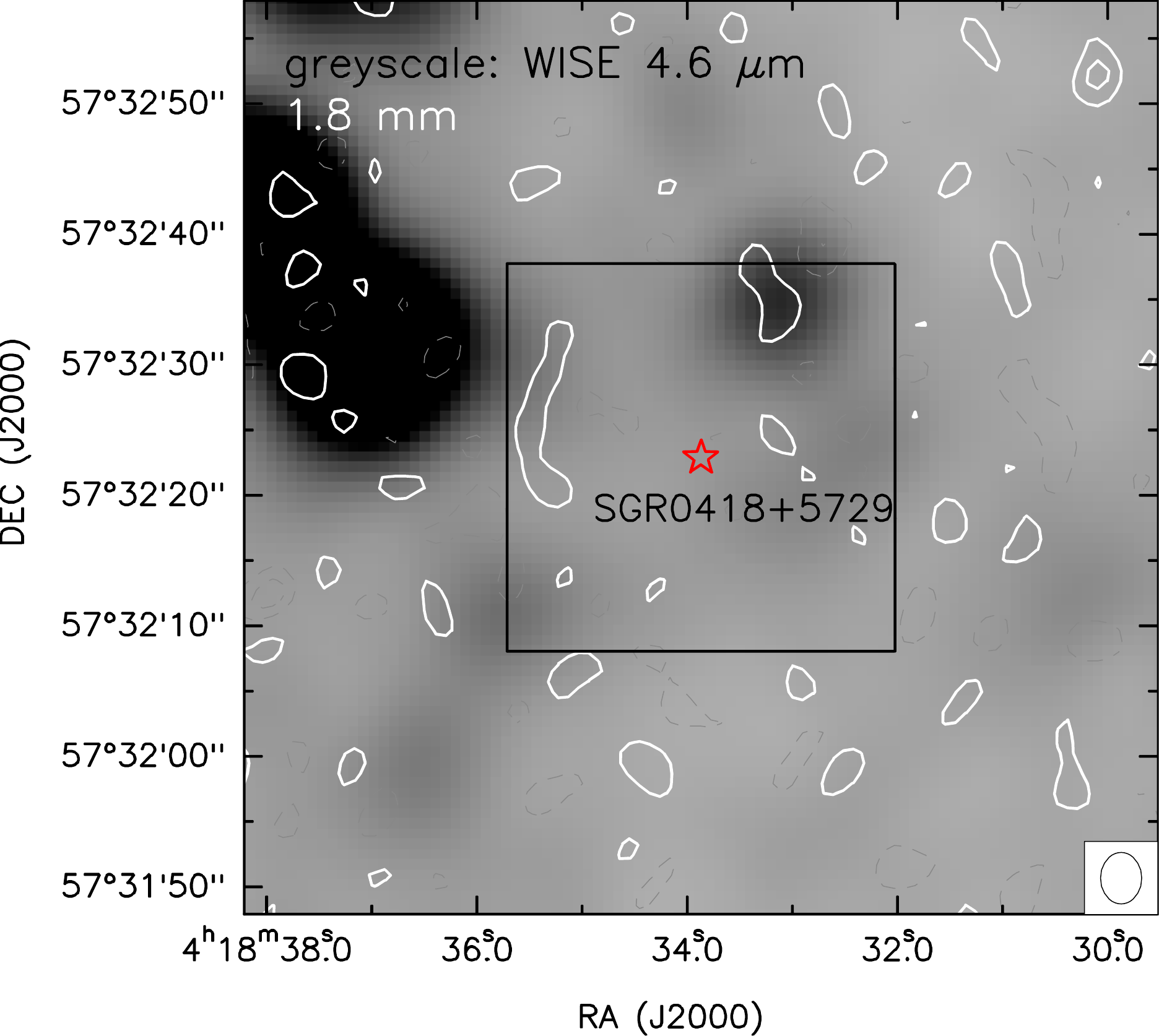}
\includegraphics[width=5.2cm]{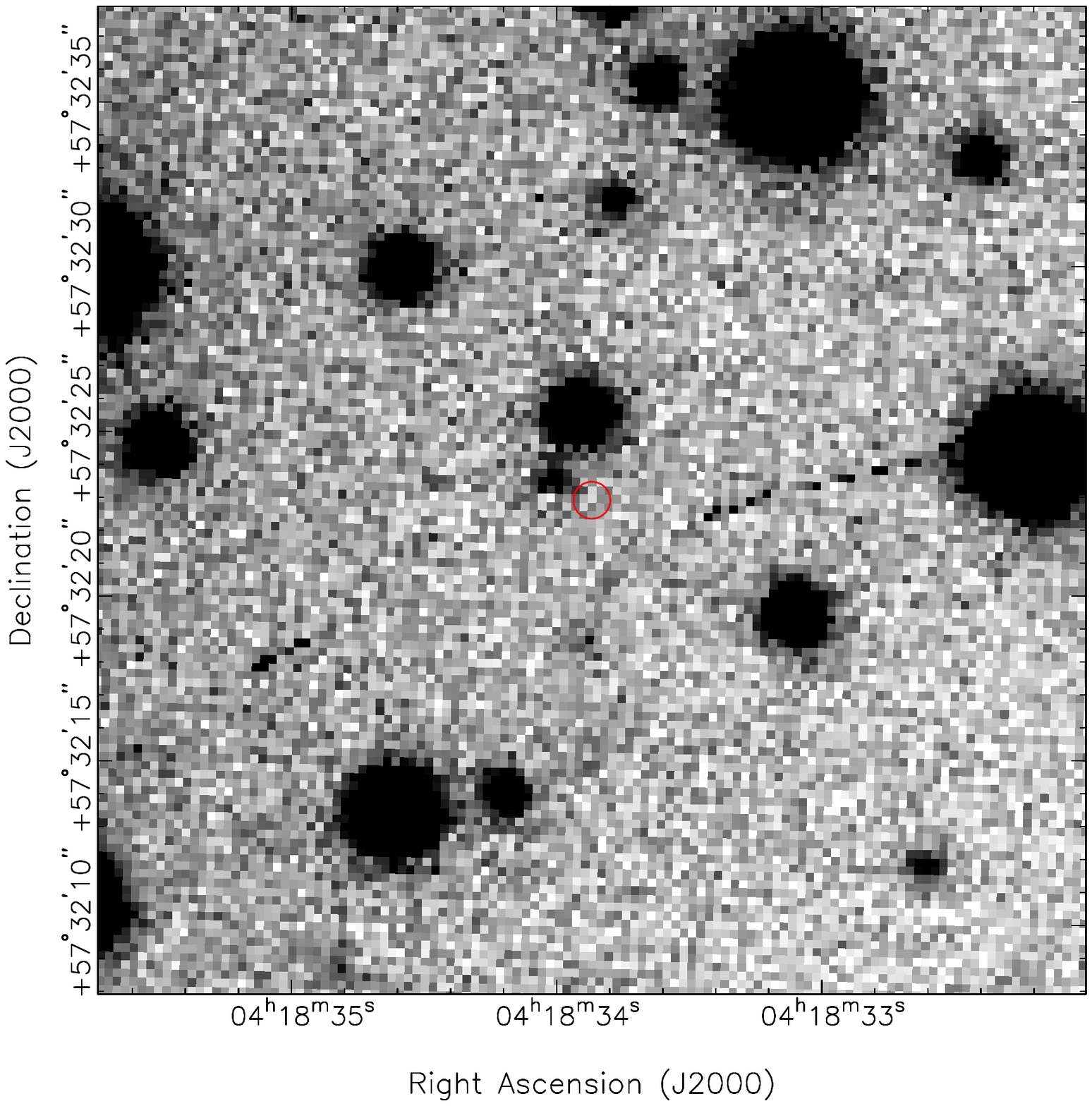}
\hspace{-0.3cm}
\includegraphics[width=6.2cm]{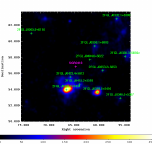}}
\caption{Left panel: Contours of the 1.8~mm Plateau de Bure emission of the field of \src. Contours are -4, -2 (dashed grey), 2, and 4 (white) times the rms noise of the map, 60 $\mu$Jy\,beam$^{-1}$, and they are over-plotted on the Wide-field Infrared Survey Explorer image at 4.6~$\mu$m. The star symbol indicates the position of \src, and its size corresponds to its positional uncertainty ($\sim1.2''$ in diameter). The synthesized beam, of $3.95''\times3.16''$, at P. A. = 0~deg, is shown in the bottom right corner. The square indicates the field of view of the $r$-map acquired with the william Herschel Telescope (central panel). Central panel: William Herschel Telescope r-band field of \src. Right panel: Fermi-LAT (0.1-100\,GeV). Diffuse subtracted TS map of the $7^{\circ} \times 7^{\circ}$ sky region centered on the magnetar position. The map is calculated for $E > 300$ MeV. The 2FGL sources are labeled in green, while \src\, in magenta.}
\label{multiband}
\end{figure*}

We did not detect continuum emission within the PdBI primary beam
towards \src\,, and obtained an upper limit of 0.24~mJy~beam$^{-1}$ at
a 4$\sigma$ level (see Figure\,\ref{multiband}).  The only detected source is at
RA=04:18:30.077, Dec=57:32:52.00, which corresponds to an offset of
($30.5''$, $29.1''$) with respect to the phase center, or a total
offset of $42.1''$. The flux density of this millimeter source is
0.34$\pm$0.06 mJy (from a Gaussian fit in the uv plane and without
correcting for the primary beam response).  In addition, we looked for
possible ``pulses'' of emission at 1.8~mm. In order to do that, we checked the
calibrated amplitude vs time for the longest track (July
15th). By averaging the visibilities in intervals of 1 minute, we found no
hints of variable emission at an upper limit of roughly $\sim10$~mJy.

We also searched the NRAO VLA Sky Survey at 21 cm and found no source within $15'$ 
of the millimeter source (Condon, et al. 1998; limiting brightness: 2.0 mJy beam$^{-1}$).


\section{William Herschel Telescope optical observations}
\label{wht}

We acquired four 300s $r$-band images of the field containing \src\,
on 2009 August 16, using the ACAM imager mounted at the 4.2 m William
Herschel Telescope on La Palma.  The average seeing was
1$^{\prime\prime}$ and airmass 1.26 . Observations of
a nearby field containing SDSS calibrated stars were obtained for the
absolute photometric calibration, while astrometry was performed
against 2MASS sources, resulting in an accuracy of $\sim$0.1" on both
RA and Dec.

No source was detected within the 95\% confidence down to a limit of
$r=24$. PSF photometry reveals that the nearest object is detected at
a magnitude $r = 22.7\pm0.1$ and center coordinates RA=04:18:34.0,
Dec=57:32:23.5. This source is consistent with the near-infrared
source reported by Wachter et al. (2009), but its distance from the
\src\, position ($\sim$1.4"), makes the association with the magnetar
rather unlikely.

\section{Fermi-LAT gamma-ray observations}
\label{fermi}

We used data from the Large Area Telescope (LAT) onboard {\em Fermi}
(Atwood et al. 2009) from 2008 August 4 until 2012 October 24. The
Fermi Science Tools {\tt SC09-28-00} package is used to analyze the
data. We selected events from the ``{\tt Source}`` class of the 
``{\tt P7.6\_P130\_BASE}'' data version within a circular region of
interest (ROI) of 10$^{\circ}$ radius centered on the position of \src,,
and in the energy range 100 MeV--100 GeV.  The good time intervals
are defined so that the ROI does not fall below the gamma-ray-bright
Earth limb (defined at 100$^{\circ}$ from the Zenith angle), and 
the source is always inside the LAT field of view, namely in a cone
angle of 66$^{\circ}$. The ``{\tt P7SOURCE\_V6}'' instrument response
functions (IRFs) are applied in the analysis.

The likelihood analysis of \src\, was performed by means of the binned
maximum-likelihood method (Mattox et al. 1996), using the official tool
{\tt gtlike} released by the \fermi-LAT collaboration. The
spectral-spatial model created for the likelihood analysis includes
the Galactic, and the Isotropic diffuse emission models, as well as
all the 2FGL sources within a radius of $15^{\circ}$ from \src. Since
there is no 2FGL source that is positionally associated to the
magnetar, we added in the spectra-spatial model a point-like source
modeled with a simple power-law with the coordinates of \src .  The
2FGL sources within 3$^{\circ}$ of \src\, (3 sources) are modeled with
the flux parameter allowed to vary, while the others 34 sources had
all their parameters fixed to the value from the 2FGL catalog.
Figure\,\ref{multiband} shows the diffuse subtracted TS map of the
$7^{\circ} \times 7^{\circ}$ region centered on \src. It was obtained
associating to each pixel (of size $0.1^{\circ} \times 0.1^{\circ}$)
the TS value calculated assuming a point-like testing source in its
center. Diffuse subtracted TS map means that the spectral-spatial
model for the null hypothesis includes only the Galactic and Isotropic
emission models, so that the point-like sources should be visible in
the map.

As no significant gamma-ray counterpart to \src\, is identified, 95\%
flux upper limit is derived using the Bayesian method developed by
Abdo et al. (2010). The 95\% flux upper limit
for $E>100$\,MeV is $F < 1.3 \times 10^{-8}$
photons~cm$^{-2}$~s$^{-1}$, including systematics.

The non detection of \src\, at energies $>100$ MeV is not surprising,
given a similar non detection of all other known magnetars (Abdo et
al. 2010).

\begin{figure*}
\hbox{
\includegraphics[width=9cm,height=7cm]{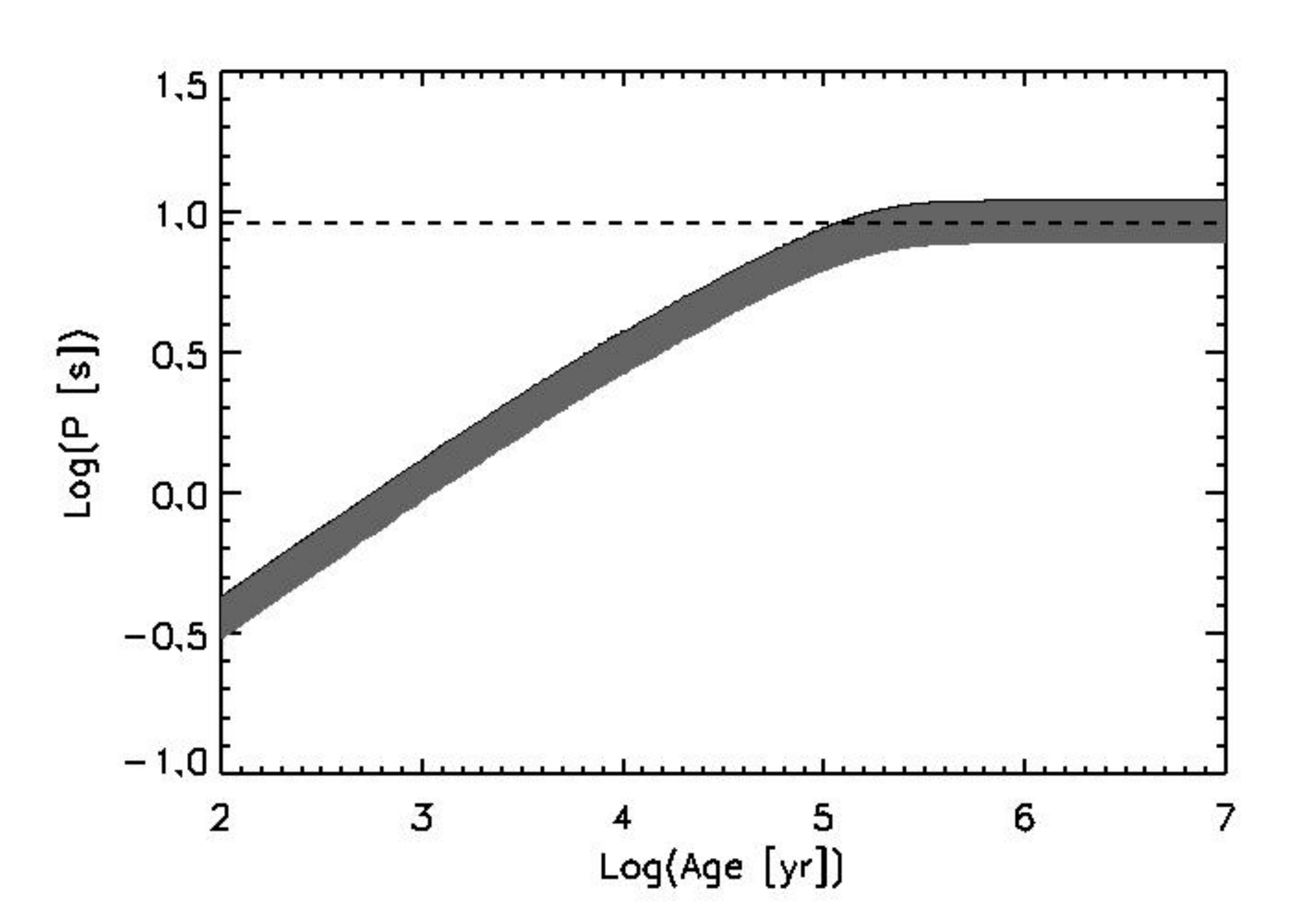}
\includegraphics[width=9cm,height=7cm]{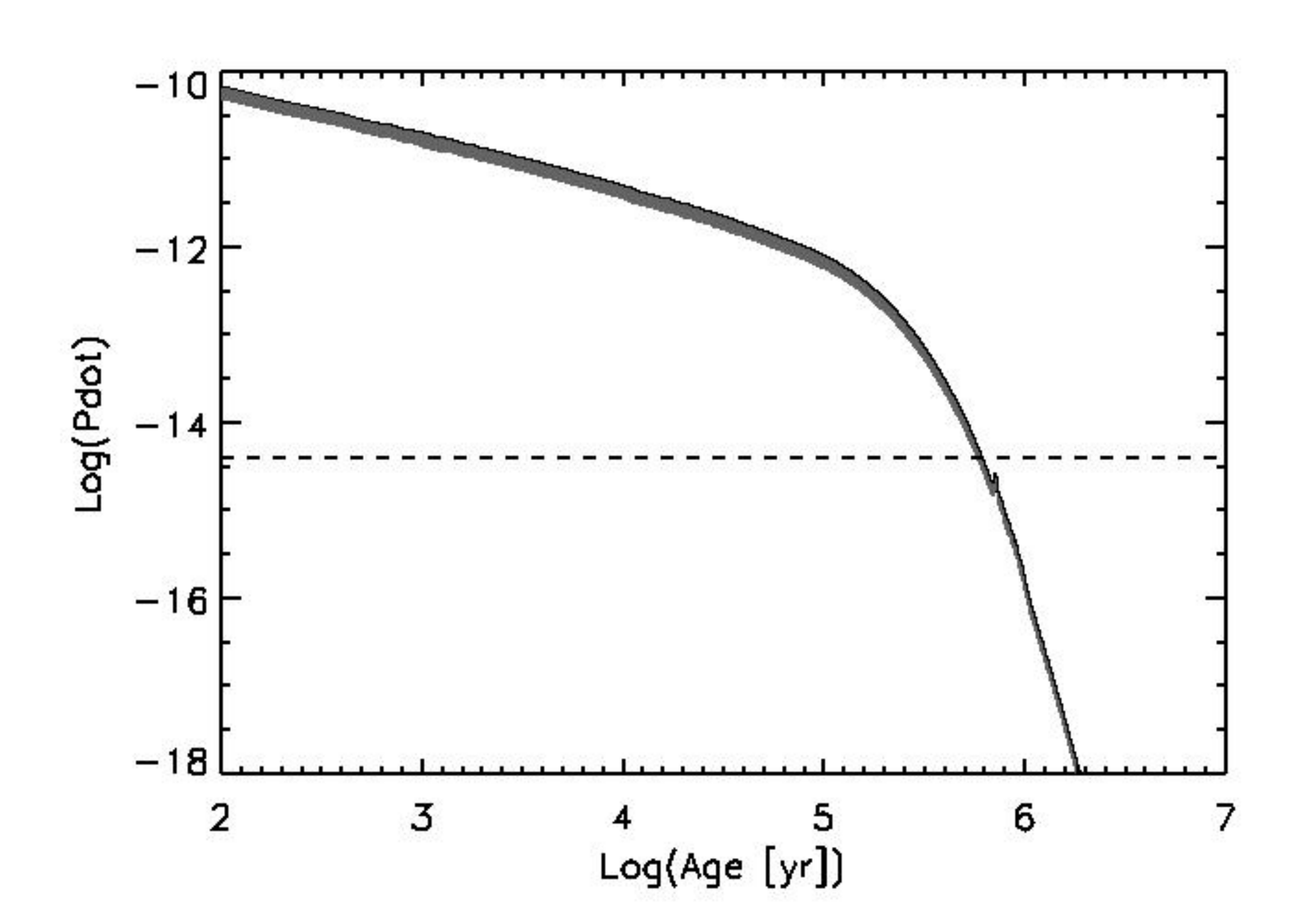}}
\hbox{
\includegraphics[width=9cm,height=7cm]{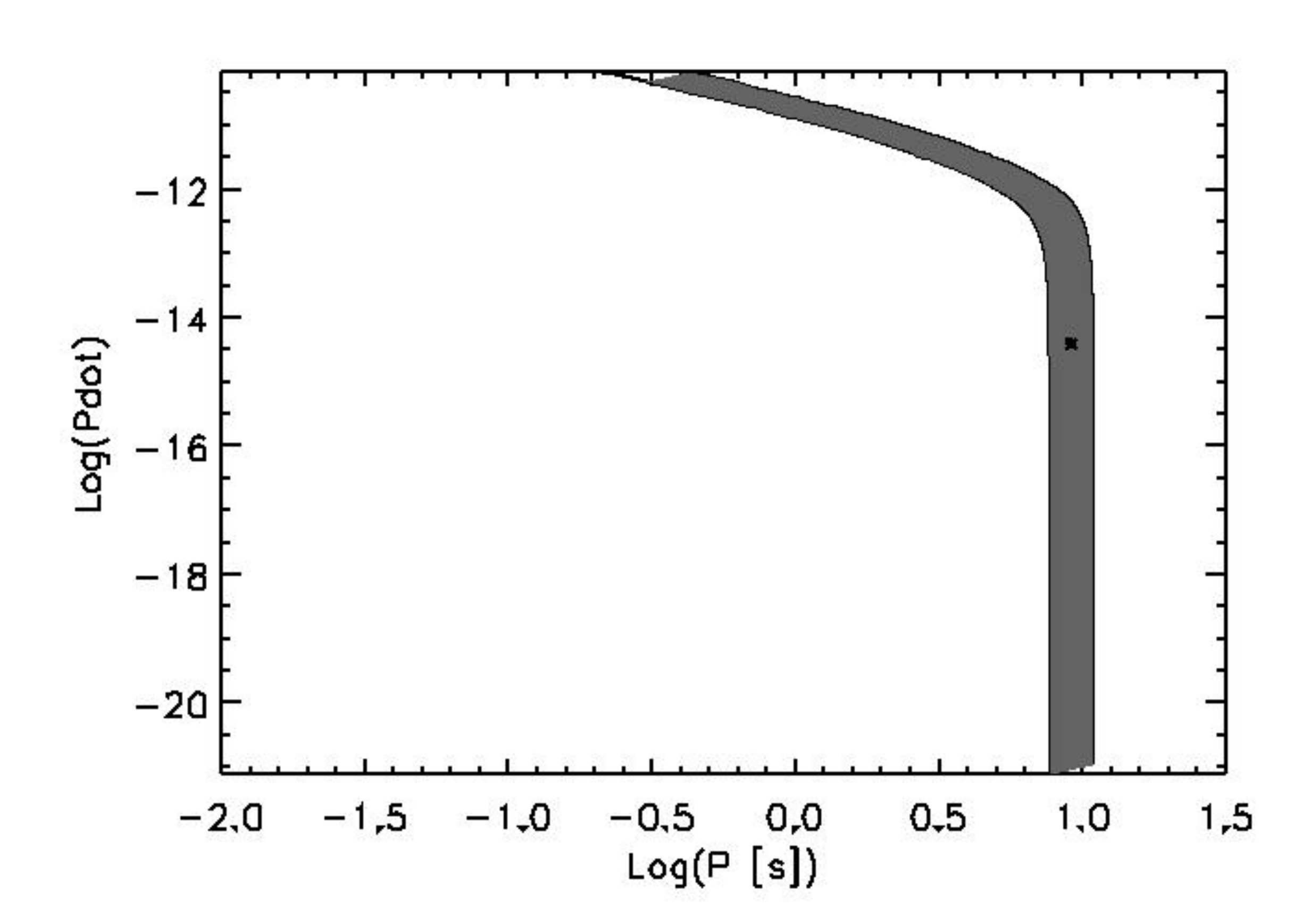}
\includegraphics[width=9cm,height=7cm]{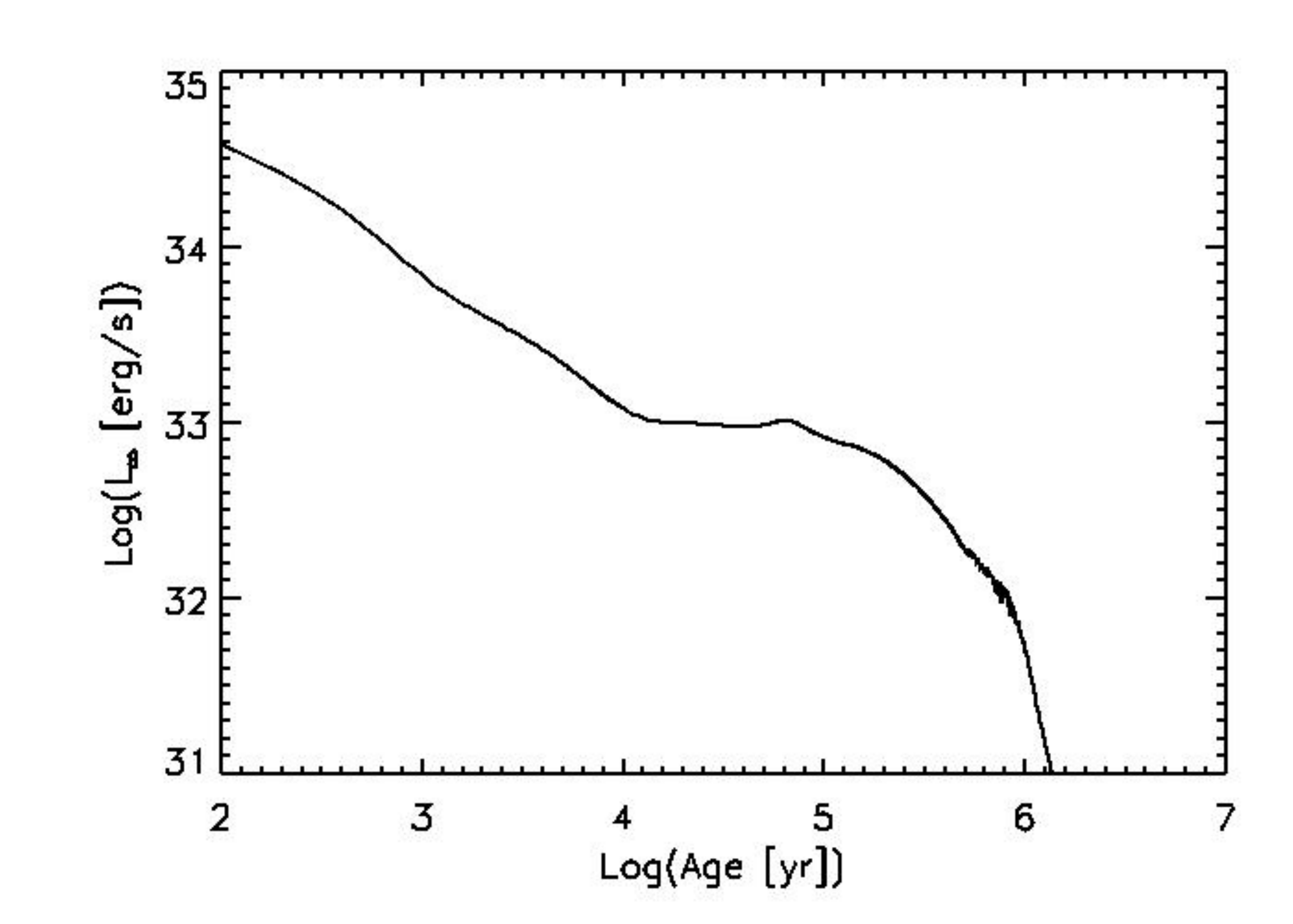}}
\caption{Magneto-thermal evolution of a neutron star with an initial poloidal
field of $B_{\rm dip}=1.5\times 10^{14}$ G: period (left-top), period
derivative (right-top), evolution in the P-$\dot{P}$ diagram (left-bottom), and bolometric thermal luminosity (right-bottom).
The gray band corresponds to the uncertainty of the
angle-dependent spin-down formula.}
\label{figure:mt}
\end{figure*}

\section{Discussion}
\label{discussion}

We have presented here a detailed X-ray study of the outburst of the
low magnetic field soft gamma repeater \src. The long term monitoring
we performed over 1200\,days allowed us to measure the period
derivative of this pulsar ($\dot{P}=4(1)\times 10^{-15} s\,s^{-1}$)
with a 3.5$\sigma$ significance (Figure\,\ref{timing}). This yields
an estimate of its dipolar magnetic field of $B_{\rm
  dip}\sim6\times10^{12}$\,G , and confirms this object as the 
magnetar with the lowest dipolar magnetic field  ever discovered.

Assuming that \src\ attained its quiescent state in the last few
observations, the X-ray quiescent emission appears dominated by a very
small spot at $kT\sim 0.3$~keV and of radius $\sim$0.16~km (assuming a
2\,kpc distance), which corresponds to a cap of semi-aperture $\sim
1^\circ$--$2^\circ$, similar to what observed in old radio pulsars.
However, in the present case the rotation power ($\dot{E}\approx
10^{29}$\ergs) is about two orders of magnitude smaller than the
observed X-ray luminosity ($\approx 10^{31}$\ergs), thus indicating a
different origin for the emission, most likely magnetic. Actually, it
is quite likely that most of the surface is at a much lower
temperature and is therefore invisible at energies between
0.5--10\,keV. This implies that the quiescent bolometric flux may be
severely underestimated (see also below).

The study of the spectral evolution during the outburst shows the
presence of a non-thermal component (probably magnetospherical)
at the beginning of the outburst, which fades
away after a few hundreds days. On the other hand, the temperature of
the small region responsible for the surface anisotropy fades from 0.9
to 0.3\,keV in a few year timescale (see Table\,1 and
Figure\,\ref{parameters}).

The pulse profile evolution during the outburst decay shows some
interesting features. As Figure\,\ref{profiles} shows, there is an
overall trend towards a simplification of the pulse, which starts with
a complex, three-peaked shape and ends with a fairly sinusoidal
pattern. Furthermore, the study of the pulse profiles as a function of
energy (Figure\,\ref{profilesenergy}) in the first outburst stages,
shows a great variability too. A large dip in an otherwise rather
sinusoidal profile is observed at energies between 1--4\,keV.

The large pulsed fraction of 40--70\% (Figure\,\ref{timing}), the
evidence of nearly phase aligned spots responsible for the 0.9\,keV
thermal emission in the early outburst phases, as well as the 0.3\,keV
emission at late times, disfavor the presence of two spots at
different temperatures, while favoring the presence of a single tiny
spot cooling down (from 0.9 to 0.3\,keV) and reducing its size (from
0.21 to 0.16\,km) during the return to quiescence. This means that the
multi-peaked pulse profile is probably due to anisotropies in the
magnetospheric electrons distribution (on top of a non-isotropic surface
thermal emission).

\subsection{\src\ as an evolved magnetar}
\label{evolution}

In \cite{turolla11} it was shown that the rotational properties of \src\, can be
reproduced if the source is an aged magnetar,
which experienced substantial field decay but still retains a
strong enough internal toroidal field. The most updated
magneto-thermal evolutionary models discussed in (Vigan\`o et al. 2013; but see also Pons et al. 2009, and Aguilera et al. 2008), confirm
this scenario. The evolution of an initial dipolar magnetic field of $B_{\rm dip}^0\sim
1.5\times
10^{14}$\,G (surface value at the pole) correctly provides the observed $P$ and
$\dot P$
at an age of $\sim 550$\,kyr, which is probably the real age of this source.

Although different combinations of the initial components of the
  magnetic field are possible, in all the models the magnetic field
  must have been large in the past ($\ga 10^{14}$ G) to explain at the
  same time the long spin period, the bright X-ray emission at this
  old age, and the flaring activity of the source. The
characteristic age overestimates the real one by almost two orders of
magnitude. In Figure~\ref{figure:mt}, we show the evolution of period,
period derivative, the source track in the P--$\dot{P}$ diagram, and
the bolometric thermal luminosity. In this scenario we estimate that
\src's mean surface temperature should be now of $\sim$0.05\,keV,
unfortunately undetectable by current X-ray observations (which are
observing only a hot tiny region on the star surface).

For the evolution of the timing properties, we assume the
magneto-dipole braking formula given by Spitkovsky (2006): $I\Omega \dot{\Omega}
\approx \frac{B_d^2R^6\Omega^4}{4 c^3} (1+\sin{\chi}^2)$, where $R$ is the NS
radius, $\chi$ is the angle between the rotational and the magnetic axis, $c$ is
the speed of light, 
$\Omega=2 \pi/P$ is the angular velocity, and $I$ is the moment of
inertia of the star. 

An alternative possibility is that the neutron
star was born with an external magnetic field close to the
present one, but its large core poloidal field
slowly diffuses out i.e. by ambipolar diffusion (Soni 2012).
Although from the timing properties alone it
is hard to discriminate between a hidden strong
crustal magnetic field, a hidden strong core field 
or an intrinsically low-B neutron
star, the magnitude of the X--ray luminosity, and the
spectral properties and light curves may be used
to distinguish the different scenarios. 

In particular, the low magnetic field scenario cannot explain the
high luminosity, large pulsed fraction, and the
flaring activity of the source. As a matter of fact,
if there is little field decay and the real age 
corresponds to the characteristic age (which in this
scenario would be needed to reach the present
period of 9\,s), no existing non-magnetic 
cooling model can account for an X-ray luminosity of
$\approx 10^{31}$ erg s$^{-1}$ at a characteristic
age of $\approx 35$ Myrs.

The scenario in which a large core field diffuses out has the same
problem: if the real age of the star is similar to its old characteristic
age, no cooling model in the literature predicts such high 
quiescent luminosity. In addition, while the timescales used in Soni (2012) are correct for normal, 
non-superfluid nuclear matter, recent work (Glampedakis, Jones \& Samuelsson 2011, MNRAS) shows that 
in the presence of superfluidity in the neutron star core,  the timescales for ambipolar diffusion are
many orders of magnitude longer, and therefore ambipolar diffusion does not
play any role during the active age of the star.

\begin{figure*}
\hbox{
\includegraphics[width=8cm]{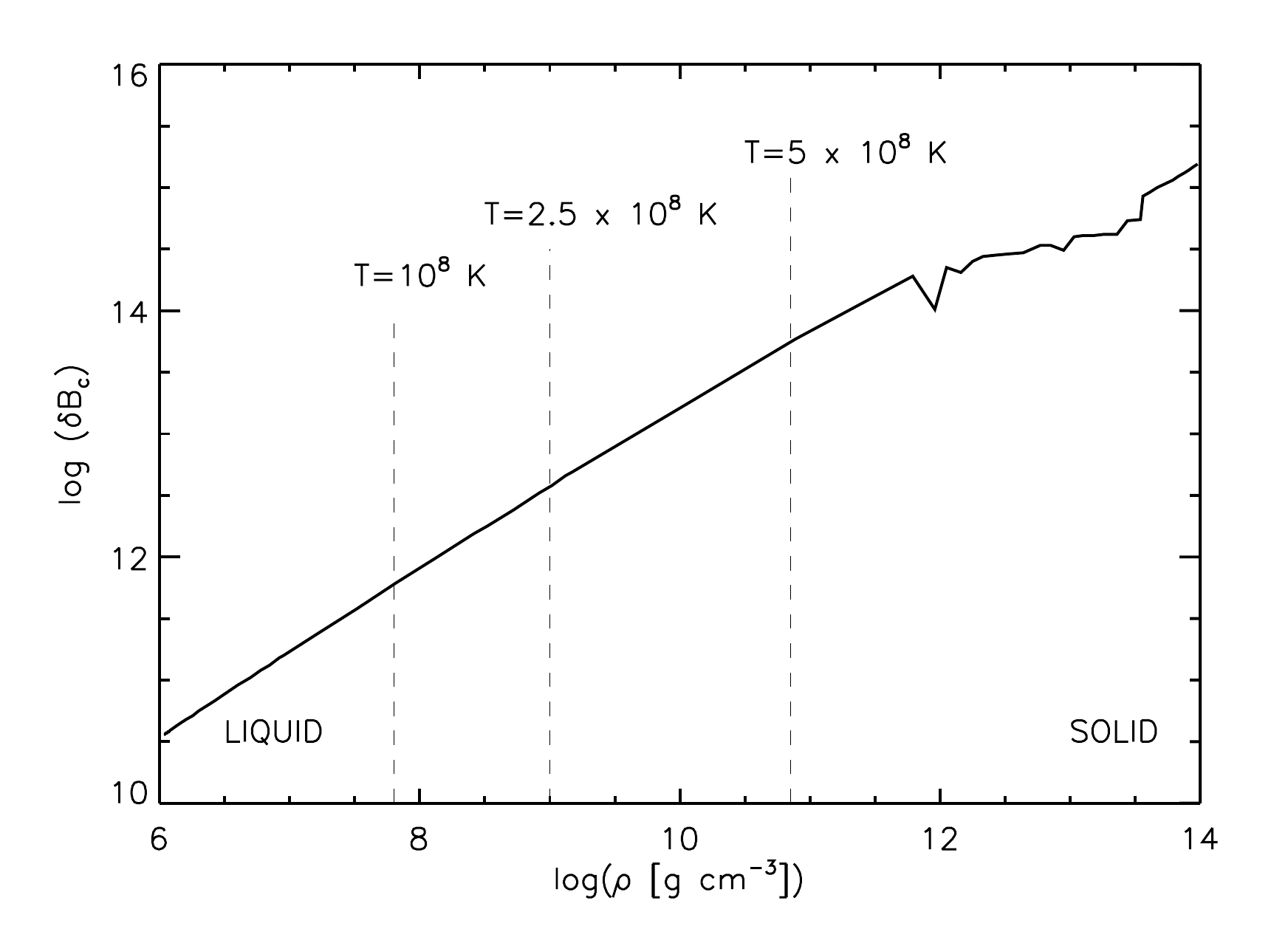}
\hspace{1cm}
\includegraphics[width=8cm]{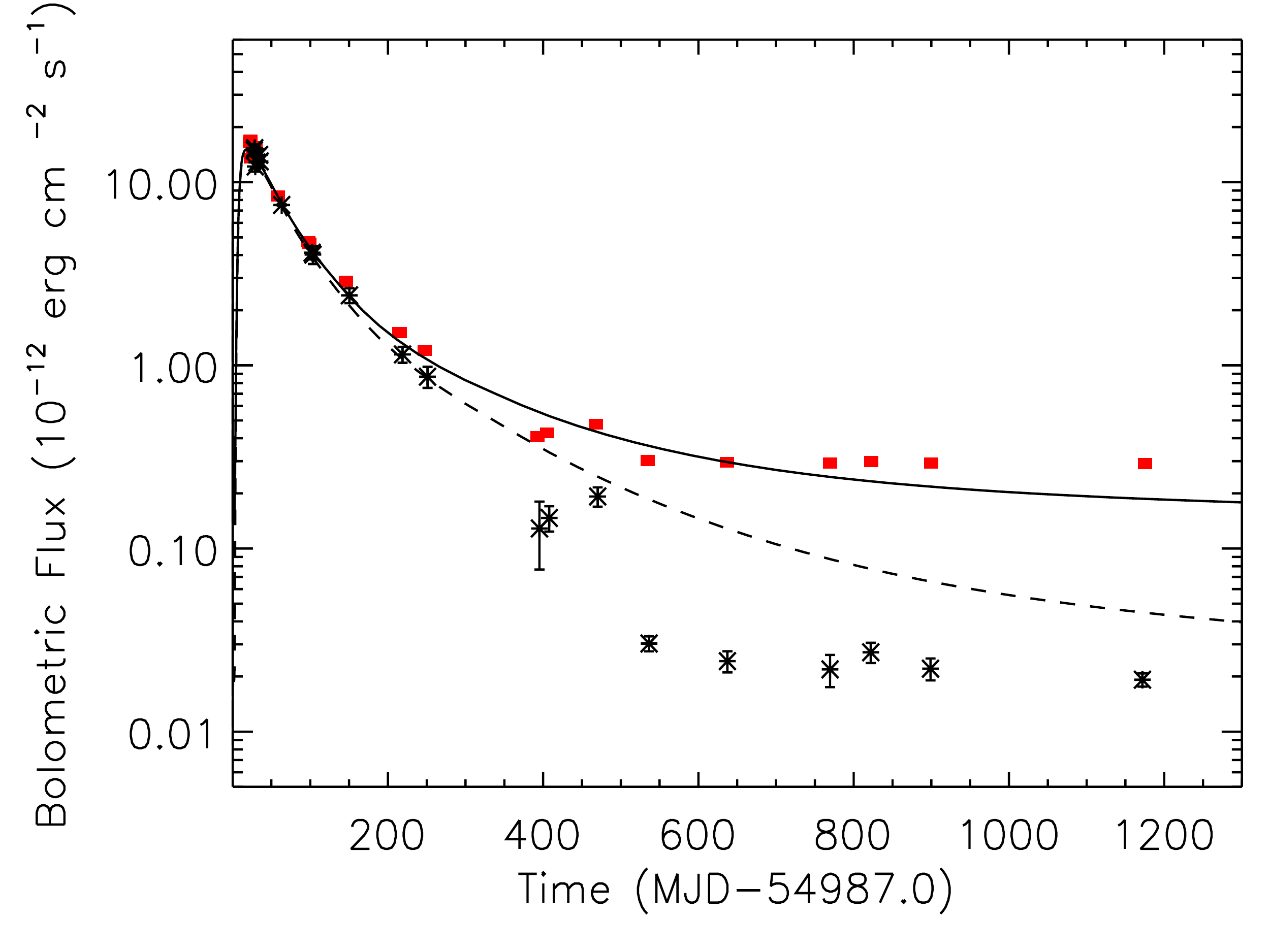}}
\caption{{\em Left panel}: Minimum variation of the magnetic field
  required to break the crust by magnetic stresses as a function of
  density. The vertical dashed lines delimit the transition from solid
  to liquid for three different temperatures. {\em Right panel}:
  Outburst modeling following Pons \& Rea (2012). Black data are the 0.5--10\,keV unabsorbed flux, while red squares are the bolometric unabsorbed flux with the addition of the flux of a thermal component at kT$=$0.05\,keV from the entire neutron star. Solid and dashed
  lines refer to the outburst model for the bolometric and 0.5--10\,keV thermal flux, 
  repectively (see text for details)}.
\label{bcrit}
\label{outburstmodel}
\end{figure*}

\subsection{\src\, outburst rate}
\label{rate}

An important question is whether a relatively low dipolar field is
consistent with the star-quake model in which the primary cause of
the outburst is an internal deposition of energy following a crust
fracture. It is often overlooked that the magnetic stress needed
to break the crust is strongly dependent on density (it is much
easier to break the outer crust than the inner crust) and that the
crust thickness grows as the temperature drops with age. In
Figure~\ref{bcrit} we show an estimate of the minimum
magnetic field variation required to induce a fracture. 
As assumed in \cite{perna11}, this estimate is obtained assuming that the crust
moves through a series of equilibrium states in which its elastic
stress balances the (time-dependent) magnetic stress. The
deviation of the magnetic field with respect to the last
unstressed configuration ($\delta B_c$) may be large enough to
break the crust when
\begin{equation}
\delta B_c \approx \left( 4 \pi  \sigma_b^{\rm max} \right)^{1/2}
\end{equation}
where $\sigma_b^{\rm max}$ is the maximum stress that a neutron star  crust
can sustain (Chu et al. 2010). For young, relatively hot magnetars
(crustal temperatures of $5\times10^8$\,K), only the inner crust is
solid, and strong field variations $\delta B_c \gtrsim 10^{14}$\,G
are required to fracture the crust. However, for old, cold neutron
stars (crustal temperatures of $10^8$\,K), the solid crust extends
down to $10^8$~g/cm$^3$, and is much easier to break, even with
variations of the magnetic field of the order of $\delta B_c
\gtrsim 10^{12}$\,G. Note also that fractures close to the inner
crust are much more energetic (because of both the higher available
elastic energy and larger volume involved) than fractures in the
low density region. In the first case, one can reach up to
$10^{44}$~erg, while in the second case, events of $\approx 10^{41}$~erg
are expected. A rough prediction of the expected outburst rate
\citep{perna11,pons11} for the solution model mentioned above
gives $\lesssim 10^{-3}$ star-quakes/yr for an object as \src. Assuming that there are
about $10^4$ neutron stars in the Galaxy with similar age, and
that a (very approximatively) 10\% of them are born as magnetars, a naive extrapolation
of this event rate to the whole neutron star population leads to
an expected low-$B$ magnetar outburst rate of $\lesssim 1$ per
year. Therefore, we expect that more and more objects of this
class will be discovered in the upcoming
years (as e.g. Swift~1822.3--1606; Rea et al. 2012; Scholz et al. 2012).

\subsection{\src\, outburst decay}
\label{rate}

By modeling the flux evolution in time we tested if the crustal
cooling model presented in \cite{pons12} can fit the flux decay of
\src , on the wave of what done for Swift~J1822.3--1606 (Rea et
al. 2012). We assume a dipolar field of $6 \times 10^{12}$ G
(equatorial), an internal toroidal field of $10^{14}$\,G (at maximum)
as inferred in \S\ref{evolution}, and an average surface temperature
of 0.05\,keV, which is the temperature we expect for the surface of
such an old magnetar (note that in the 0.5--10\,keV band we are only
seeing a tiny hot spot). The best modeling was found by injecting $2.5
\times 10^{26}$~erg~cm$^{-3}$ in a thin layer in the outer crust
between 4.5$\times10^9$ and $10^{10}$\,g~cm$^{-3}$, and in the region
contained within a cone with axis in the direction of the magnetic
pole, and aperture a$\approx 0.4$~rad, for a total energy deposition
of $2.5\times 10^{41}$~erg (compatible with typical magnetar
outbursts; Pons \& Rea 2012). The evolution of the bolometric, and of
the 0.5--10 keV flux is shown in Figure\,\ref{outburstmodel} (solid
and dashed lines, respectively). We have shown with red squares
  how the observed flux decay would appear when adding the
  contribution of a blackbody component at 0.05\,keV, mimicking the
  entire neutron star surface. It is clear that crustal cooling can
  easily explain the decay only if this further component is taken
  into account. In particular, the solid line is fitting the red points because the entire neutron star surface is taken into account, while it is not in the observed black data, which in fact cannot be reproduced by the dashed line. This is indicative of the difficulty of comparing theoretical cooling curves with data obtained in a certain energy band.
We also note that no theoretical model predicts a surface temperature
as high as 0.3~keV on a timescale of years, unless a continuous energy
release is assumed (e.g. by long-lived internal currents). We finally
mention that all the previous considerations are based on the
(implicit) assumption that the blackbody temperature is a measure of
the physical temperature of the emitting region. If this turned out
not to be the case, e.g. because the spectrum is thermal but not
Planckian so that a color correction is required, the physical surface
temperature may be smaller than the measured blackbody
temperature.

An alternative model to the crustal cooling scenario consists in the
presence of currents flowing into the magnetosphere through a
gradually shrinking magnetic bundle heating the neutron star surface
from the top.  In particular, the deepest available \xmm\ observation
of \src, performed two months after the outburst onset, reveled that
the 0.5--10 keV spectrum of \src\ is best reproduced by a blackbody
component plus an additional non-thermal component, or by a resonant
cyclotron scattering model. This suggests the presence of twisted
magnetic field lines, at least in the first outburst
stages. Furthermore, the limited spatial extent of the heated region
($< 1$\,km) is suggestive of a scenario in which the twist is confined
within a small part of the magnetosphere, a thin current-carrying bundle,
or j-bundle \cite[][]{beloborodov09}.  As the j-bundle untwists during
the outburst decay, the spectrum becomes more and more blackbody-like,
as indeed observed. 

\begin{figure}
\vbox{
\includegraphics[width=8cm,height=5cm]{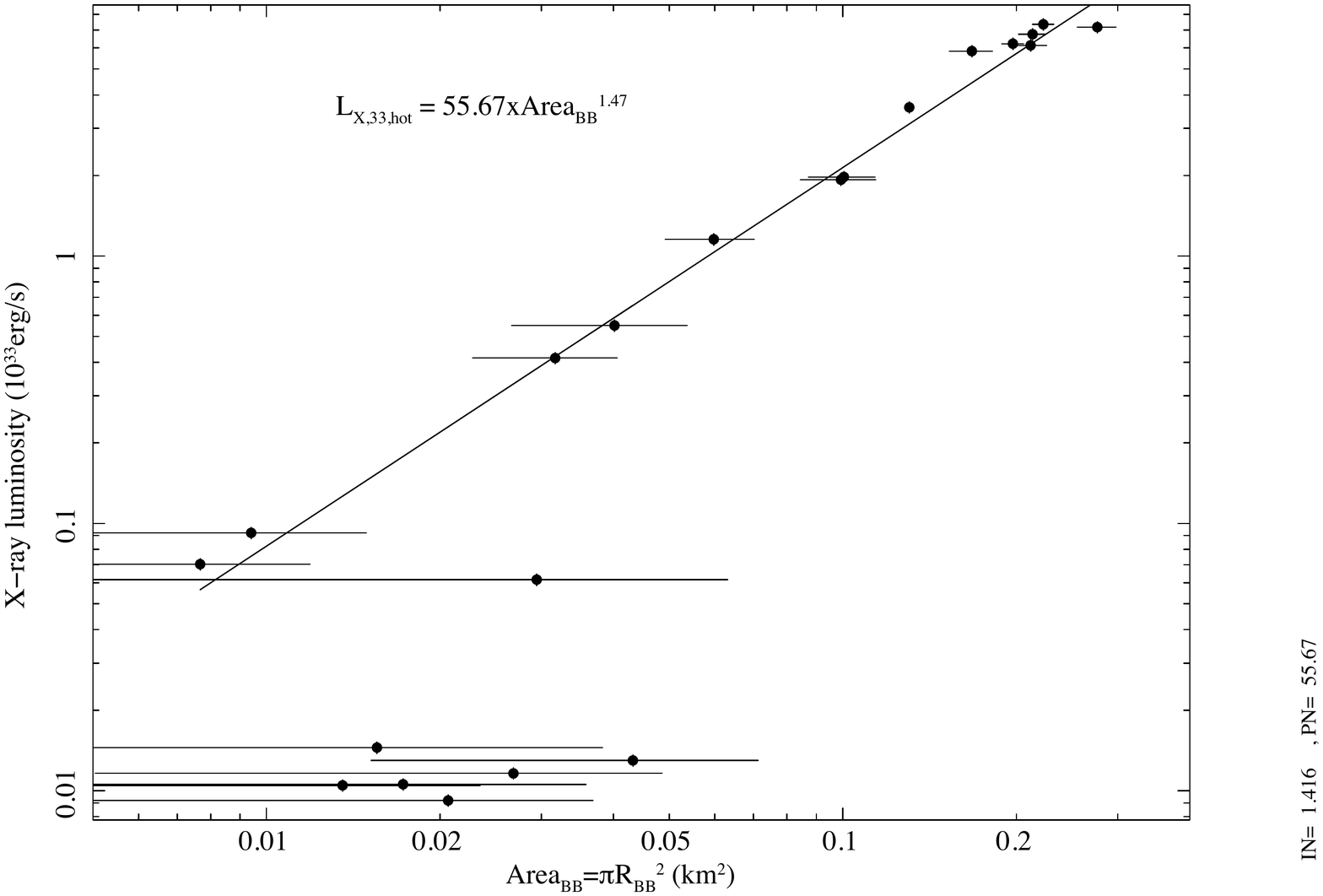}
\includegraphics[width=8cm,height=5cm]{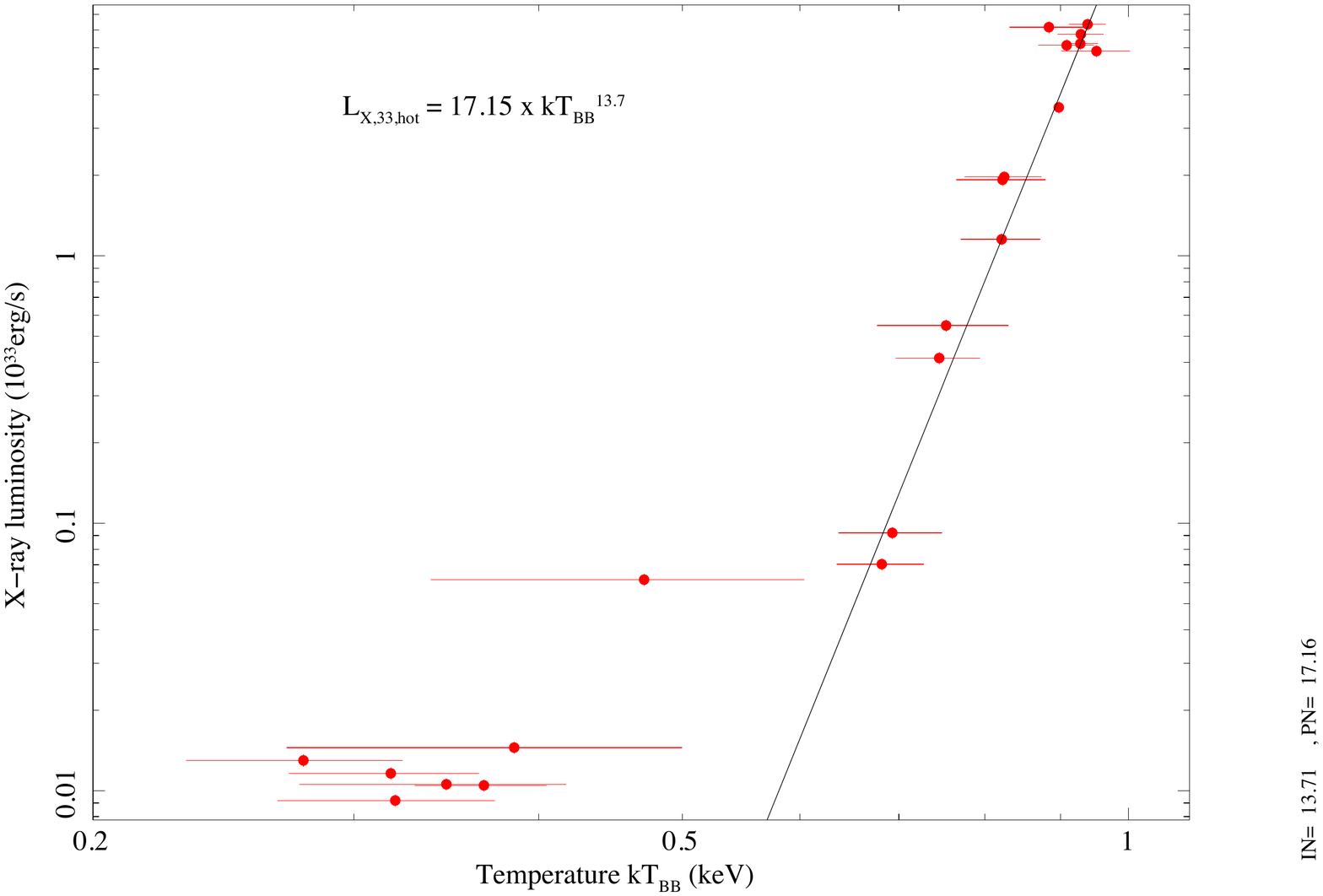}}
\caption{X-ray luminosity evolution as a function of the blackbody emitting area and temperature (see also Table\,1; we assume a 2\,kpc distance).}
\label{luminosity}
\end{figure}

Resonant cyclotron scattering from a thin, decaying j-bundle appears
also capable of explaining (qualitatively) the spectrum and its evolution
during the first outburst stages, but whether it can explain the
double-peaked pulse profiles is unclear. However, this scenario has
some further difficulties: 1) the total luminosity produced by
currents in the bundle is, for such a low-$B$ and a small thermal spot,
well below the one observed at early times, at least if the spot is at
the polar cap \cite[see again][and also
  \citealt{turolla11}]{beloborodov09}; 2) the timescale for the twist
decay is much shorter ($< 1$~yr) than what implied by the long
outburst of \src , and 3) an approximate relation between the emitting
area and the luminosity exists ($A\sim L^2$; Beloborodov 2009) if most
of the luminosity is produced by current dissipation, but \src\ data
show a somewhat flatter dependence when the first stages of the
outburst are fitted (see Figure\,\ref{luminosity}).

In summary, the crustal cooling model, when including also a
  possible hidden contribution from the entire neutron star surface, appears
  favourable in explaining the outburst decay of \src . However, it is
  likely that a combination of crustal cooling and magnetospheric
  untwisting bundle can be operating at the first stages of the
  outburst.

\subsection{Constraints on the presence of a fossil disk surrounding SGR~0418+5729}

A fallback disk around \src\, was suggested by Alpar et al
(2011) as a way to aid the spin-down of the pulsar and explain the 9\,s periodicity of this source.
As an alternative, our results show (see Figure\,\ref{evolution}) that both the thermal and the timing properties of this
source can be reproduced for a pulsar age of $\sim 550$~kyr by
properly accounting for magnetic field evolution and dissipation,
which also imply that the neutron star was born with a much higher dipolar field
than the one measured today (see \S\ref{evolution}). Hence, in principle, the timing properties of this
source would not necessarily require an additional spin down torque by a disk.
However, given the suggestion that fallback disks around isolated neutron stars might be common
(Michel 1988; Chevalier 1989; Lin et al. 1991), it is worthwhile to
use the current multi-band upper limits to set constraints on the presence of a fallback
disk around \src.

Since the pulsar is currently spinning down, any disk-magnetosphere interaction must probably occur in
the propeller regime, with the pulsar transferring angular momentum to
the disk. For this condition to be satisfied, the inner boundary of
the disk, located at about the magnetospheric radius $R_m=2.5\times
10^8[\dot{M}/(10^{16}\;{\rm g}{\rm s}^{-1})]^{-2/7} (M_{\rm
NS}/M_\odot)^{-1/7}[B/(10^{12}{\rm G})]^{4/7}$, must be equal to or
larger than the corotation radius $R_{\rm co}=(GM_{\rm
NS})^{1/3}\Omega^{2/3}$. The strongest constraint on the disk emission
is obtained when the inner radius of the disk obtains its minimum
value, i.e. $R_{\rm in}=R_m=R_{\rm co}$. For the outer radius, we
assume $R_{\rm out}=10^{14}$~cm.  We found that larger values do not
result in appreciably larger emission at the frequency of interest,
and hence this value allows to set the tightest constraint on the disk 
emission.


With the inner and outer disk radii fixed as discussed above, the
emission spectra from the fallback disk is computed using the model of
Perna et al. (2000). The disk is assumed to be optically thick and
geometrically thin, and the anisotropy in the X-ray luminosity from
the source (which irradiates the disk) is neglected, since it is found
to be of second order (Perna \& Hernquist 2000). The disk is assumed
to be still "active", i.e. viscously accreting (see Menou et al. 2001). The disk emission is the result of both viscous
dissipation and re-radiation of the pulsar X-ray luminosity.  In order
for the magnetospheric radius not to exceed the corotation radius, the
accretion rate must be limited to $\dot{M}\lsim
10^{15}$~g~s$^{-1}$. With this value, the disk luminosity in the {\it mm}
band is dominated by reprocessing of the pulsar X-ray luminosity.
At 166~GHz, the predicted flux is about 0.01~mJy for a face-on disk, below
the measured limit of 0.24~mJy. Hence the presence of a fossil disk cannot
be ruled out by the current {\it mm} measurements. Even adding the contribution from the whole surface of the star by
a putative thermal component at 0.05~keV, would bring the predicted
{\it mm} flux just around the measured flux limit for a face-on disk.  

The field around \src\, was also observed with the Grantecan and  {\em Hubble}
telescopes (Esposito et al. 2010; Durant et al. 2011). In particular, the latter observations were performed in two wide filters, the optical, with
a pivot wavelength of 5921~\AA, and in the NIR, with pivot wavelength
of 11534~\AA.  The source was not detected down to the flux of $f_O < 2.3 \times 10^{-31}$~erg~s$^{-1}$~cm$^{-2}$~Hz$^{-1}$ and $f_{NIR} < 4.4 \times 10^{-31}$~erg~s$^{-1}$~cm$^{-2}$~Hz$^{-1}$, respectively.
We found that this optical limit (nor the Grantecan or WHT limit) is not sufficiently constraining
for a disk with the properties described above (the predicted emission
for a face-on disk is about a factor of four below the limit).
On the other hand, in the NIR, the observational limit is already able
to rule out a face-on disk, which would yield an emission about
twice larger than the measured flux limit. However, a disk inclined with respect to the observer by an angle $\cos\theta\lsim 0.5$ would still be allowed by the observations (although falling short in explaining the X-ray bursts of this object).

\subsection{Conclusions}

At the time of writing, in the ATNF pulsar catalogue (Manchester et
al. 2005) 138 isolated radio pulsars have a dipolar magnetic field
larger than that inferred for \src. Our results imply that some of
these objects might hide a strong toroidal component of the internal
field, not measurable via the pulsar timing properties. A hint for
such strong fields might be a high surface temperature, hotter than
what would be predicted by standard cooling models at the pulsar
age. However, only a few of those pulsars have had dedicated X-ray
observations, and the shallow surveys do not suffice to detect such
emission (expected to be as luminous as $L_{X}\sim10^{31}$\ergs
). Furthermore, our calculation of the outburst rate of a low magnetic
field magnetar also suggests that roughly once a year a quiet
neutron star might turn on with magnetar-like activity.

On the other hand, if indeed a large number of neutron stars is hiding
a strong magnetic field component, there would be important
consequences also for other branches of astrophysics. In particular,
it would imply that supernova explosions should generally produce
strong magnetic fields, and that most massive stars are either
producing fast rotating cores during the explosion to activate a
dynamo, or are strongly magnetized themselves. Furthermore, in this
scenario a non-negligible fraction of gamma-ray bursts might be due to
the formation of magnetars, and the gravitational wave background
produced by magnetar births should then be larger than predicted so
far (important for future instruments as Advanced-LIGO).

\acknowledgements  We are indebted to the \chandra, \swift, \xmm, GBT, WHT and PdBI support teams, with a special thank to Michael Bremer, for the extraordinary job in planning the PdB observations presented in this paper. We thank S. Mereghetti and H. Tong for their valuable comments on the manuscript. NR is supported by a Ramon y Cajal Research Fellowship, and by grants AYA2009-07391, AYA2012-39303, SGR2009-811, TW2010005 and iLINK 2011-0303.  JAP and DV acknowledge support from the the grants AYA 2010-21097-C03-02 and Prometeo/2009/103. RT and SM are partially funded through  an INAF 2011 PRIN grant. A.P. is supported by a JAE-Doc CSIC fellowship co-funded with the European Social Fund under the program `Junta para la Ampliaci\'on de Estudios', by the Spanish MICINN grant AYA2011-30228-C03-02 (co-funded with FEDER funds), and by the AGAUR grant 2009SGR1172 (Catalonia).


\end{document}